\documentclass[reprint,amsmath,amssymb,aps,superscriptaddress]{revtex4-1}
\usepackage{graphicx}
\usepackage{dcolumn}
\usepackage{bm}
\usepackage{lmodern}
\usepackage{mathtools}
\usepackage{dsfont}
\usepackage{mathrsfs}
\usepackage{amsthm,amsfonts,mathtools}
\usepackage{mathbbol}
\usepackage{hyperref}
\usepackage{xcolor}
\usepackage{amsmath}
\usepackage[normalem]{ulem}
\usepackage{tcolorbox}
\tcbuselibrary{skins, breakable}

\newcommand{\Tr}{\mathrm{Tr}}
\newcommand{\Id}{\mathds{I}}

\theoremstyle{plain}

\theoremstyle{definition}
\newtheorem{defn}{Definition}

\newcommand{\ie}{\textit{i.e.,\ }}
\newcommand{\eg}{\textit{e.g.,\ }}

\renewcommand{\phi}{\varphi}

\begin{document}

\preprint{APS/123-QED}

\title{A unified framework for data-driven Bell inequalities}

\author{C. E. Lopetegui-González}\email{carlos-ernesto.lopetegui-gonzalez@lkb.upmc.fr}
\affiliation{Laboratoire Kastler Brossel, Sorbonne Université, CNRS, ENS-Université PSL, Collège de France, 4 place Jussieu, Paris F-75252, France}
\author{G. Massé}
\affiliation{Sorbonne Université, CNRS, LIP6}
\author{E. Oudot}
\affiliation{Sorbonne Université, CNRS, LIP6}
\author{U. I. Meyer}
\affiliation{Sorbonne Université, CNRS, LIP6}
\author{F.~Centrone}
\affiliation{ICFO--Institut de Ciencies Fotoniques, The Barcelona Institute of Science and Technology}
\affiliation{Universidad de Buenos Aires, Instituto de Física de Buenos Aires (IFIBA)}
\author{F. Grosshans}
\affiliation{Sorbonne Université, CNRS, LIP6}
\author{P. E. Emeriau}
\affiliation{Quandela, 7 rue Léonard de Vinci, 91300 Massy, France}
\author{U. Chabaud}
\affiliation{DIENS, \'Ecole Normale Sup\'erieure, PSL University, CNRS, INRIA, 45 rue d’Ulm, Paris, 75005, France}
\author{M. Walschaers}\email{mattia.walschaers@lkb.upmc.fr}
\affiliation{Laboratoire Kastler Brossel, Sorbonne Université, CNRS, ENS-Université PSL, Collège de France, 4 place Jussieu, Paris F-75252, France}

\date{\today} 

\begin{abstract}

Although the original EPR paradox was formulated in terms of position and momentum, most studies of these phenomena have focused on measurement scenarios with only a discrete number of possible measurement outcomes. Here, we present a framework for studying non-locality that is agnostic to the dimension of the physical systems involved, allowing us to probe purely continuous-variable, discrete-variable, and hybrid continuous/discrete non-locality. Our approach allows us to find the optimal Bell inequality for any given measurement scenario and quantifies the amount of non-locality that is present in measurement statistics. This formalism unifies the existing literature on continuous-variable non-locality and allows us to identify new states in which Bell non-locality can be probed through homodyne detection. Notably, we find the first example of two-mode continuous-variable non-locality that cannot be mapped to a binary Bell inequality. Moreover, we provide several examples of simple hybrid DV-CV entangled states that could lead to near-term violation of Bell inequalities. 

\end{abstract}
\maketitle
\section{Introduction}
The inconsistency of the predictions of quantum theory with local realism was first pointed out by Einstein, Podolsky and Rosen (EPR) in \cite{EPR1935}. For a long time the subject of philosophical debates, it was the work of Bell \cite{Bell_1964} that transformed this inconsistency into a quantifiable, experimentally testable claim, namely the violation of a Bell inequality.  Several experiments have been performed since the advent of Bell's inequalities to probe the non-local features of quantum mechanics, with an increasing level of refinement \cite{Aspect1982,Zeilinger2015, hensen_loophole-free_2015,Shalm_2015,Rosenfeld_2017,Li_2018}. Although the original EPR paradox was phrased in continuous variables (CV), the work by Bell and all subsequent successful experiments dealt with discrete variable (DV) systems. \par 
Beyond its fundamental interest, non-locality has emerged as a key resource for applications such as device-independent quantum information processing \cite{Vidick_2014}, where the violation of a Bell inequality serves as a security guarantee. To deploy such protocols in practice, optical implementations with room-temperature detectors are desirable. Homodyne detection of the quadratures of an electromagnetic field \cite{Bachor2019} offers such detection capabilities for CV quantum information processing on traveling modes of light \cite{braunstein_quantum_2005}. However, violating Bell inequalities in such systems remains a big challenge, despite extensive theoretical efforts \cite{Gilchrist1998,Munro1999,Fred_2003,Nha2004,RGP1,RGP2,Cavalcanti_2007,Acin2009,Etesse_2014,Enky_Gael_2024}. \par
Most of the proposals for performing Bell tests on CV systems rely on imposing a predefined binning strategy that maps the continuum of possible outcomes to either $+1$ or $-1$, to then apply the Clauser--Horne--Shimony--Holt (CHSH) inequality \cite{CHSH}. In \cite{Acin2009}, using similar binning strategies, the Mermin--Klyshko inequality \cite{Mermin1990,KLYSHKO1996} was used to probe non-locality for more than two modes, which allows one to obtain larger normalized Bell inequality violations than what is possible with CHSH \cite{Abramsky2017}. Recent efforts have been devoted to automatizing the binning procedure \cite{Enky_Gael_2024}, leading to considerable improvements in the expected violation of the local bounds. Nevertheless, the procedure is constrained to states with a limited number of photons. \par 
A genuinely CV Bell inequality has only been proposed in \cite{Cavalcanti_2007}, which does not rely on any discretization of the continuum of measurement outcomes, but rather on a non-linear combination of their moments. The downside of this approach is that at least ten modes are required to obtain a violation of the local bound. Therefore, its relevance for an experimental implementation remains limited, despite the theoretical appeal of this inequality.  Thus, the quest for genuinely CV Bell inequalities of practical relevance, tracing back to the initial EPR paradox, remains open. 
\par 
All the above approaches start from a Bell inequality without any guarantee that it will be practical for the measurement scenario at hand. Here, we take a radically different approach and introduce a general framework which instead starts from experimental measurement data and finds the optimal Bell inequality associated with the data. To do so, we also discretize the outcomes, but not as a predefined physical post-processing to force a specific Bell inequality such as CHSH. Instead, binning is used as a convergent numerical relaxation of an underlying infinite-dimensional optimization program: increasing the number of bins yields a hierarchy of programs whose outputs monotonically approach the continuous optimum. 
\par
From a technical standpoint, our framework builds upon a recently proven CV version \cite{PE_2022} of the Fine--Abramsky--Brandenburger theorem \cite{Fine1982,Abramsky_2011}, which shows that CV Bell non-locality is a specific instance of contextuality \cite{Abramsky_2011}. This connection provides a well-defined quantifier of the amount of non-locality of a given experiment, via its contextual fraction \cite{Abramsky2017}. For local measurement scenarios, the contextual fraction explains which fraction of the observed measurement statistics cannot be explained with a local hidden-variable model. It is equivalent to the maximal normalized Bell inequality violation possible with the available data. Both the value of the contextual fraction and the optimal Bell inequality can be obtained through a linear optimization program. In the CV setting, it becomes an infinite linear program, which cannot be solved exactly. In \cite{PE_2022}, a convergent hierarchy of relaxations of this infinite linear program was provided. This hierarchy consists of positive semidefinite optimization programs (SDPs), based on the moments of the empirical data. However, after an in-depth numerical exploration of this approach, we find it to be unfruitful for detecting Bell non-locality in CV quantum experiments. In Appendix \ref{app:SDP_definition}, we present a description of the hierarchy and a discussion of our findings. 
For this reason, we introduce a different relaxation of the infinite linear program to compute the contextual fraction. The relaxation is constructed by performing instead a fine-grained binning of the probability distributions of the measurement outcomes. This results in a hierarchy of finite linear optimization programs, whose results converge to the infinite-dimensional one by increasing the number of bins. In \cite{PE_2022}, it was proven that the contextual fraction decreases monotonically with such binning, which guarantees that the amount of non-locality obtained for a finite number of bins provides a lower bound on the true non-locality of the experimental data at hand. 
As a result, this approach allows us to develop a framework that, given access to experimental data, quantifies the amount of non-locality and provides the corresponding optimal Bell inequality. 

We apply our framework to several experimentally-motivated scenarios, with our main results summarized in Section \ref{sec:summary_of_results}. The rest of the paper is then organized as follows. In Section \ref{sec:general_setting}, we formalize the general setting we consider. In Section \ref{sec:finite_program_def}, we introduce the hierarchy of finite linear programs which we implement. Later, in Section \ref{sec:results}, we present different examples of CV quantum states and measurement settings for which the program detects Bell non-locality. For each example, we present the corresponding optimal Bell inequality. In Section \ref{sec:results_hybrid} we show the potential of the method for detecting Bell non-locality on hybrid DV-CV quantum systems. Finally, in Section \ref{sec:discussion_results_and_open_questions}, we discuss the implications of the results obtained and highlight interesting open questions. 
\par 
 

\section{Summary of results}\label{sec:summary_of_results}
In this section we provide a high-level summary of the main results reported in this work. 

\begin{figure*}[htbp]
    \centering
    \includegraphics[width=0.8\linewidth]{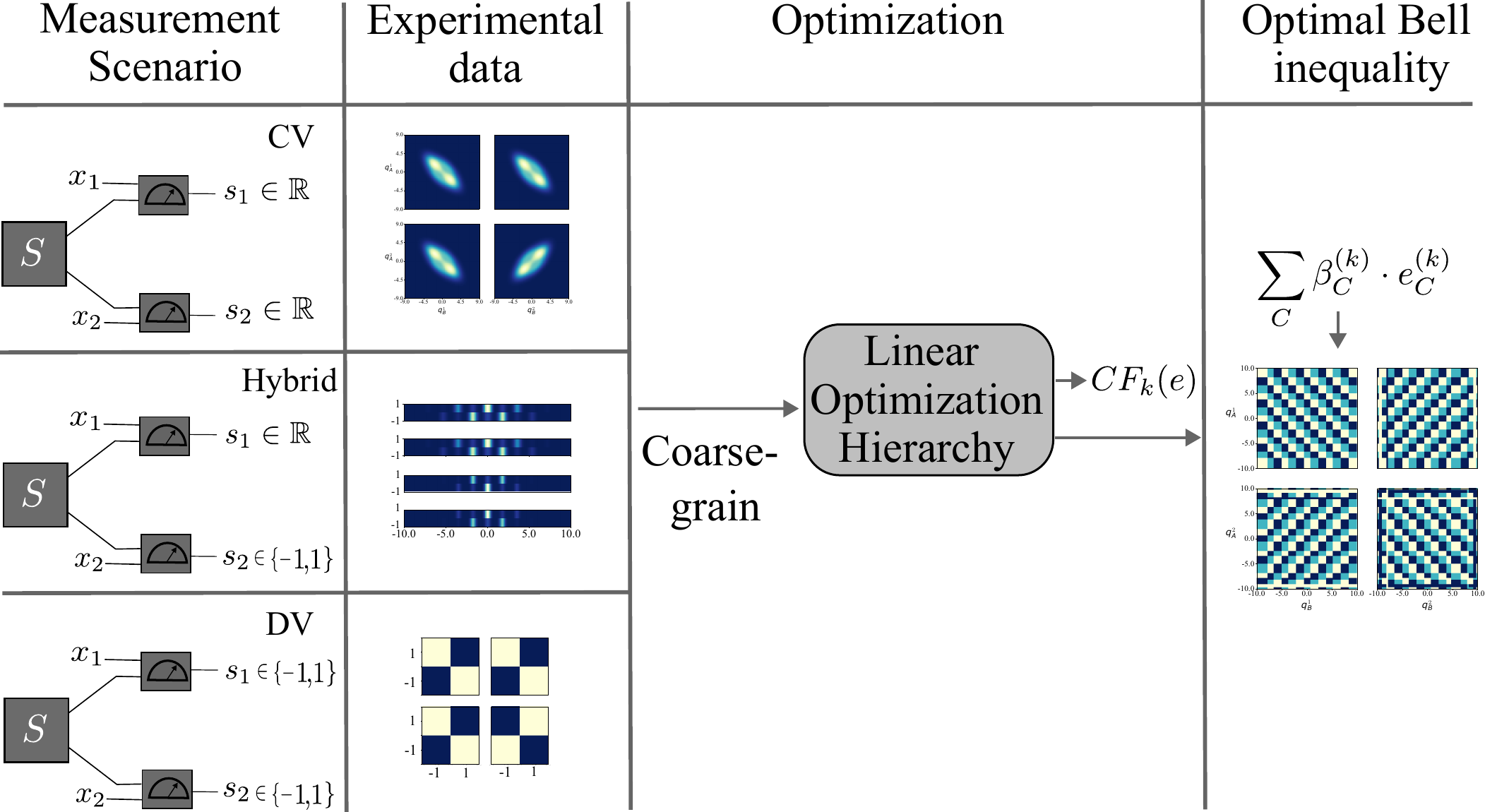}
    \caption{General workflow for obtaining data-driven Bell inequalities. We consider a measurement scenario, in which several parties (only two shown for simplicity) share a state and perform measurements labeled by $x_i$. The experimental data is collected into a so called empirical model $e=\{e_C=p(s_1,s_2|x_1,x_2)\}$, where $C=\{x_1,x_2\}$ denotes a \textit{context}, or set of jointly measurable observables. The methodology is agnostic to the dimension of the underlying system, which allows us to consider discrete and continuous outcome spaces alike. Given the empirical distributions for each context, we can to implement a hierarchy of linear optimization programs,
    to which we input increasingly fine-grained discretizations of the statistics, with $N_b(k) = 2^k$ bins at level $k$ of the hierarchy. 
    This hierarchy will return lower bounds on the contextual fraction $CF(e)$, as well as parametrized Bell inequalities (nontrivial only if the obtained contextual fraction is not zero). This consists of piecewise constant functions $\beta_C^{(k)}$.}
    \label{fig:general_workflow}
\end{figure*}

In Fig.~\ref{fig:general_workflow} we represent the workflow for obtaining data-driven Bell inequalities. Although motivated by the search for CV Bell inequalities, the framework we develop is agnostic to the underlying dimensionality of the data. Therefore, it allows us to unify, in a single coherent approach, the derivation of optimal Bell inequalities for CV, DV, and hybrid CV/DV experiments.  \par 
We consider a physical system of $M$ parties, where each party is characterized by either discrete or continuous degrees of freedom. Each party can perform a measurement labeled by a parameter $x_i$, for $i\in \{1,\ldots, M\}$. The result of the experiment is a collection of the joint statistics of simultaneously measurable observables, $e=\{p(s_{1},\ldots, s_{M}|x_1,\ldots,x_M): \forall x_1, \ldots,  x_M\}$, which in the contextuality literature is referred to as empirical model for the given state and measurement scenario. Rather than probing the non-locality of this model with a predefined Bell inequality, we feed it to an optimization scheme that is guaranteed to return the optimal Bell inequality for it. When considering continuous degrees of freedom we cannot consider an exact representation of the outcome statistics. Instead, we consider discrete representations of the data, either through its moments or through increasingly fine-grained binning. It is the latter alternative that allows us to witness non-locality in CV scenarios. On the other hand, discrete degrees of freedom can be treated exactly, which permits us to concentrate the computational resources on the continuous part when considering  hybrid measurement scenarios. 
\par 
The main contributions of our work are summarized in the following: 
    
    \subsubsection{A unified framework}
    We develop a unified approach for deriving data-driven Bell inequalities for both CV and DV systems. This generalizes previously existing methods for DV systems to the continuous scenario in a controlled way. The resulting method is universally relevant for experimental certification of non-locality. Indeed, given a parametrized family of states available to a given experiment, and a specification of the accessible measurements, our framework allows us to perform an exhaustive numerical exploration to probe the optimal construction for observing non-locality under the experimental constraints. 
    \par
    Continuous degrees of freedom cannot be treated exactly in our framework, but this is also the case in actual experiments. Therefore, at least for relatively small systems, we can consider exactly the histograms of experimental data that come out of experiments and find optimal Bell inequalities for them. 
    
    \subsubsection{Insights into CV non-locality}
    
     We performed an extensive numerical exploration of different families of CV systems. We employed both the hierarchy of moment-based programs, and the hierarchy of linear programs based on binning of the outcome statistics. A first surprising insight was that no relevant Bell inequality was found using the hierarchy of moment-based programs proposed in \cite{PE_2022} (see Appendix~\ref{app:SDP_definition}). \par 
     Despite using binning as a mere proxy to produce increasingly fine-grained approximations of the outcome statistics, all optimal Bell inequalities obtained are of a discrete nature (see Section \ref{sec:results}). This validates the optimality of many previous approaches in the literature \cite{Gilchrist1998,Munro1999,Fred_2003,RGP1,RGP2,Acin2009,Etesse_2014,Enky_Gael_2024} (see Appendix~\ref{app:previous_examples}). Moreover, this raises fundamental questions about the nature of non-locality in homodyne detection scenarios.\par 
     Inspired by bosonic encodings, we unveil a family of new examples displaying CV Bell non-locality. While the onset of non-locality with homodyne detection on ideal encoded states is a straightforward consequence of the construction of the codes, we show that very low quality states suffice to observe significant non-locality. Contrary to the ideal case, this is a non-trivial result, which provides perhaps the most interesting perspective so far for a near- to intermediate-term realization of non-locality with homodyne detection measurements (see Section \ref{sec:bosonic_encoding}). Using bosonic encoding of qutrit states, we report the first example of a two-mode bosonic state for which the optimal Bell inequality is effectively different from the CHSH Bell inequality \cite{CHSH}. The resulting inequality can be shown to be equivalent to a qutrit Bell inequality first reported in \cite{CGLMP2002}, known as the CGLMP inequality (see section~\ref{sec:qutrit_GKP}). Finally, we construct three-mode examples that display a highly non-local behavior, with a contextual fraction of up to $60 \%$ (see section~\ref{sec:multimode_GKP}).
    \subsubsection{Non-locality in hybrid CV/DV systems}
     We consider hybrid CV/DV states and show that simpler states, compared to the CV case, unlock significant levels of non-locality (see section~\ref{sec:results_hybrid}). The states we consider may be within the reach of near-term experiments on hybrid systems. In particular, we show that states like the ones prepared in \cite{Hacker2019} display a significant degree of non-locality under homodyne detection of the continuous degree of freedom. These results highlight the relevance of hybrid platforms for quantum information processing.
     \par 
In the rest of the paper we detail the derivation of the results we have outlined in this section. 

\section{Measurement scenario}
\label{sec:general_setting}
In this section, we present the general setting that we are dealing with. We place emphasis on the link between Bell non-locality experiments and the more general language of contextuality. \par
In Fig.~\ref{fig:general_workflow}, we present the general scheme for a Bell experiment with two parties. We consider the broader scenario with more than two parties, but represent only two for clarity. Alice and Bob each have access to a subsystem of a state $\hat \rho$, prepared in a black-box fashion, about which they have no information. Each of them can perform a measurement, chosen among the sets  $x_1\in M_A=\{a_1,...,a_{n_A}\}$ and $x_2\in M_B=\{b_1,...,b_{n_B}\}$ respectively. The main case of interest for us in the CV setting is when Alice and Bob perform homodyne measurements of the quadratures of a harmonic oscillator, for example, the amplitude and phase quadratures of a mode of an electromagnetic field \cite{Bachor2019}. In this case, the different settings correspond to different choices of the measured quadrature, and the outcome space is continuous and unbounded, $o_{i}\in \mathbb R$. In Appendix \ref{app:CV}, we provide a brief introduction to CV quantum information, with a focus on quantum optics. \par
Due to measurement incompatibility, an arbitrarily precise simultaneous measurement of different local settings is not possible, and the results of the experiment are the different probability densities (histograms) $p(o_1,o_2|x_1,x_2)$ for each possible pair of compatible measurements $\{x_1,x_2\}$. The non-signaling principle implies that for spatially separated experimenters, the choice of measurement setting by one of them should not influence the statistics of outcomes observed by the other, \ie 
\begin{equation}\label{eq:no_signaling}
\begin{split}
p(o_{1}|x_1)&=\int_{\mathbb R} do_{2} \quad \!\! \!\!p(o_{1},o_{2}|x_1,x_2)\\
p(o_{2}|x_2)&=\int_{\mathbb R} do_{1} \quad \!\!\!\! p(o_{1},o_{2}|x_1,x_2),
\end{split}
\end{equation} 
for all different choices of measurement settings $x_1,k$. A similar expression can be written for different choices of measurements by Alice. \par
The central question in the analysis of Bell non-locality is to determine whether the statistics obtained from the experiment are consistent with local realism, which would imply that it can be described by a local hidden-variable model (LHVM) of the form
\begin{equation}
    p(o_1,o_2|x_1,x_2)=\int_{\Lambda}d \mu(\lambda) p(o_1|x_1,\lambda)p(o_2|x_2,\lambda),
\end{equation}
where $\mu(\lambda)$ defines a probability measure for the hidden variable $\lambda \in \Lambda$. Note that this assumes that all correlations observed in an experiment are of a classical nature, and the state vector description comes from a lack of knowledge about an underlying statistical process. \par
The question about the existence of an LHVM has been addressed either by methods that directly try to construct it or, in a more historical approach, through Bell inequalities \cite{Bell_1964,review_Bell_nonlocality}. Convexity of the set of local behaviors, \ie of the set of experiments that could be described by an LHVM, guarantees that for every behavior outside this set, there is an inequality that can optimally discriminate it.\par 
An equivalent description can be obtained through the lens of the broader concept of contextuality \cite{Abramsky_2011, Kochen_Specker}. The notion of contextuality that we deal with is due to Kochen and Specker \cite{Kochen_Specker}.

The kind of setting considered by contextuality includes those depicted in Fig.~\ref{fig:general_workflow}, but is not restricted to it. It can include non-local measurements, or multiple sequential local measurements. More formally, any contextuality experiment is described by a measurement scenario $\langle X,\mathcal M, \it{\bm O}\rangle$, where $X$ specifies the set of measurement labels, $\mathcal M$ is the set of contexts -- jointly measurable observables -- and $\it{\bm O}$ the measurable space of measurement outcomes. The description compatible with Bell non-local experiments of the type described in Fig.~\ref{fig:general_workflow} is given by 
\begin{itemize}
    \item $X=\{a_1,...,a_{n_A},b_1,...,b_{n_B}\}$
    \item $\mathcal M= \{\{a_i,b_j\}: i,j\}$
    \item $O_x=\langle \mathbb R,\mathcal B_{\mathbb R}\rangle$, where $\mathcal B_{\mathbb R}$ is the $\sigma$-algebra generated by all the open sets on the real line.
\end{itemize}

Typically, we will consider $O_x$ defined by a bounded subspace of the reals, which is consistent with the finite-energy constraint of experiments. \par
The set of histograms that are accessible through the experiment $e_C=p(o_{1},\ldots,o_{M}|x_1,\ldots,x_M)$, which we here call $e_C$ to simplify the notation, defines an \textit{empirical model}, on the measurement scenario $\langle X,\mathcal M, \it O\rangle$. The non-signaling constraints described in Eq.\eqref{eq:no_signaling}, are a specific instance of the compatibility conditions $e_{C\cap C^{'}|C}=e_{C\cap C^{'}|C^{'}}$, which imply that the statistics of subsets of a context are independent of the remaining set of compatible measurements performed. \par
\begin{defn}\label{def:contextual_model}
An empirical model $e=(e_C)_{C\in \mathcal M}$, on a measurement scenario $\langle X,\mathcal M, \it{\bm O}\rangle$, is extendable (or non-contextual) if there is a probability measure $\mu$ on the space $\it{\bm O}_X= \prod_{x\in X} O_x$, such that $\mu|_C=e_C$, for every $C\in \mathcal M$.    
\end{defn}
In the specific case that we consider, what this implies is that the measurement statistics obtained from the Bell experiment admit a local description if there is a probability distribution over all observables measured, $\mu(\mathbf o_X)$, such that for every context $C=\{a_i,b_j\}$, 
\begin{equation*}
\begin{split}
&p(o_{1},o_{2}|a_i,b_j)=e_{C}(\mathbf o_C) =\int_{O_{X\backslash C}} d \mathbf o_{X\backslash C}  \mu (\mathbf o_X),
\end{split}
\end{equation*}
where $\mathbf o_X=\{o_{a_1},\ldots,o_{a_{n_A}},o_{b_1},\ldots,o_{b_{n_B}}\}\in O_X$ can be seen as a global assignment of values to all possible measurements in $X$, and $\mathbf o_X=\mathbf o_{X\backslash C} \oplus \mathbf o_C$, where $\mathbf o_C\in O_C$ corresponds to assignments of values to all measurements in a given context, and $\mathbf o_{X\backslash C}\in O_{X\backslash C}$ are the values of the remaining variables. \par  
An empirical model that is not extendable is called \textit{contextual}. Notice that if a model is extendable, it implies that the same statistics could have resulted from an underlying stochastic process that samples all observables jointly. The contradiction between contextual assignments and global assignments of values to each observable is at the core of contextuality. In non-locality experiments the existence of an LHVM is equivalent to the existence of an extendable or non-contextual description \cite{Fine1982,Kochen_Specker,Abramsky_2011}. Contextual data in this case can equally be called Bell non-local. Recently, this connection was formally proven for CV measurement scenarios \cite{PE_2022}. \par
If an empirical model is non-contextual, its global description can be found by solving the following system of linear equations 
\begin{equation}\label{eq:equality_problem}
    \begin{cases}
        \text{Find } &\mu:  \\
        \text{s.t. } &\mu\geq 0 \\
        & \mu|_C=e_C.
    \end{cases}
\end{equation}
This equation only has a solution when the model is local (or non-contextual). If instead we replace the equality constraints by $\mu|_C\leq e_C$, many possible solutions exist, and in the case of a local model, the solution of Eq.~\eqref{eq:equality_problem} corresponds to the optimal value of the optimization 
\begin{equation}\label{eq:optimization_first_statement}
    \begin{cases}
        \max &\mu(O_X):\\
        \text{s.t.} & \mu\geq 0\\
        & \mu|_C\leq e_C,
    \end{cases}
\end{equation}
where $\mu(O_X)$ is the integral of $\mu$ over the whole outcome space, which is $1$ for a normalized distribution. This is consistent with considering the statistics to be of the form 
\begin{equation}
    e=\lambda e_L +(1-\lambda) e_{NL},
\end{equation}
where $e_L$ is a local model, and $e_{NL}$ a non-local one. The maximum possible value of $\lambda$ for which such a  decomposition makes sense is called the non-contextual fraction \cite{Abramsky2017,PE_2022}, also known as local content in the non-locality literature \cite{{Elitzur_1992,Cabello_2002,Barrett2006,Aolita_2012}}. Intuitively, the non-contextual fraction quantifies the portion of the data that can be \textit{explained} through an extended probability distribution. An extendable model has a non-contextual fraction equal to one. Formally, it is defined as follows. 
\begin{defn}\label{def:contextual_fraction}
    The non-contextual fraction $NCF(e)$ associated to an empirical model $e$ on a measurement scenario $\langle X,\mathcal M,\it O\rangle$, is given by 
    \begin{equation*}
        \sup\{\mu(O_X)\,|\,\mu\geq 0, \forall C\in \mathcal M: \mu|_C\leq e_C\}.
    \end{equation*}
    Because all $e_C$ are well-defined probability distributions, the non-contextual fraction $NCF(e)\in \left[0,1\right]$. The contextual fraction is defined as $CF(e)=1-NCF(e)$. 
\end{defn}
The contextual fraction can be understood as the part of the data that cannot be explained with a local model. Defining the contextual fraction through an optimization over a space of functions allows one to formulate a linear optimization program to compute it \cite{Abramsky2017}, which in the case of CV systems is infinite \cite{PE_2022}. In the following section we discuss the said program and its relaxations. 

\section{Quantifying non-locality}
\label{sec:finite_program_def}
Definition \ref{def:contextual_fraction} allows us to define a (possibly infinite) linear optimization program to compute the {(non-)}contextual fraction, as a well-defined quantifier of (non-) contextuality, and in turn of non-locality. In this section we formalize the structure of the optimization problem, and the information we can extract from it. This section is technical and a reader interested in the analysis of the main results may skip to the following sections.\par 

We rephrase here the optimization in Eq.~\eqref{eq:optimization_first_statement} for completeness:
\begin{equation}\tag{\ref{eq:optimization_first_statement}}
(P)\begin{cases}\begin{aligned}\text{Find} \quad & \mu\in  F_{\pm}(\bm{\mathrm{O}_X})\\\max \quad & \mu(\mathrm{O}_X)\\\textrm{s.t.} \quad &   \mu|_{C}\leq e_{C} \quad\forall C\in \mathcal{M}\\& \mu\geq 0.\end{aligned}\end{cases}
\end{equation}
Here $\mathrm F_{\pm}$ represents the set of signed distributions, which constitutes a well-defined vector space. The output of this program is the non-contextual fraction $NCF(e)$, associated to the empirical model $e$, over the corresponding measurement scenario. If a non-contextual model for $e$ exists, then $\mu$ will be a well-defined probability distribution over $\mathrm O_X$, and the output of the program will be a maximal non-contextual fraction of 1. As for any optimization program, a dual program can be defined \cite{Boyd_Vandenberghe_2004}. 
The derivation of this program can be found in \cite{PE_2022}. It can be written conveniently in the following form 
\begin{equation}\label{eq:program_B}
    (B)\begin{cases}        \begin{aligned}\text{Find} \quad & (\beta_{C})_{C\in \mathcal M}\in \mathrm \prod_{C\in \mathcal M}C_{0}(\mathrm O_C,\mathrm R)\\\max \quad & \sum_C \int_{\mathrm O_C} \beta_C d e_C\\\textrm{s.t.} \quad & \sum_{C} \beta_C \leq 0 \text{ on } \mathrm{O}_{X}\\& \forall C \in \mathcal M, \beta_C\leq |\mathcal M|^{-1} \bm{1} \text{ on } \mathrm O_C. \end{aligned}    \end{cases}
\end{equation}
The output of this program is directly the contextual fraction $CF(e)$. Moreover, this program offers a direct way to recover a Bell inequality for the given empirical model, provided it is contextual. A local model $e=\{e_C\}$ can be extended to some $\mu$ over $\mathrm O_X$, which implies that the integral and summations in the objective function can be exchanged, and we have, following the first constraint in the program,
\begin{equation*}
    \sum_C \int_{\mathrm O_C} \beta_C d e_C=\int_{\mathrm O_X}d\mu \left(\sum_C\beta_C\right)\leq 0. 
\end{equation*}
 Consequently, any solution $\{\beta_C\}$ parametrizes a Bell inequality of the form 
\begin{equation}\label{eq:Bell_inequality}
    B(e_C)=\sum_C \int_{\mathrm O_C} \beta_C d e_C\leq 0.
\end{equation}
 If the empirical model is contextual (non-local) and $CF(e)>0$, then the set of functions $\beta_C^{*}$, obtained through Eq.~\eqref{eq:program_B} for $e$, parametrizes the optimal Bell inequality for the given empirical model. For a maximally contextual model 
there is a strategy which allows us to assign $\beta_C(x_C)=1/|\mathcal M|$, for every $x_C$ for which $e_C(x_C)\neq 0$. This implies that the output of the program is
\begin{equation*}
    \sum_{C\in \mathcal M} \int_{\mathrm O_C} \beta_C d e_C=\sum_{C\in \mathcal M} \frac{1}{|\mathcal M|}\int_{\mathrm O_C}de_C=1, 
\end{equation*}
which is indeed the maximum possible value of the contextual fraction.  \par 
The programs described in this section are infinite linear programs, and thus cannot be solved exactly. Nevertheless, there are ways of relaxing them into hierarchies of finite programs which incrementally approximate the solution of the infinite ones. In \cite{PE_2022}, a relaxation into a hierarchy of positive-semidefinite programs (SDPs) \cite{Boyd_Vandenberghe_2004}, based on the moments of the empirical distributions, was proposed. In Appendix \ref{app:SDP_definition}, we present it for completeness. The Bell inequalities that can be obtained at any level of this hierarchy correspond to polynomial Bell inequalities, \ie all $\beta_C$ in \eqref{eq:Bell_inequality} are polynomials of the observables, and thus the Bell inequality can be evaluated in terms of the moments of the distribution, without need for discretization of the data. During the course of our work, as a first attempt to obtain genuinely CV Bell inequalities, we implemented the hierarchy of moment-based SDPs. The implementation can be found on the GitHub repository \cite{repo}. After extensive numerical search, it was not possible to find any CV empirical behavior for which a polynomial Bell inequality existed. Naturally, if we could probe arbitrarily high orders in the hierarchy, we would find Bell inequalities at some point; yet, up to the practical limit \ie polynomials of degree at most $\sim 18$, no working example was found. On the other hand, as discussed in Appendix \ref{app:SDP_definition}, the program was able to obtain a Bell inequality for non-physical empirical models like the continuous extensions of Popescu-Rohrlich's box \cite{popescu_quantum_1994}, described in \cite{Aolita_2018}. However, when optimizing over the set of quantum states, no state violated these Bell inequalities. \par 
Hereafter, we follow a different approach to relax the infinite linear program. We approximate the continuous distributions as histograms, using an arbitrary coarse-graining of the outcome space. This relaxes the program into a finite linear optimization program, which we describe in the following section. 

\subsection{Finite version of the infinite LP}
To formalize the construction of a hierarchy we can define the action of a binning procedure as a mapping $h$ from a measurement scenario $\langle X, \mathcal M, \mathbf O\rangle$, to a different measurement scenario $\langle X, \mathcal M, \mathbf O^\prime\rangle$, where only the outcome space has changed. Here $\mathbf O=\prod_{x\in X} \mathbf O_x$, is the global continuous outcome space, with outcome spaces for each variable given by $\mathbf O_x=(K_x, \mathcal B_{K_x})$, with $K_x\subseteq \mathbb R$. On the other hand, the coarse-grained outcome space is $O_x^\prime=\{1,\ldots , N_b\}$, where $N_b$ is the number of bins into which we have discretized the outcomes. To each outcome $i\in O_x^\prime$, we associate an interval $E_i\subseteq K_x \subseteq \mathbb R$, such that $\bigcup_{i\in O_{x}^\prime}E_i=K_x\subseteq \mathbb R$. The discretized empirical models $e_C^\prime$ are obtained as
\begin{equation}\label{eq:discretizing_e}
    e_C^\prime(s^\prime\in O_C^\prime)=\int_{E_{s'}} de_C.
\end{equation}
$O_C^\prime$ is the outcome space for each context, given by $O_C^\prime=\prod_{x\in C} O_x^\prime$, so that $s^\prime \in \mathbb N^{|C|}$ specifies the interval $E_{s^\prime}=\prod_{x\in C} E_{s^\prime_{x}}\subseteq K_{C}$. \par 

\begin{figure*}[htbp]
\centering
\includegraphics[width =0.8\linewidth]{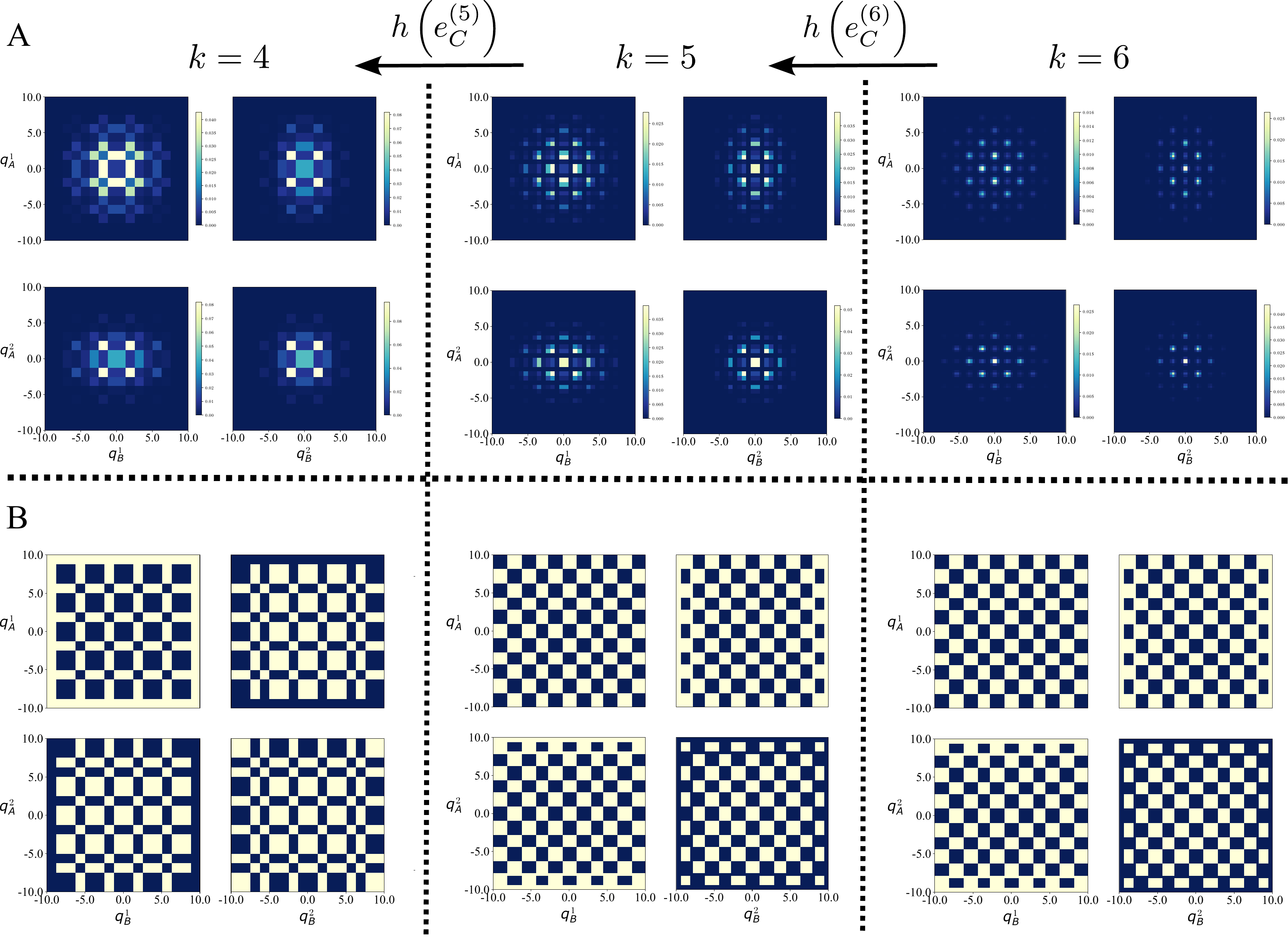}
\caption{ (A) Schematics of the coarsening of the empirical distributions for implementing the hierarchy of programs $P_{2^k}$ and $B_{2^k}$. For a given empirical model, we show the steps from level $k=6$, for which there are $N_b=64$ bins per quadrature to level $k=4$ for which $N_b=16$. (B) Optimal Bell inequalities obtained at each level of the hierarchy. In this case, the optimal Bell inequalities at levels $k=5$ and $k=6$ are the same. }
\label{fig:binning_hierarchy}
\end{figure*}

For a given empirical model and binning choice, the primal program becomes 
\begin{equation}\label{eq:primal_LP}
\left(P_{N_b}\right)\begin{cases}\begin{aligned}\text{Find} \quad & \mu\in \mathbb R^{\overbrace{N_b \times \cdots \times N_b}^{|X|}}\\\max \quad & \mathbf 1 \cdot \mu\\\textrm{s.t.} \quad &  \mu|_{C}\leq e_{C}^{\prime}\quad\forall C\in \mathcal{M}\\& \mu\geq 0,\end{aligned}\end{cases}
\end{equation}
where all inequalities are element-wise, and $\mathbf 1 \cdot \mu$ corresponds to the sum of all elements of $\mu$, which is equal to one if $\mu$ corresponds to a finite distribution. The dual program, directly written in terms of $\beta_C$, so that it returns the contextual fraction, is given by 
\begin{equation}\label{eq:dual_LP}
    (B_{N_b})\begin{cases}        \begin{aligned}\text{Find} \quad & (\beta_{C})_{C\in \mathcal M}\in \mathrm \bigoplus_{C\in \mathcal M}\mathbb R^{\overbrace{N_b \times \cdots \times N_b}^{|C|}}\\\max \quad & \sum_{C\in \mathcal M}\beta_C \cdot e_C^{\prime}\\\textrm{s.t.} \quad & \sum_{C\in \mathcal M} P_{X}(\beta_C) \leq \vec 0 \\&  \beta_C\leq |\mathcal M|^{-1} \bm{1} \quad \forall C \in \mathcal M. \end{aligned}    \end{cases}
\end{equation}
Here, $P_{X}(\beta_C)$ projects $\beta_C$ from a context vector with $N_b^{|C|}$ elements to a global vector with $N_b^{|X|}$ elements. For an arbitrary number of bins, this program provides a lower bound on the exact contextual fraction, since the contextual fraction is decreasing under classical processing like binning \cite{PE_2022}. On the other hand, to build up a well-defined hierarchy of programs, with a monotonic convergence, each level of the hierarchy must be a coarse-grainning of the next level. We can choose to do it by first fixing $A$, such that outcomes are constrained to $O_{x}= \left[-A,A\right]$, which can be done without loss of generality, as long as we make sure to leave only an $\epsilon \ll 1$ fraction of the data outside, with the guarantee of an error $\mathcal O(\epsilon)$ on the contextual fraction \cite{PE_2022}. We then choose $N_b(k)=2^k$, \textit{i.e.}, we double the number of bins at each level, and define 
\begin{equation}
    E_{j}^{(k)}=\left[-A+(j-1)\times \frac{2 A}{2^k},-A+j\times \frac{2 A}{2^k}\right).
\end{equation}
This guarantees that $e_{C}^{(k-1)}$ can be obtained from $e_{C}^{(k)}$ by coarse-graining, which is shown in Fig.~\ref{fig:binning_hierarchy}.  Any other choice of incremental binning that respects this could be considered.  

For the hierarchy to be sound, the following two statements have to hold:
\begin{enumerate}
    \item $\forall k\in\mathbb N^{+}$, we have $CF_k\geq CF_{k-1}$; 
    \item $\lim_{k\to\infty}(CF_k)=CF$. 
\end{enumerate}
Here, $CF_k$ is the contextual fraction obtained at level $k$ in the hierarchy, \ie by solving $B_{2^k}$. The proof of the monotonicity of the contextual fraction under binning as a classical operation in \cite[Section 5.2]{PE_2022} directly implies that the two statements above hold. 
\par

Before presenting the results that we obtain by implementing this hierarchy of programs, there are several points to discuss. First, the numerical implementation is restricted because of computational limitations, with some strict constraints imposed by the number of parameters that commercial optimization solvers can handle. In our implementation, which can be found in \cite{repo}, this corresponds, in the case in which Alice and Bob each perform two measurements, to a maximum of 214 bins per quadrature. For a case in which we deal with three modes and two measurements, the limit is around $32$ bins, as well as for three measurements per party in a two-party experiment. That being said, for the case of two parties and two measurements per mode, we find that the description is sufficiently fine-grained to recover all the smooth features in empirical distributions, and as such, to recover any smooth features that there could be in an optimal Bell inequality. In fact, we observe that in many relevant cases, the hierarchy already converges at order $k=7$. Moreover, the functions $\beta_C$ that parametrize the Bell inequalities are not constrained to specific values, but can take any arbitrary values as long as they collectively satisfy the constraints on the program. We emphasize that this is a radically different approach to all previous attempts to find linear Bell inequalities for CV systems \cite{Gilchrist1998,Munro1999,Fred_2003,RGP1,RGP2,Acin2009,Enky_Gael_2024}.\par  
Each level of the hierarchy of programs in Eq.\eqref{eq:primal_LP} and Eq.\eqref{eq:dual_LP} is equivalent to the programs described in \cite{Abramsky2017} that constitute the basis of the CV extension discussed in \cite{PE_2022} that inspired our work. Similar optimization programs for the analysis of DV non-locality  were previously introduced in  \cite{Kaszlikowski2000} and discussed in \cite{review_Bell_nonlocality}. This connection is a direct consequence of the equivalence between local-hidden-variable models, deterministic hidden-variable models, and non-contextual descriptions \cite{Fine1982, Abramsky_2011}. The advantage of the formulation of Eq.\eqref{eq:dual_LP} is two fold first it apply to any probability distribution coming from a CV, DV or hybrid system, second using the contextual fraction as an objective function allow for a quantitative comparison between all cases and provide a continuous optimization for simulated probabilities.  
 \par 
 In the framed box, we describe the general methodology for probing non-locality and contextuality using our numerical programs, provided access to some experimental data. This description highlights the fact that the method is data-driven: instead of probing the non-locality of the data with a specific Bell inequality, the optimal inequality is found for each different experiment.    \par

\begin{tcolorbox}[
  enhanced,
  breakable,
  colback=blue!10,
  colframe=black,
  boxrule=0.8pt,
  title=Witnessing Bell non-locality:,
  coltitle=white,
  colbacktitle=blue!40!black,
  fonttitle=\bfseries,
  attach boxed title to top left={
    xshift=2mm,
    yshift*=-\tcboxedtitleheight
  },
  boxed title style={
    boxrule=0pt
  }
] 
\begin{enumerate}
    \item Organize the different measurement labels for each mode (the method is agnostic to the actual measurement, but the contexts have to be matched appropriately), as $X_i=\{a^{(i)}_1,...,a^{(i)}_{n_i}\}$. 
    \item The set of all possible choices of measurement settings defines the set of all contexts
    \begin{equation*}
        \qquad \mathcal M=\{(a^{(1)}_{k_1},\dots,a^{(M)}_{k_M}):k_i \in \{1,\ldots,n_i\}\}
    \end{equation*}
    \item Gather all the experimental histograms for each context into an empirical model $e=\bigcup_{C\in \mathcal M} e_C$, where $e_C$ is the empirical distribution for context $C$. 
    \item Starting from $k=1$, bin each empirical distribution into $N_b=2^k$ bins for each measurement label (one can also allow for different choices of $N_b$ for each measurement, \eg for hybrid systems). 
    \item Solve the primal program ($P_{N_b}$) or dual program ($B_{N_b}$), giving the set of empirical distributions as inputs. Repeat the process for incremental values of $k$ until convergence is reached, or the computational resources are no longer enough. 
    \item If the contextual fraction is non-zero, recover the Bell inequality from the optimal variables of the program. From the dual  $(B_{N_b})$, this is obtained directly in the form of the filter functions $\beta_C\in \mathbb R^{N_b ^{|C|}}$. 
\end{enumerate}

\end{tcolorbox}

Having introduced our numerical framework, we discuss in section~\ref{sec:results} several examples of states and measurements for which the hierarchy succeeds in finding the optimal Bell inequality.
Furthermore, the hierarchy of programs defined in this section unifies the treatment of continuous and discrete degrees of freedom into a single optimization framework. Therefore it offers a powerful tool to analyze hybrid DV-CV experiments. It allows us to treat the DV system exactly, while focusing computational resources on the CV one. For instance, if we consider a qubit system for the DV part, there is practically no constraint on the number of bins to consider on the CV side. In Section \ref{sec:results_hybrid} we discuss several examples of hybrid states that can lead to a very large Bell inequality violation with homodyne detection on the CV side. 

\section{Results for continuous-variable systems}
\label{sec:results}

In this section we present several examples of quantum states of bosonic systems for which we are able to witness non-locality in homodyne-detection measurement scenarios. Though our framework can be used to consider arbitrary measurements, in this section we restrict to homodyne detection, which is the most common measurement in CV quantum optical experiments. \par 
We will analyze both two- and three-mode systems, with a special focus on two-mode examples, which are more easily tractable numerically. All the measurement scenarios that we consider are of the form 
\begin{equation}
    \begin{split}
        &X=\{\phi_{1}^{(1)},\ldots, \phi_1^{(n_A)},\phi_2^{(1)},\ldots,\phi_{M}^{(n_M)}\},\\
        &\mathcal M= \{\left(\phi_{1}^{(1)},\ldots, \phi_{M}^{(k_M)}\right):k_i \in \{1,\ldots,n_i\}\}\\
        &O_{\phi}=\mathbb R \quad \forall \phi \in X,
    \end{split}
\end{equation}
where $\phi_{m}^{(j)}$ represents a choice of quadrature to measure in mode $m$
\begin{equation}
    \hat q^{(\phi)}_{m}=\hat q_{m}\cos \phi -\hat p_m \sin \phi.
\end{equation}
An introduction to CV quantum information, with a focus on quantum optics, is provided in Appendix~\ref{app:CV}.\par 
As we discussed in the introduction, the majority of previous approaches to witness non-locality with homodyne detection statistics have relied on dichotomizing the outcome statistics \cite{Gilchrist1998,Munro1999,Fred_2003,RGP1,RGP2,Acin2009,Etesse_2014,Enky_Gael_2024}. That is, they map different segments of the real line to either $+1$ or $-1$. For example, in \cite{RGP1}, sign binning was proposed to probe the non-locality of photon-subtracted states. This consists of mapping the quadrature outcome $q_{\phi}$ to $a_{\phi}=\text{sign}(q_{\phi})$. More complicated value assignments were considered in other works, but they still ultimately map to binary observables that could be treated with standard Bell inequalities like CHSH \cite{CHSH}. In Appendix~\ref{app:previous_examples} we show evidence that all these approaches are actually optimal, and considering more continuous features does not result in any improvement for any of these examples. Indeed, using our approach, which allows for quasi-continuous Bell inequalities, we observe a convergence towards highly discrete Bell inequalities in all those cases, which can always be mapped to the CHSH inequality. In the rest of this section we will report our novel findings, mostly based on connections to bosonic qubit encoding. 




\subsection{Finding needles in the haystack}\label{sec:bosonic_encoding}
As it turns out, finding states that display non-locality under homodyne detection and in particular for the case in which each party is allowed to make two measurements is a challenging task. Looking for such states involves a search over a huge parameter space. One would be tempted to think of this as an optimization problem: we could define a parametrized family of states and search for the one(s) providing the largest contextual fraction. However, the corresponding optimization landscape is mostly flat, i.e., for most states the contextual fraction is zero. This makes it impossible to find good directions to move along in the parameter space. In other words, we are looking for a needle in a haystack. \par
That being said, it is still possible to find interesting states. Notably, bosonic codes, \ie CV states used to encode qubits, offer a proxy to search for Bell inequality violations. Among those codes, GKP states, introduced in  \cite{GKP_2001}, are particularly well suited to find Bell inequalities based on homodyne detection, given that logical Pauli measurements are performed by homodyne detection (see Appendix~\ref{app:GKP_states}). The logical Pauli $Z$ measurement corresponds to measuring the amplitude quadrature $\hat q$, while the logical $X$ measurement corresponds to measuring the phase quadrature $\hat p$. On the other hand, non-Pauli measurements, like $\cos \theta X+\sin \theta Z$ require non-linear operations before performing homodyne detection, which we do not allow in our measurement scenarios. \par 
To find a GKP entangled state that should violate a Bell inequality by measuring only amplitude and phase quadratures, we search for the qubit entangled state that maximally violates the CHSH Bell inequality under Pauli measurements $X,Z$ (other states could be constructed by considering other combinations of Pauli measurements). One such state is 
\begin{equation}\label{eq:entangled_GKP}
    |\psi_{qubit}\rangle=\frac{\left(|11\rangle-|00\rangle-(1+\sqrt{2}) (|01\rangle+|10\rangle)\right)}{2\sqrt{2+\sqrt{2}}}.
\end{equation}
If instead we fix the measurement scenario to be composed of $X,Y$ (corresponding to measuring the quadratures $\hat p$ and $\hat q_{\pi/4}$), the corresponding state has a simpler description, related to the Bell state by a single-mode qubit rotation
\begin{equation}\label{eq:second_choice_of_GKP}
|\psi_{qubit}^{(2)}\rangle=\frac{|00\rangle +e^{-i \pi/4}|11\rangle}{\sqrt{2}}.
\end{equation}

Corresponding CV states can be obtained by replacing the logical codewords $|0\rangle,|1\rangle$ by approximate finite-energy GKP states, given by weighted superpositions of displaced squeezed vacuum states of the form
\begin{equation}\label{eq:finite_energy_GKP}
    \!\!|j\rangle_{\Delta}\!\!=\!\!\frac{1}{\mathcal N_j} \!\!\sum_{k=-\infty}^{\infty}\!\!\!\! e^{-\frac{1}{2}\Delta^2 \left[(2k +j) \frac{\alpha}{2}\right]^2} \hat D\!\left((2 k+j)\frac{\alpha}{2}\right)\!\hat S(r_{\Delta}) |0\rangle_b,
\end{equation}
where $r_{\Delta}=-\log( \sqrt{2}\Delta)$ is the amount of squeezing per peak, and $|0\rangle_b$ is the vacuum state of a single bosonic mode. The damping parameter $\Delta$ provides a control knob for the quality of the code states. The smaller the value of $\Delta$, the better the encoding of the GKP state (narrower peaks, and effectively more of them). Complementarily, we can see it as a control on the complexity of the state. Decreasing the amount of squeezing per-peak (and equivalently increasing the damping on the coefficients) implies that the resulting state is simpler and more likely within the reach of near-term experiments. \par 
\begin{figure}[htbp]
    \centering
    \includegraphics[width=\linewidth]{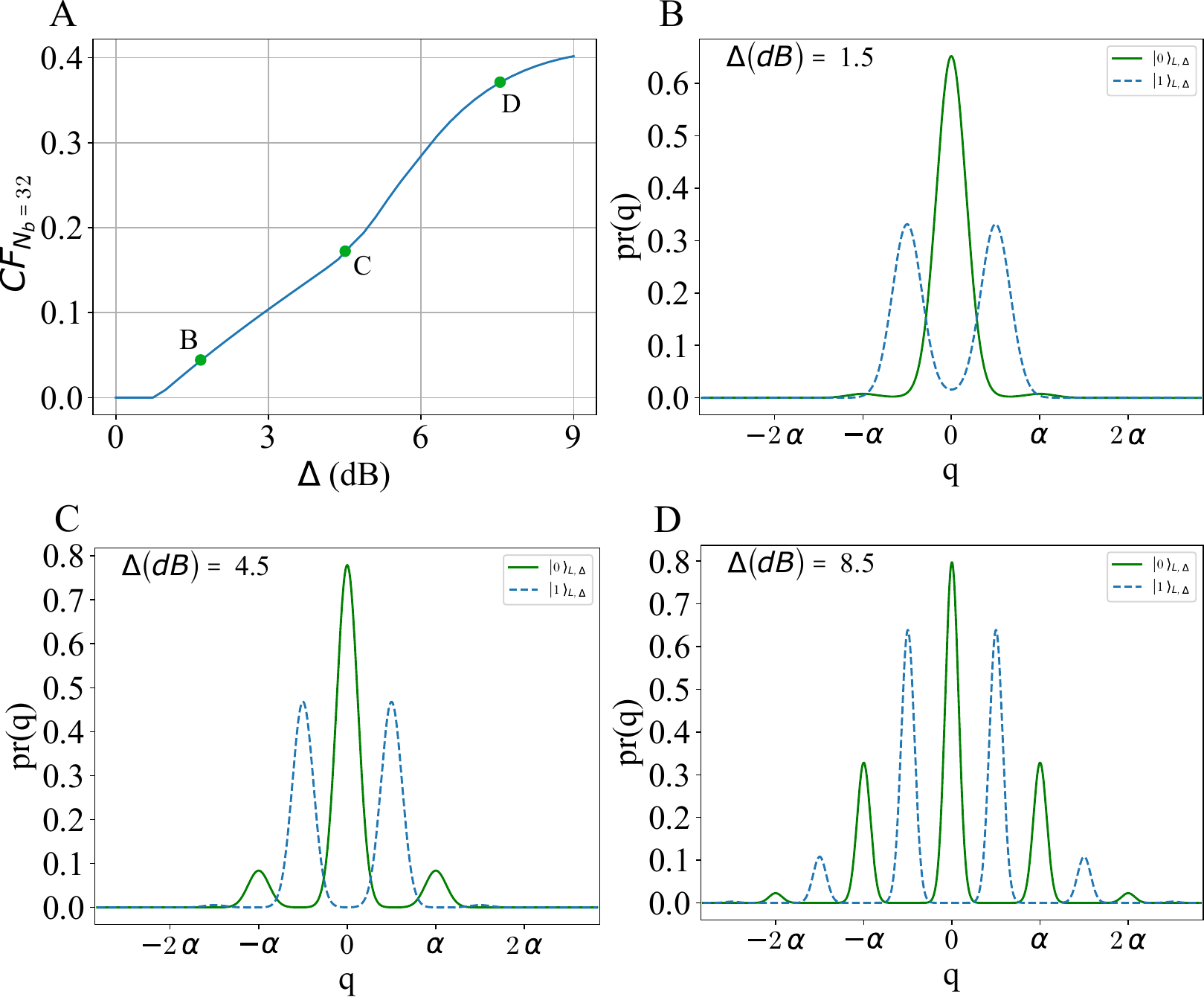}
    \caption{(A) Contextual fraction obtained as a function of the parameter $\Delta$ that governs the quality of the GKP encoding, for the entangled GKP state described in Eq.\eqref{eq:entangled_GKP}. We have considered in this plot the hierarchy at the level $k=5$, \ie considering $N_b = 32$ bins per quadrature. This suffices to capture most of the contextual behavior. (B), (C) and (D) represent the position quadrature probability densities associated to the GKP codewords for different levels of squeezing per peak, obtained from Eq.~\eqref{eq:finite_energy_GKP}. The corresponding contextual fractions are highlighted in A with green dots.}
    \label{fig:CF_vs_Delta}
\end{figure}

In Fig.~\ref{fig:CF_vs_Delta} we show the behavior of the contextual fraction --equivalently, the amount of non-locality-- of an entangled GKP state of the form in Eq.~\eqref{eq:entangled_GKP}, as a function of the parameter $\Delta$ governing the quality of the encoding. We show the value of $\Delta$ in decibels ($dB$), which is customary in the quantum optics literature. It is related to $\Delta$ in Eq.~\eqref{eq:finite_energy_GKP} by 
\begin{equation*}
    \Delta^2=\frac{10^{-\frac{\Delta_{dB}}{10}}}{2}.
\end{equation*}
 We consider the spacing between peaks to be $\alpha=2\sqrt{\pi}$. A different spacing would result in a shifted curve, but the qualitative behavior would be the same. In Appendix~\ref{app:entangled_GKP} we show a similar analysis for the entangled state in Eq.\eqref{eq:second_choice_of_GKP}.   \par
\begin{figure}[htbp]
\centering
\includegraphics[width =\linewidth]{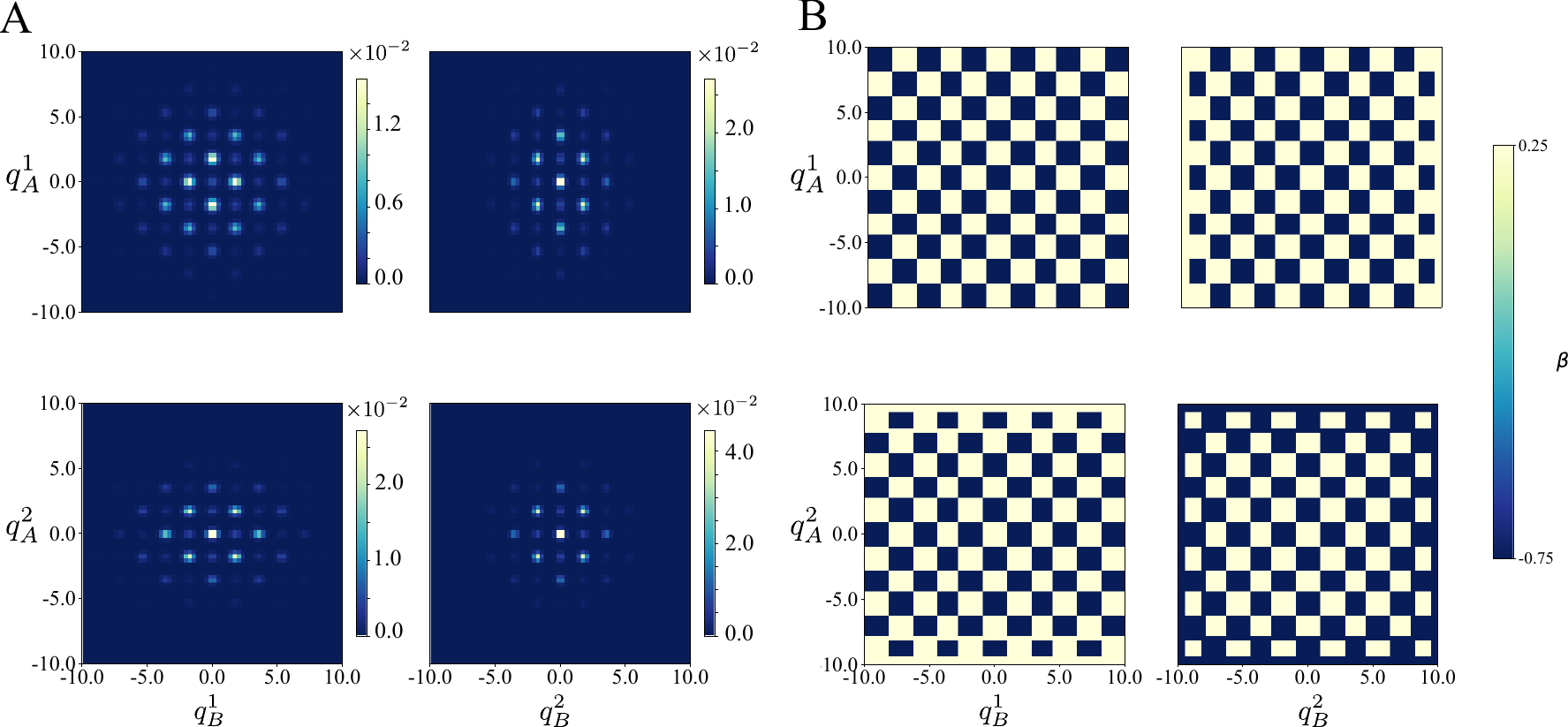}
\caption{(A) joint histograms corresponding to the measurement scenario given by the homodyne settings $\phi_{A,B}\in\{0,\pi/2\}$, on an entangled GKP state, given by equation \eqref{eq:entangled_GKP}. The approximate codewords are given by superpositions of $5$ and $6$ displaced squeezed states, for the $|0\rangle$ and $|1\rangle$ states respectively. (B) Filter functions $\beta_C$ that parametrize the optimal Bell inequality for the data at hand.   }
\label{fig:GKP_distribution_and_inequality}
\end{figure}

A first feature we can observe from Fig.~\ref{fig:CF_vs_Delta} is that for high quality states the amount of non-locality approaches $0.4$. This value is close to the bound for qubit systems in a measurement scenario where two parties perform two measurements, known as Tsirelson's bound \cite{tsirelson_quantum_1987}. The exact value of the bound is $CF_{T}=\sqrt{2}-1\approx 0.414$. In Fig.~\ref{fig:GKP_distribution_and_inequality}A we present the statistics corresponding to a measurement scenario where both parties measure position and momentum quadratures only on the state given in Eq.~\eqref{eq:entangled_GKP}, using finite-energy GKP encoding with $\Delta=0.25$ ($\Delta_{dB}\approx 9.1 dB$). To the right, in Fig.~\ref{fig:GKP_distribution_and_inequality}B we represent the filters parameterizing the optimal Bell inequality for this empirical behavior. The figure is made with $64$ bins per quadrature, for which we obtain $CF_{GKP}=0.412$, meaning that $41.2 \%$ of the data cannot be explained by a non-contextual model. Considering only $32$ bins the value obtained is $0.403$, for $16$ bins it is already $0.357$, while $8$ bins are not enough to capture the contextual behavior. \par 

The underlying GKP states used in Fig.~\ref{fig:GKP_distribution_and_inequality} are similar to the ones shown in Fig.~\ref{fig:CF_vs_Delta}D, which are very good approximations of ideal logical codewords. Though they lead to a significant non-local behavior under a straightforward measurement scenario, such states are very hard to engineer. In particular, in optical experiments, the best states reported so far, on the single-mode level, have no more than $3$ peaks, with very small inter-peak resolution \cite{Xanadu_2025}. Nevertheless, the main take-away from Fig.~\ref{fig:CF_vs_Delta}A is precisely that very low quality \textit{GKP states} (no longer worthy of the name), like those represented in Fig.~\ref{fig:CF_vs_Delta}B and Fig.~\ref{fig:CF_vs_Delta}C, are enough to witness non-locality. A remarkable example is given by the codewords in Fig.~\ref{fig:CF_vs_Delta}B, which are effectively a single squeezed peak and a cat state. Inspired by this, we can explicitly fix the encoding to be of the form 
\begin{equation}\label{eq:crapy_GKP}
    \begin{split}
        & |0\rangle=\hat S(r_{\Delta})|0\rangle_{b}\\
        & |1\rangle = \frac{\hat D(\alpha/2)+\hat D(-\alpha/2)}{\sqrt{2}} \hat S(r_{\Delta})|0\rangle_b,
    \end{split}
\end{equation}
where $|0\rangle_b$ corresponds to the vacuum state of the bosonic field, $\hat S(r_{\Delta})$ is the squeezing operator with parameter $r_\Delta=-\log (\sqrt{2} \Delta)$, and $\hat D(\alpha/2)$ is the displacement operator, that implements a displacement by $\alpha/2\in \mathbb R_+$ along the amplitude quadrature. The fact that these states are separately easy to prepare in optical experiments (depending on the value of $\alpha$) does not imply that preparing the entangled state given by the expansion in Eq.~\eqref{eq:entangled_GKP} is a simple task. Nevertheless, it is a more reasonable first step than preparing the highly multi-peak states of Fig.~\ref{fig:GKP_distribution_and_inequality}. 
\par 
\begin{figure}[htbp]
\centering
\includegraphics[width =\linewidth]{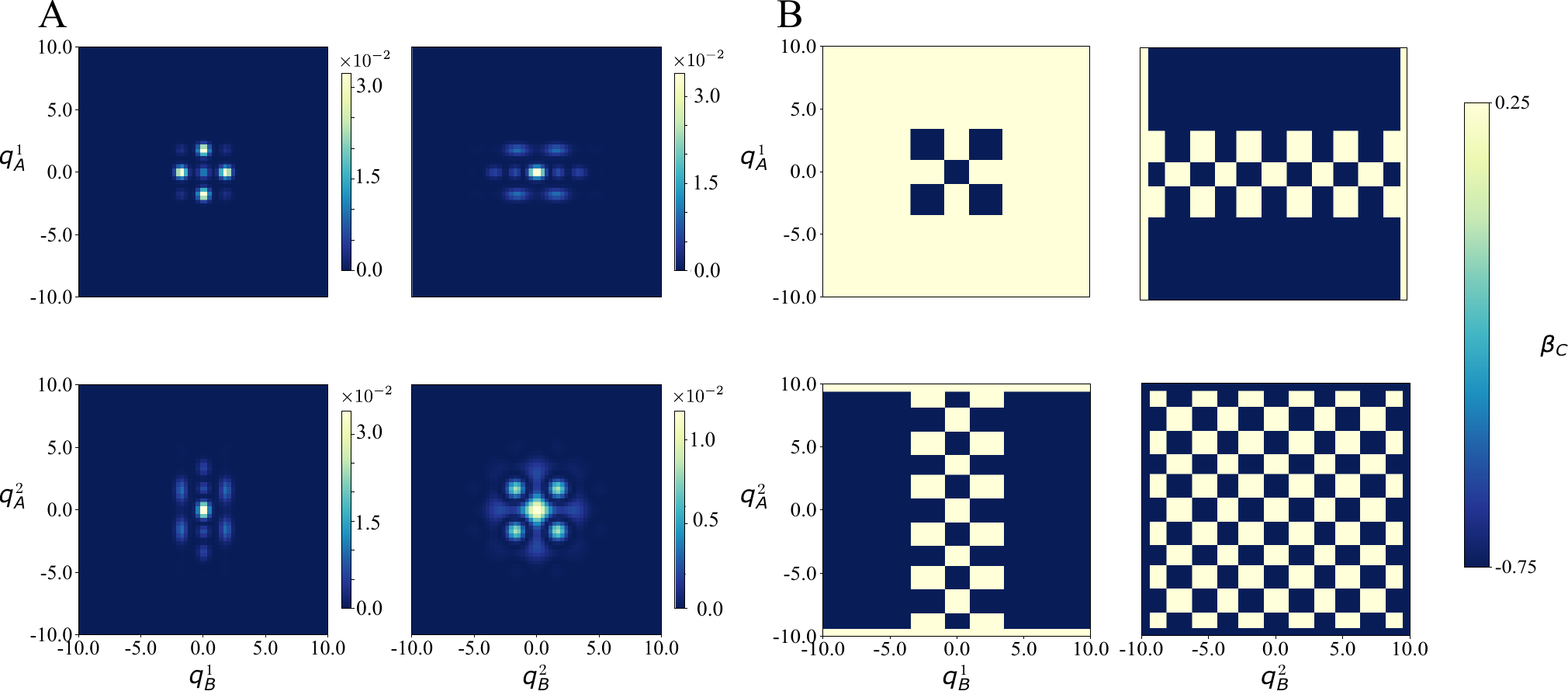}
\caption{(A) joint histograms corresponding to the measurement scenario given by the homodyne settings $\phi_{A,B}\in\{0,\pi/2\}$, on an entangled state of the form given in Eq.~\eqref{eq:entangled_GKP}. The approximate codewords employed are given in Eq.~\eqref{eq:crapy_GKP}, for $\Delta=0.28$, which corresponds to $\Delta_{dB}\approx 7.95 dB$. (B) Filter functions $\beta_C$ that parametrize the optimal Bell inequality for the empirical model at hand.   }
\label{fig:crappy_GKP_empirical_distribution_and_bell_filters}
\end{figure}

\begin{figure}[h]
\centering
\includegraphics[width =0.8\linewidth]{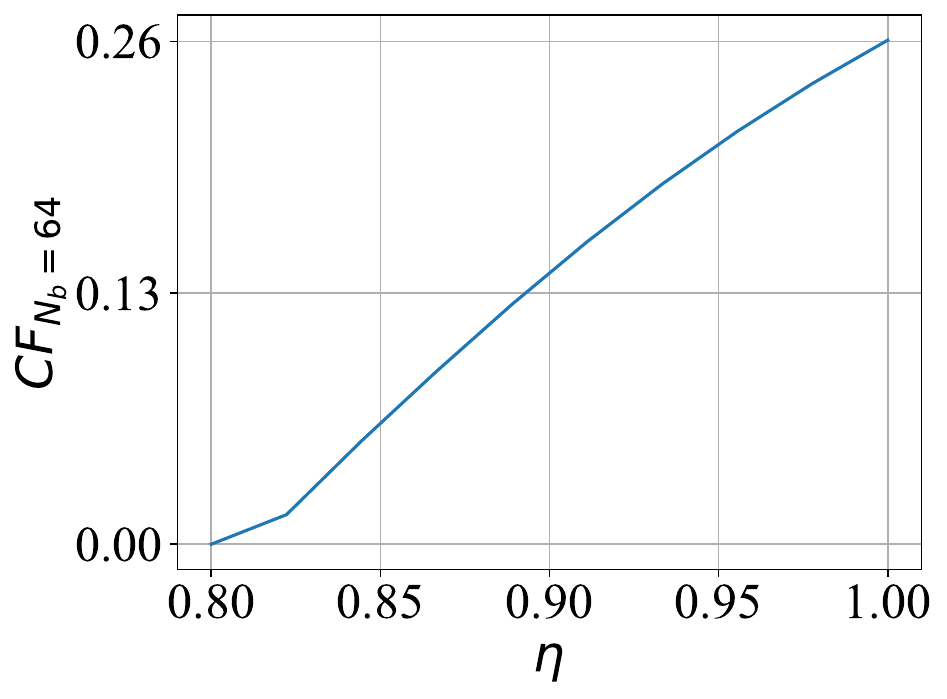}
\caption{Behavior of the contextual fraction under the effect of losses, for the joint statistics of position and momentum quadrature measurements on the entangled state in Eq.~\eqref{eq:entangled_GKP}, with approximate code states defined according to equation \eqref{eq:crapy_GKP} for $\Delta=0.28$. On the x axis we plot the transmissivity $\eta$ of the optical channel. The contextual behavior survives up to around $18\%$ losses. We consider a symmetric loss channel, where both modes undergo the same amount of losses. }
\label{fig:crappy_GKP_loss_resilience}
\end{figure}

In Fig.~\ref{fig:crappy_GKP_empirical_distribution_and_bell_filters} we can observe the statistics corresponding to the choice of $\Delta=0.28$ in Eq.~\eqref{eq:crapy_GKP}. To the right of the figure are the Bell filters $\beta_C$ that parametrize the optimal Bell inequality for this empirical behavior. The contextual fraction obtained is $CF_{cat}=0.26$, at level $k=6$ of the hierarchy of linear programs, \ie considering $N_b=64$ bins per quadrature. In Fig.~\ref{fig:crappy_GKP_loss_resilience}, we show the behavior of the contextual fraction when both of the modes go through lossy optical channels with the same loss rate $\eta$. We observe that a non-zero contextual fraction survives up to around $18\%$ losses, which is well within the range of state-of-the-art experiments. If instead we employ $32$ bins per quadrature, the tolerance reduces to around $14 \%$. Note that we consider an ideal preparation and only apply losses subsequently---depending on the actual realistic preparation scheme for the state in Eq.~\eqref{eq:entangled_GKP} the performance may differ. The simulation of the effect of losses and all the statistics of the resulting states were computed using the open source software MrMustard \cite{MrMustard}. \par 
Though the states that lead to the statistics studied in Fig.~\ref{fig:crappy_GKP_empirical_distribution_and_bell_filters} are not easy to produce in an experimental scenario, they are one step closer than the ones based on high quality GKP states. In particular, we can see them as superpositions of nine two-mode Gaussian states, which fits within the frameworks described in \cite{Braccini2025,Fede2026}. By appropriate coupling to finite-dimensional systems like qubits, it is possible, at least in principle, to create this kind of state. In the near-term this is doable in microwave cavities coupled to superconducting qubits, where single-mode GKP states are routinely prepared \cite{Wang2016,Touzard2018,Kudra2022,Loke2026}. Ion trap platforms \cite{fluhmann_encoding_2019,Fontbote2026}, or emerging acousto-optical systems \cite{Yang_2024} offer promising frameworks as well. Nevertheless, for these kinds of experiments, fast and efficient homodyne detection remain a challenge, with recent advances in superconducting platforms still far from the performance required for a Bell experiment \cite{Strandberg_2024}. An alternative would be to perform the full tomography of the state and obtain the marginals for the quadratures of interest, but this would no longer amount to a proper Bell experiment. \par 

\subsection{``Qutrit" GKP entangled states}\label{sec:qutrit_GKP}

Motivated by the qudit Bell inequalities introduced in \cite{CGLMP2002}, we now look for a family of bosonic states for which a Bell inequality with more than two values would be required, thus going beyond CHSH. 
The original proposal of Gottesman, Kitaev and Preskill in \cite{GKP_2001} included the encoding of arbitrary finite-dimensional quantum states into the infinite-dimensional Hilbert space of a harmonic oscillator. The ideal GKP codewords for a qudit of dimension $d$ are given by 
\begin{equation}\label{eq:qutrit_GKP_codewords}
    |j\rangle_d\propto \sum_{k\in \mathbb Z} |(kd +j)\frac{\alpha}{d}\rangle_q. 
\end{equation}

\begin{figure}[h]
\centering
\includegraphics[width =0.8\linewidth]{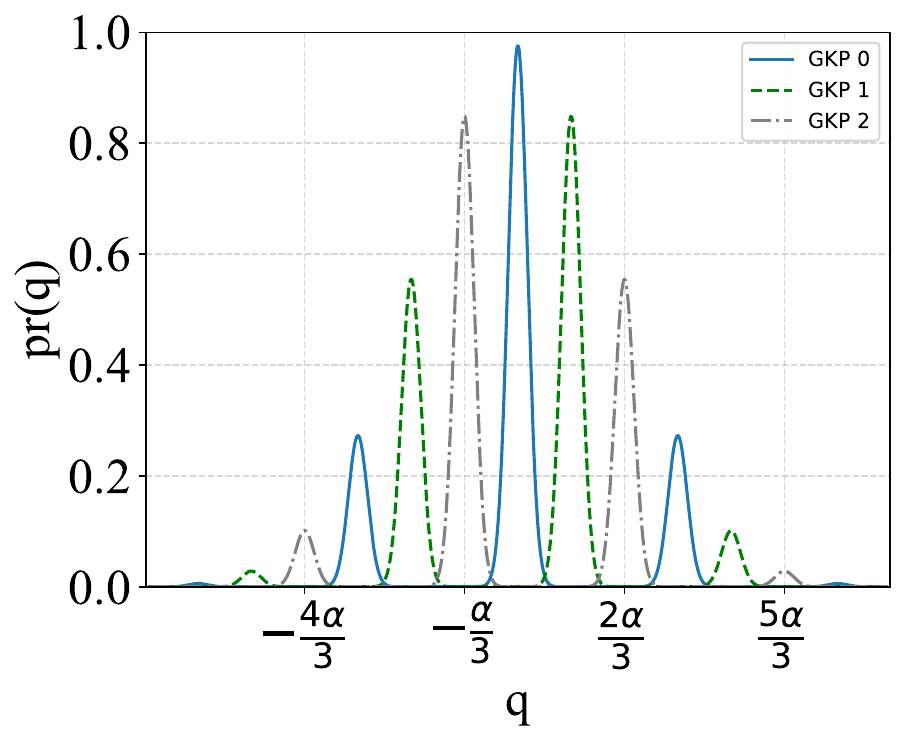}
\caption{Probability densities for the $q$ quadrature of the GKP codewords used to approximate a qutrit entangled state. }
\label{fig:qutrit_GKP_code_words}
\end{figure}

Finite-energy approximations are constructed using an expansion similar to that in Eq.~\eqref{eq:finite_energy_GKP}, in terms of finitely squeezed states, rather than position eigenstates. The resulting codewords for $d=3$, considering a squeezing of around $8.68 dB$ per peak, and a spacing $\alpha=\sqrt{6\pi}$, are represented in Fig.~\ref{fig:qutrit_GKP_code_words}. \par

For determining the relevant entangled states we follow a strategy similar to that outlined in the previous section for the qubit case. The Bell inequality introduced in \cite{CGLMP2002} is provided in a way which is consistent with the treatment of the finite LP program introduced in Section \ref{sec:finite_program_def}. It is thus not provided in operator form. We can nevertheless obtain an explicit operator form  following the procedure that we describe in Appendix~\ref{app:qutrit}. After obtaining the corresponding Bell operator, we optimize it for the case in which both parties perform generalized qutrit Pauli $X$ and $Z$ measurements. The resulting state involves a non-trivial superposition of qutrit codewords. The exact form of the state can be found in Appendix~\ref{app:qutrit}. \par 

\begin{figure}
\centering
\includegraphics[width =\linewidth]{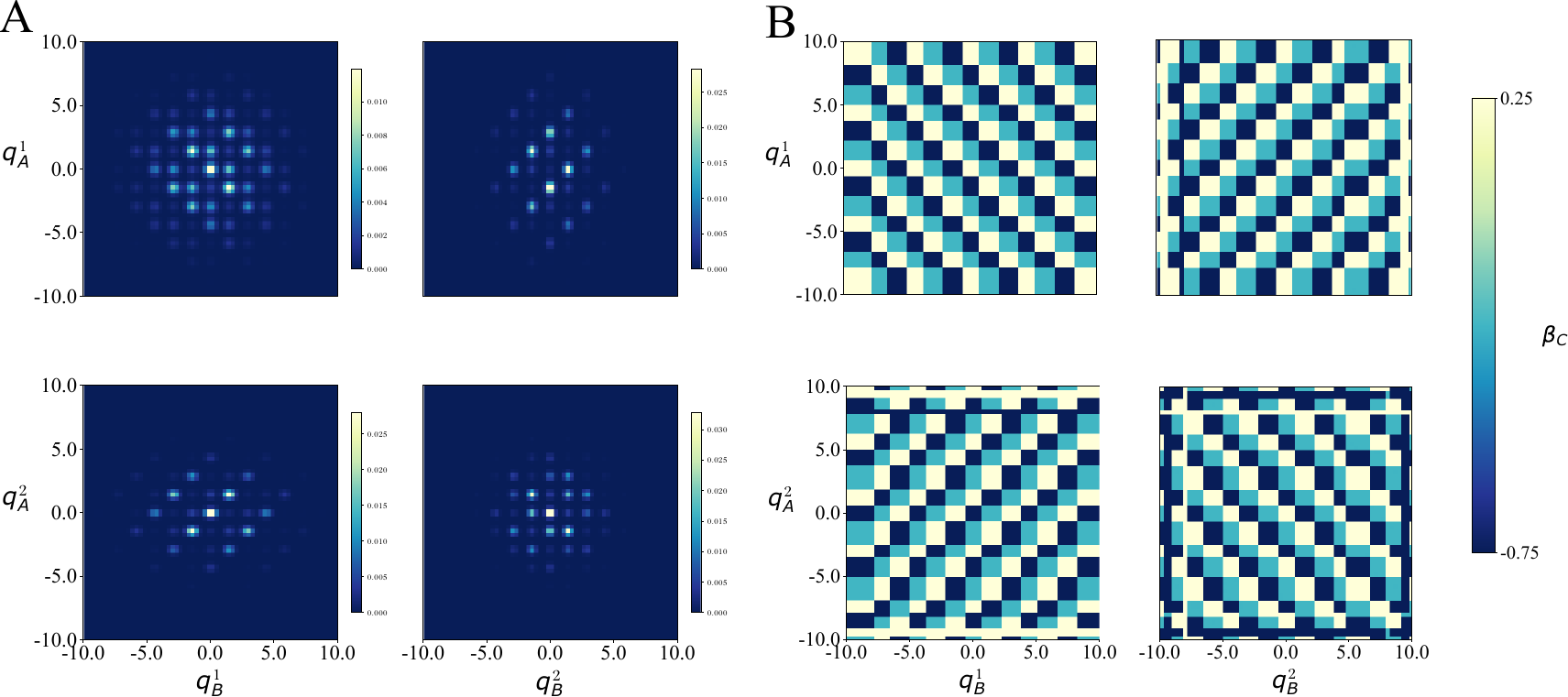}
\caption{To the left: joint histograms corresponding to the measurement scenario given by the homodyne settings $\phi_A,B\in\{0,\pi/2\}$, on an entangled GKP qutrit state, according to the description in Appendix \ref{app:qutrit}. The approximate codewords are plotted in Fig.~\ref{fig:qutrit_GKP_code_words}. To the right: Filter functions $\beta_C$ that parametrize the optimal Bell inequality for this empirical model. A contextual fraction of $28\%$ was obtained for this statistics.  }
\label{fig:qutrit_GKP_histograms_and_filters}
\end{figure}

We generate the entangled state according to the description presented in Appendix~\ref{app:qutrit} using the codewords provided in Fig.~\ref{fig:qutrit_GKP_code_words}. We consider the measurement scenario in which both Alice and Bob perform measurements of the amplitude and phase quadratures ($q$ and $p$, \ie $\phi_{A(B)}=\{0,\pi/2\}$). The corresponding statistics can be observed in Fig.~\ref{fig:qutrit_GKP_histograms_and_filters}A. In panel (B) of the same figure we show the optimal Bell inequality for this empirical model, parametrized by the filters $\beta_C$. The plots were made using $64$ bins per quadrature, which is enough to capture all the relevant features of the distribution. The contextual fraction observed is around $28 \%$. Using a smaller number of peaks still leads to a significant amount of non-locality. \par

The Bell inequality encoded by the filters plotted in Fig.~\ref{fig:qutrit_GKP_histograms_and_filters} offers, for the first time, a practical Bell inequality for CV states that is not equivalent to CHSH. It is, nevertheless, equivalent to another discrete Bell inequality, the CGLMP qutrit Bell inequality presented in \cite{CGLMP2002}. This, once again, highlights an apparent discrete structure of Bell non-locality in CV systems under homodyne detection. Further simplifications of the codewords, of the kind described in Eq.~\eqref{eq:crapy_GKP}, show that much simpler states than the ones leading to the data analyzed in Fig.~\ref{fig:qutrit_GKP_histograms_and_filters} still provide significant non-local behavior. 

\subsection{Three-mode states}\label{sec:multimode_GKP}
Considering multimode scenarios using the tools we have developed suffers from a limitation due to the exponential increase in computational requirements with respect to the number of modes, which makes it impossible to consider a very fine-grained binning of the histograms. In particular, if we consider three modes, and two measurements per mode, it becomes hard to consider more than 16 bins. On the other hand, as we will show in this section, 16 bins suffice to capture a very large contextual behavior in properly chosen states. \par

Composite systems of at least three parties offer a playground to realize strong non-locality (strong contextuality), characterized by a contextual fraction of $1$. This is the main idea behind the seminal paper by Greenberger, Horne and Zeillinger (GHZ) \cite{GHZ1990}.  In \cite{Abramsky2017}, these states were considered using tools similar to ours. In particular, they show that the empirical model obtained by performing Pauli $X$ and $Y$ measurements on each party of a GHZ state 
\begin{equation}\label{eq:GHZ}
    |\mathrm{GHZ}_3\rangle=\frac{|000\rangle+|111\rangle}{\sqrt{2}},
\end{equation}
is maximally contextual, \ie the contextual fraction is $1$. In \cite{Acin2009} a three-mode GHZ state in a CV regime was considered, encoded via even and odd cat states. In that case, a violation that is equivalent to a contextual fraction of around $10 \%$ is obtained. Here we consider a similar construction, but pick the encoded states to be approximate GKP codewords. We consider codewords of the form described in Eq.\eqref{eq:finite_energy_GKP} with a squeezing per peak of around $8.68 dB$ ($\Delta=0.26$). In each mode, we perform homodyne measurements of the $p$ quadrature and of the quadrature $q_{\pi/4}=\frac{p+q}{\sqrt{2}}$ ($\phi_{j}=\{\pi/2,\pi/4\}$). A modular measurement of these quadratures reproduces respectively Pauli $X$ and Pauli $Y$ measurements in the GKP encoding. Under this measurement scenario, we are able to certify a contextual fraction of $60.5 \%$ on the resulting experimental data considering only $16$ bins per quadrature.  If a larger number of bins could be considered, a larger contextual fraction would probably be obtained. Nevertheless, this suffices to witness contextuality that is higher than is possible to observe with the CHSH inequality in two modes, for which the limit is $\approx 41\%$. \par

\begin{figure}[htbp]
\centering
    \includegraphics[width=0.5\textwidth]{images/3D_GKP_2_X_4.pdf}
    \caption{(A) Slices of the 3D empirical models corresponding to the GKP encoded GHZ state in a measurement scenario in which each party measures along the quadratures $\hat q$ and $\hat q_{\pi/4}$. We show five slices of the form $p(q_A, \cdot, \cdot)$. (B) Slices of the 3D Bell filters $\beta_C(q_A, q_B, q_C)$ that parametrize the optimal Bell inequality for the data above. The Bell inequalities have a discrete structure similar to that in the two-mode case. In this case, the contextual fraction obtained was $\approx 0.6$.}
    \label{fig:three_mode_GHZ_emp_model_and_inequality}
\end{figure}
In Fig.~\ref{fig:three_mode_GHZ_emp_model_and_inequality}A we show slices of the empirical model we considered. In panel (B) we show slices of the corresponding optimal Bell inequality. We can observe that, as for the two-mode examples we have considered, the corresponding Bell inequality is of a discrete nature. For a direct handling of the inequalities we refer to the implementation of these examples \cite{repo}. \par 
Using the encoding described in \eqref{eq:crapy_GKP}, with a cat-state-like encoding the qubit state $|1\rangle$ and a squeezed state encoding the  qubit state $|0\rangle$, a contextual fraction of $8.5\%$ is obtained.

\begin{figure*}[t]
\centering
\includegraphics[width =0.8\linewidth]{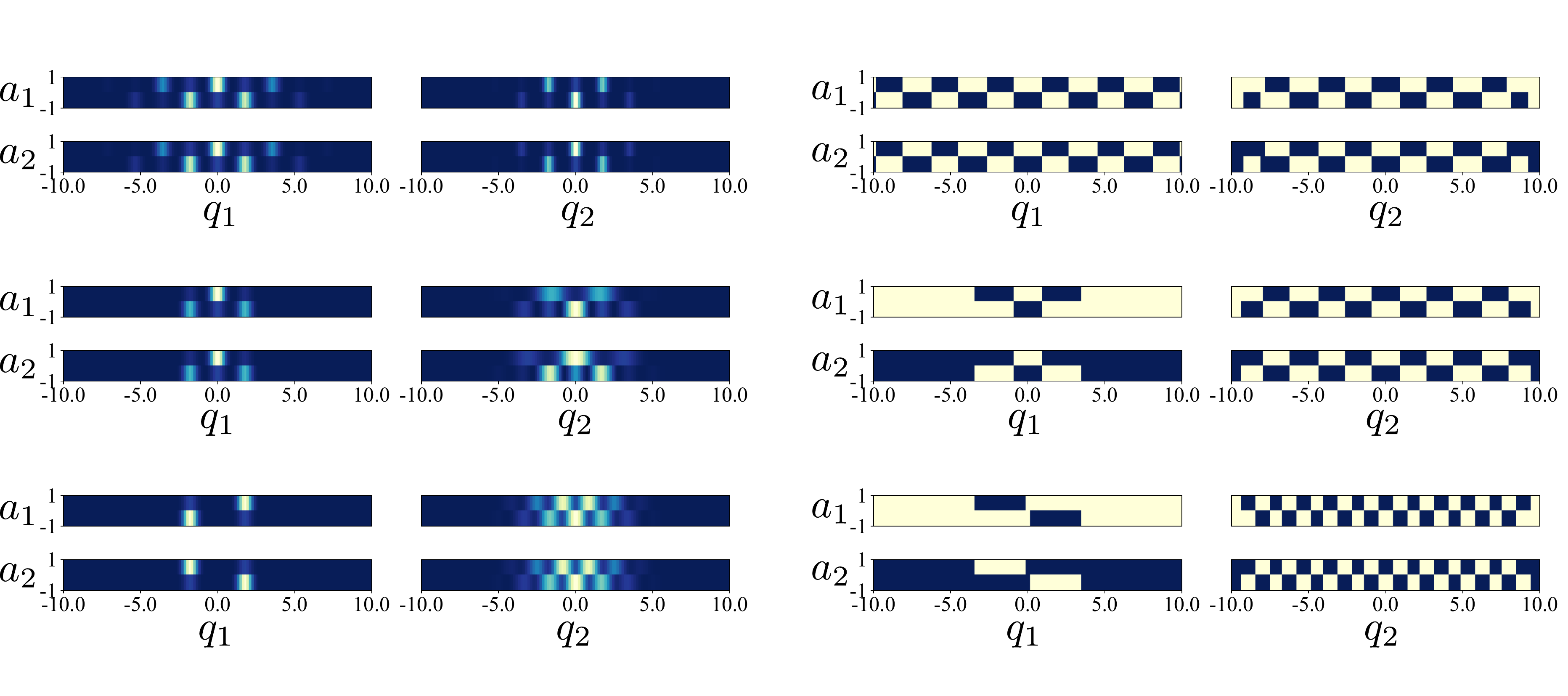}

\caption{From top to bottom the three different experimental data we considered for probing Bell non-locality on hybrid DV-CV systems, according to the measurement scenario described in the main text. To the left there are all the histograms that define the corresponding statistics. To the right, the filter functions $\beta_C$ that parametrize the optimal Bell inequality for each case. From top to bottom, contextual fraction values of $40\%$, $34.1\%$ and $15.6 \%$ were obtained. All color bars were omitted because only the features of the histograms and Bell inequalities are relevant here.  }
\label{fig:hybrid_cases}
\end{figure*}

\section{Results for hybrid systems}\label{sec:results_hybrid}
Hitherto we have explored experimental scenarios in which the states held by all parties involved are described in a CV setting, and homodyne detection is performed. Nevertheless, the approach that we have developed is agnostic to the actual measurements being performed, and is well-suited to deal with observables with discrete outcomes. In particular, it can be used to study hybrid systems, which are composed of both DV and CV subsystems \cite{Hybrid_tutorial}. The relevance of hybrid systems for quantum information processing has been increasingly recognized in recent years due to their relevance for the development of quantum technologies \cite{Hybrid_tutorial, Andersen2015}. Hybrid approaches are behind the huge success of platforms like circuit quantum electrodynamics \cite{Kudra2022}, optomechanical systems \cite{Yang_2024}, and ion-trapped systems \cite{fluhmann_encoding_2019} in creating exotic non-Gaussian states. Moreover, hybrid encoding of quantum information in optical systems appears to be a promising route for a scalable development of hybrid quantum networks \cite{Cavailles2018,Laurat2020,deGliniasty2024spinopticalquantum}.\par

Our analysis is platform-independent, in the sense that we consider the DV  subsystem in an agnostic way, assuming that one can perform arbitrary measurements on the Bloch sphere. This allows us to consider simpler states than in the purely CV setting, given that we need not maximize Bell inequalities violations specifically for Pauli measurements. In particular, the Bell state
\begin{equation}\label{eq:Bell_state}
|\psi\rangle=\frac{|0\rangle_{DV}|0\rangle_{CV}+|1\rangle_{DV}|1\rangle_{CV}}{\sqrt{2}},
\end{equation}
maximizes the CHSH Bell inequality when measuring $\{\frac{X+Z}{\sqrt{2}},\frac{X-Z}{\sqrt{2}}\}$ on one side and $\{X,Z\}$ on the other side. This motivates the use of a GKP encoding on the CV side. The probability distribution for any pair of observables will then be an object of dimension $ 2\times N_b$, where $N_b$ is the number of bins used to discretize the quadrature statistics of the CV subsystem. The wave vector associated to a pair of observables $\hat A$ on the qubit part and $\hat q_\phi$ on the CV side, is given by 
\begin{equation}\label{eq:hybrid_wavefunction}
    \psi(a,q)=\frac{\langle a|0 \rangle_{DV} \psi_0(q|\phi)+\langle a|1 \rangle_{DV} \psi_1(q|\phi)}{\sqrt{2}},
\end{equation}
where $|a\rangle$ is the eigenvector of $\hat A$ corresponding to the outcome (eigenvalue) $a$, and $\psi_j(q|\phi)$ are the wave functions of the CV state encoding the logical state $|j\rangle$, along the quadrature $\hat q_{\phi}$. The expression  in Eq.~\eqref{eq:hybrid_wavefunction} is generally easy to compute, and from that expression we can construct the distributions which define the empirical model to probe Bell non-locality. \par

We consider three different examples for the CV encoding: 
\begin{enumerate}
    \item Same codewords as in Fig.~\ref{fig:GKP_distribution_and_inequality}, \ie finite-energy GKP states with $\Delta\approx 0.25$.
    \item Same codewords as in equation \eqref{eq:crapy_GKP}, \ie a squeezed vacuum state encoding the $|0\rangle$, and a squeezed cat state for the $|1\rangle$. 
    \item Encoding as even and odd squeezed cat states, given by 
\begin{equation}
    \begin{split}
        &|0\rangle_{CV}=\frac{|\alpha,r\rangle+|-\alpha,r\rangle}{\mathcal N(\alpha,r)}\\
        &|1\rangle_{CV}=\frac{|\alpha,r\rangle-|-\alpha,r\rangle}{\mathcal N(\alpha,r)},
    \end{split}
\end{equation}
where $|\pm \alpha,r\rangle=\hat D(\pm\alpha/2)\hat S(r)|0\rangle$ is a displaced squeezed state. We consider $\alpha=2\sqrt{\pi}$, $r=0.9$.
\end{enumerate}  
In Fig.~\ref{fig:hybrid_cases} we show the empirical models corresponding to each of these possible choices. To the right of this figure we show the corresponding optimal Bell inequalities. All of these examples produce data with significant contextual fractions. For the specific parameters considered, the values are $40\%$, $34.1\%$ and $15.6\%$, respectively. The states corresponding to the third example have been previously prepared in \cite{Hacker2019}, in a hybrid optical-atomic system. In that experiment the optical system was addressed using homodyne detection, and is therefore ideal as a test-bed for our method. \par 

Here we have mainly analyzed the effect of different possible encodings of the CV side, fixing the measurement to homodyne detection. Nevertheless, a more complete analysis requires considering physically motivated restrictions on the possible measurements on the DV side and the inclusion of other choices of measurement on the bosonic side. The latter has been considered, for example, in \cite{Moradi_2024} for the analysis of hybrid optical systems. The versatility of our approach permits us to analyze any possible extension without any modification on the workflow described in Fig.~\ref{fig:general_workflow}. Moreover, an analysis of the effect of imperfections, informed by the specific choice of hybrid platform, can also be performed within our framework. We leave such explorations for future work directly applied to specific experimental implementations. \par

\section{Discussion and outlook}
\label{sec:discussion_results_and_open_questions}
On the technical level, our approach has the strength of being general, suitable for dealing with experimental data, and agnostic to the details of the underlying physics of the system. However, the scaling of our implementation with the number of modes is currently a bottleneck. Given the limitations of commonly used optimization solvers, if we want to study four-mode systems, we cannot consider more than ten bins per quadrature. This is a significant limitation when studying CV systems. For DV systems, or hybrid systems, the scaling improves significantly. Notably, for qubits, up to $32$ parties can be considered. An alternative to such discretization schemes is provided by the moment-based approach from \cite{PE_2022}. Nevertheless, as we already discussed, it does not seem to capture CV non-local behaviors within the order of moments that we can efficiently consider. Yet, we cannot exclude the possibility of finding a similar decomposition technique that could both reduce the dimensionality of the problem and capture the non-local behaviors, which is something we are actively exploring. Studying coarse-grained observables, like one does in spin systems \cite{PRXQuantum.2.030329}, or considering trigonometric moments rather than standard moments, may offer such alternatives. 
\par
Beyond the technical interest that the results obtained in this work may bear, they offer several new insights into the non-locality of homodyne-detection-based experiments. Our results question the existence and relevance of a genuinely continuous type of non-locality. Throughout all the examples that we have analyzed, a common feature emerges for all the optimal Bell inequalities: they are parametrized by discrete functions, \ie functions that evaluate to a finite number of values (2 or 3 in the cases we found). All the behaviors obtained can thus be simulated by experiments on DV systems. This hints at a discrete nature of homodyne-based non-locality. Given the extensive numerical search that we have performed, we conjecture that this may indeed be a general fact. However, as we have no formal proof,  this clearly provides an interesting open question for further research. 
\par 
Given that in all the examples considered the contextual fraction never increases by adding more than two homodyne settings per mode, we can furthermore conjecture that no advantage can be obtained from considering larger numbers of settings. We have not been able to prove (or disprove) this claim, and we also leave it as an open question. \par

Since Bell non-locality can be formulated as contextuality for Bell measurement scenarios, the study of CV contextuality may also improve our understanding of CV non-locality. However, we must emphasize that resource characterizations of contextuality do not automatically transfer to Bell non-locality. In particular, equivalences between contextuality and Wigner-function negativity established for broader measurement sets do not directly imply an analogous criterion for Bell non-locality under local homodyne measurements. Indeed, Wigner negativity can occur even for separable states (e.g., Werner-type mixtures built from non-Gaussian components), showing that negativity is not sufficient for Bell non-locality. Thus, connecting Bell non-locality to other physical resources remains an open challenge in the CV setting.\par
Overall, the search for states and homodyne measurement settings that provide non-local behavior has proven intensive, and in many ways feels like looking for a needle in a haystack. Indeed, we only managed to find robust Bell inequality violations through the use of bosonic codes, constructed explicitly to encode information efficiently in homodyne detection statistics. Interestingly though, once an example was found based on such constructions, it was sufficiently robust to allow for large departures from ideal encodings. This implies that many states may have been overlooked in our search, and reflects the many current gaps in the understanding of the emergence of non-locality in homodyne detection measurement scenarios.\par  
Finally, the methodology presented here is extremely versatile. The fact that it is agnostic to the actual dimension of the underlying system makes it universally relevant for continuous, discrete, and hybrid experiments. This data-driven approach returns the optimal Bell inequality, should one exist, for any input empirical statistics. Therefore, provided a restricted measurement scenario and physically motivated constraints on the family of preparable states, it is possible to perform a search for the optimal regime for Bell inequality violations. We expect it to be relevant for implementations across different experimental platforms, some of which we are currently exploring.

\section*{Acknowledgements}

We acknowledge fruitful discussions with Robert Booth, Damian Markham, Irénée Frérot, Lucas Porto and Antonio Ac\'in. C.E.L.G. and M.W. acknowledge support through the ANR JCJC project NORDIC (ANR-21-CE47-0005). C.E.L.G., M.W., and F.G. were supported by Plan France 2030 through the project OQuLus (ANR-22-PETQ-0013). U.I.M. was supported by Plan France 2030 through the project EPiQ (ANR-22-PETQ-0007). 
E.O. and G.M. acknowledge funding by the French national quantum initiative managed by Agence Nationale de la Recherche in the framework of France 2030 through the projects DIQKD (ANR-22-PETQ-0009) and QCommTestbed (ANR-22-PETQ-0011).
U.C. and M.W. acknowledge funding from the European Union’s Horizon Europe Framework Programme (EIC Pathfinder Challenge project Veriqub) under Grant Agreement No.\ 101114899. F.C. acknowledges funding from the European Union (EQC, 101149233). 
U.I.M. and F.G. acknowledge funding from the European Union’s Horizon Europe Framework Programme (CLUSTEC) under Grant Agreement No.\ 101080173.
\par 
\textit{Related work:} Shortly after the appearance of the first draft of this manuscript, results similar to those reported in Section \ref{sec:multimode_GKP} were presented in \cite{yang2026}, using entirely different methods. Further analysis of non-locality with GKP and cat-state encodings has recently been presented in \cite{Salek2026,Erkilic2026}, with an emphasis on the use of non-Gaussian measurements, which allows them to obtain performances similar to those reported in Fig.~\ref{fig:CF_vs_Delta}, but using GKP Bell states.  

\bibliography{biblio.bib}

@article{review_Bell_nonlocality,
  title = {Bell nonlocality},
  author = {Brunner, Nicolas and Cavalcanti, Daniel and Pironio, Stefano and Scarani, Valerio and Wehner, Stephanie},
  journal = {Rev. Mod. Phys.},
  volume = {86},
  issue = {2},
  pages = {419--478},
  numpages = {60},
  year = {2014},
  month = {Apr},
  publisher = {American Physical Society},
  doi = {10.1103/RevModPhys.86.419},
  url = {https://link.aps.org/doi/10.1103/RevModPhys.86.419}
}

@article{Kaszlikowski2000,
  title = {Violations of Local Realism by Two Entangled $\mathit{N}$-Dimensional Systems Are Stronger than for Two Qubits},
  author = {Kaszlikowski, Dagomir and Gnaci\ifmmode \acute{n}\else \'{n}\fi{}ski, Piotr and \ifmmode \dot{Z}\else \.{Z}\fi{}ukowski, Marek and Miklaszewski, Wieslaw and Zeilinger, Anton},
  journal = {Phys. Rev. Lett.},
  volume = {85},
  issue = {21},
  pages = {4418--4421},
  numpages = {0},
  year = {2000},
  month = {Nov},
  publisher = {American Physical Society},
  doi = {10.1103/PhysRevLett.85.4418},
  url = {https://link.aps.org/doi/10.1103/PhysRevLett.85.4418}
}

@misc{Salek2026,
      title={Robust CHSH Self-Testing with Finite-Energy GKP States}, 
      author={Farzin Salek and Masahito Hayashi},
      year={2026},
      eprint={2608.11122},
      archivePrefix={arXiv},
      primaryClass={quant-ph},
      url={https://arxiv.org/abs/2608.11122}, 
}

@misc{Erkilic2026,
      title={Near-Optimal Bell Nonlocality with GKP and Entangled Cat States}, 
      author={Özlem Erkılıç and Aritra Das and S. Nibedita Swain and Simon Devitt and Timothy C. Ralph},
      year={2026},
      eprint={2609.27640},
      archivePrefix={arXiv},
      primaryClass={quant-ph},
      url={https://arxiv.org/abs/2609.27640}, 
}

@article{Nha2004,
  title = {Proposed Test of Quantum Nonlocality for Continuous Variables},
  author = {Nha, Hyunchul and Carmichael, H. J.},
  journal = {Phys. Rev. Lett.},
  volume = {93},
  issue = {2},
  pages = {020401},
  numpages = {4},
  year = {2004},
  month = {Jul},
  publisher = {American Physical Society},
  doi = {10.1103/PhysRevLett.93.020401},
  url = {https://link.aps.org/doi/10.1103/PhysRevLett.93.020401}
}

@misc{yang2026,
      title={Robust {B}ell Nonlocality from {G}ottesman-{K}itaev-{P}reskill States}, 
      author={Xiaotian Yang and Santiago Zamora and Rafael Chaves and Ulrik L. Andersen and Jonatan Bohr Brask and A. de Oliveira Junior},
      year={2026},
      eprint={2601.16189},
      archivePrefix={arXiv},
      primaryClass={quant-ph},
      url={https://arxiv.org/abs/2601.16189}, 
}

@misc{repo,
  author       = {Lopetegui-González, Carlos Ernesto},
  title        = {{CV}\_contextuality},
  version      = {},
  year         = {2026},
  note         = {Repository of the code used in this article, archived by Software Heritage as \texttt{swh:1:dir:\allowbreak5351f8e6\allowbreak9770cdb2\allowbreak8cc2b81b\allowbreak d7b4bcbd\allowbreak965b349d}},
  url          = {https://archive.softwareheritage.org/swh:1:dir:5351f8e69770cdb28cc2b81bd7b4bcbd965b349d}
}

@article{PE_2022,
	title = {Continuous-variable nonlocality and contextuality},
	volume = {391},
	issn = {0010-3616, 1432-0916},
	url = {http://arxiv.org/abs/1905.08267},
	doi = {10.1007/s00220-021-04285-7},
	pages = {1047--1089},
	number = {3},
	journal = {Commun. Math. Phys.},
	author = {Barbosa, Rui Soares and Douce, Tom and Emeriau, Pierre-Emmanuel and Kashefi, Elham and Mansfield, Shane},
	date = {2022-05},
    year = {2022}
}

@article{PRXQuantum.2.030329,
  title = {Inferring Nonlinear Many-Body Bell Inequalities From Average Two-Body Correlations: Systematic Approach for Arbitrary Spin-$j$ Ensembles},
  author = {M\"uller-Rigat, Guillem and Aloy, Albert and Lewenstein, Maciej and Fr\'erot, Ir\'en\'ee},
  journal = {PRX Quantum},
  volume = {2},
  issue = {3},
  pages = {030329},
  numpages = {20},
  year = {2021},
  month = {Aug},
  publisher = {American Physical Society},
  doi = {10.1103/PRXQuantum.2.030329},
  url = {https://link.aps.org/doi/10.1103/PRXQuantum.2.030329}
}

@article{popescu_quantum_1994,
	title = {Quantum nonlocality as an axiom},
	volume = {24},
	issn = {1572-9516},
	url = {https://doi.org/10.1007/BF02058098},
	doi = {10.1007/BF02058098},
	pages = {379--385},
	number = {3},
	journal = {Foundations of Physics},
	shortjournal = {Found Phys},
	author = {Popescu, Sandu and Rohrlich, Daniel},
    year = {1994},
	date = {1994-03-01},
	langid = {english},
}

@article{EPR1935,
  title = {Can Quantum-Mechanical Description of Physical Reality Be Considered Complete?},
  author = {Einstein, A. and Podolsky, B. and Rosen, N.},
  journal = {Phys. Rev.},
  volume ={47},
  issue ={10},
  pages = {777-780},
  year = {1935},
  month = {May},
  publisher = {American Physical Society},
  doi = {10.1103/PhysRev.47.777},
  url = {https://link.aps.org/doi/10.1103/PhysRev.47.777}
}

@article{Bell_1964,
  title = {On the Einstein Podolsky Rosen paradox},
  author = {Bell, J. S.},
  journal = {Physics Physique Fizika},
  volume = {1},
  issue = {3},
  pages = {195--200},
  numpages = {6},
  year = {1964},
  month = {Nov},
  publisher = {American Physical Society},
  doi = {10.1103/PhysicsPhysiqueFizika.1.195},
  url = {https://link.aps.org/doi/10.1103/PhysicsPhysiqueFizika.1.195}
}

@article{Kochen_Specker,
 ISSN = {00959057, 19435274},
 URL = {http://www.jstor.org/stable/24902153},
 author = {Simon Kochen and E. P. Specker},
 journal = {Journal of Mathematics and Mechanics},
 number = {1},
 pages = {59--87},
 publisher = {Indiana University Mathematics Department},
 title = {The Problem of Hidden Variables in Quantum Mechanics},
 urldate = {2025-11-12},
 volume = {17},
 year = {1967}
}

@article{CHSH,
  title = {Proposed Experiment to Test Local Hidden-Variable Theories},
  author = {Clauser, John F. and Horne, Michael A. and Shimony, Abner and Holt, Richard A.},
  journal = {Phys. Rev. Lett.},
  volume = {23},
  issue = {15},
  pages = {880--884},
  numpages = {0},
  year = {1969},
  month = {Oct},
  publisher = {American Physical Society},
  doi = {10.1103/PhysRevLett.23.880},
  url = {https://link.aps.org/doi/10.1103/PhysRevLett.23.880}
}

@article{Mermin1990,
  title = {Extreme quantum entanglement in a superposition of macroscopically distinct states},
  author = {Mermin, N. David},
  journal = {Phys. Rev. Lett.},
  volume = {65},
  issue = {15},
  pages = {1838--1840},
  numpages = {0},
  year = {1990},
  month = {Oct},
  publisher = {American Physical Society},
  doi = {10.1103/PhysRevLett.65.1838},
  url = {https://link.aps.org/doi/10.1103/PhysRevLett.65.1838}
}

@article{KLYSHKO1996,
title = {The Bell theorem and the problem of moments},
journal = {Physics Letters A},
volume = {218},
number = {3},
pages = {119-127},
year = {1996},
issn = {0375-9601},
doi = {https://doi.org/10.1016/0375-9601(96)00444-6},
url = {https://www.sciencedirect.com/science/article/pii/0375960196004446},
author = {D.N. Klyshko}
}

@article{tsirelson_quantum_1987,
	title = {Quantum analogues of the Bell inequalities. The case of two spatially separated domains},
	volume = {36},
	issn = {1573-8795},
	url = {https://doi.org/10.1007/BF01663472},
	doi = {10.1007/BF01663472},
	pages = {557--570},
	number = {4},
	journal = {Journal of Soviet Mathematics},
	shortjournal = {J Math Sci},
	author = {Tsirel'son, B. S.},
	date = {1987-02-01},    year= {1987},
	langid = {english},
}

@article{CGLMP2002,
  title = {Bell Inequalities for Arbitrarily High-Dimensional Systems},
  author = {Collins, Daniel and Gisin, Nicolas and Linden, Noah and Massar, Serge and Popescu, Sandu},
  journal = {Phys. Rev. Lett.},
  volume = {88},
  issue = {4},
  pages = {040404},
  numpages = {4},
  year = {2002},
  month = {Jan},
  publisher = {American Physical Society},
  doi = {10.1103/PhysRevLett.88.040404},
  url = {https://link.aps.org/doi/10.1103/PhysRevLett.88.040404}
}

@article{GHZ1990,
    author = {Greenberger, Daniel M. and Horne, Michael A. and Shimony, Abner and Zeilinger, Anton},
    title = {Bell’s theorem without inequalities},
    journal = {American Journal of Physics},
    volume = {58},
    number = {12},
    pages = {1131-1143},
    year = {1990},
    month = {12},
    issn = {0002-9505},
    doi = {10.1119/1.16243},
    url = {https://doi.org/10.1119/1.16243}
}

@article{Aolita_2018,
  title = {Continuous-variable supraquantum nonlocality},
  author = {Ketterer, Andreas and Laversanne-Finot, Adrien and Aolita, Leandro},
  journal = {Phys. Rev. A},
  volume = {97},
  issue = {1},
  pages = {012133},
  numpages = {10},
  year = {2018},
  month = {Jan},
  publisher = {American Physical Society},
  doi = {10.1103/PhysRevA.97.012133},
  url = {https://link.aps.org/doi/10.1103/PhysRevA.97.012133}
}

@article{Aspect1982,
  title = {Experimental Test of Bell's Inequalities Using Time-Varying Analyzers},
  author = {Aspect, Alain and Dalibard, Jean and Roger, G\'erard},
  journal = {Phys. Rev. Lett.},
  volume = {49},
  issue = {25},
  pages = {1804--1807},
  numpages = {0},
  year = {1982},
  month = {Dec},
  publisher = {American Physical Society},
  doi = {10.1103/PhysRevLett.49.1804},
  url = {https://link.aps.org/doi/10.1103/PhysRevLett.49.1804}
}

@article{Zeilinger2015,
  title = {Significant-Loophole-Free Test of Bell's Theorem with Entangled Photons},
  author = {Giustina, Marissa and Versteegh, Marijn A. M. and Wengerowsky, S\"oren and Handsteiner, Johannes and Hochrainer, Armin and Phelan, Kevin and Steinlechner, Fabian and Kofler, Johannes and Larsson, Jan-\AA{}ke and Abell\'an, Carlos and Amaya, Waldimar and Pruneri, Valerio and Mitchell, Morgan W. and Beyer, J\"orn and Gerrits, Thomas and Lita, Adriana E. and Shalm, Lynden K. and Nam, Sae Woo and Scheidl, Thomas and Ursin, Rupert and Wittmann, Bernhard and Zeilinger, Anton},
  journal = {Phys. Rev. Lett.},
  volume = {115},
  issue = {25},
  pages = {250401},
  numpages = {7},
  year = {2015},
  month = {Dec},
  publisher = {American Physical Society},
  doi = {10.1103/PhysRevLett.115.250401},
  url = {https://link.aps.org/doi/10.1103/PhysRevLett.115.250401}
}

@article{hensen_loophole-free_2015,
	title = {Loophole-free Bell inequality violation using electron spins separated by 1.3 kilometres},
	volume = {526},
	rights = {2015 Springer Nature Limited},
	issn = {1476-4687},
	url = {https://www.nature.com/articles/nature15759},
	doi = {10.1038/nature15759},
	pages = {682--686},
	number = {7575},
	journal = {Nature},
	author = {Hensen, B. and Bernien, H. and Dréau, A. E. and Reiserer, A. and Kalb, N. and Blok, M. S. and Ruitenberg, J. and Vermeulen, R. F. L. and Schouten, R. N. and Abellán, C. and Amaya, W. and Pruneri, V. and Mitchell, M. W. and Markham, M. and Twitchen, D. J. and Elkouss, D. and Wehner, S. and Taminiau, T. H. and Hanson, R.},
	urldate = {2025-11-10},
	date = {2015-10},
year={2015},
	langid = {english},
	note = {Publisher: Nature Publishing Group},
}

@article{Shalm_2015,
  title = {Strong Loophole-Free Test of Local Realism},
  author = {Shalm, Lynden K. and Meyer-Scott, Evan and Christensen, Bradley G. and Bierhorst, Peter and Wayne, Michael A. and Stevens, Martin J. and Gerrits, Thomas and Glancy, Scott and Hamel, Deny R. and Allman, Michael S. and Coakley, Kevin J. and Dyer, Shellee D. and Hodge, Carson and Lita, Adriana E. and Verma, Varun B. and Lambrocco, Camilla and Tortorici, Edward and Migdall, Alan L. and Zhang, Yanbao and Kumor, Daniel R. and Farr, William H. and Marsili, Francesco and Shaw, Matthew D. and Stern, Jeffrey A. and Abell\'an, Carlos and Amaya, Waldimar and Pruneri, Valerio and Jennewein, Thomas and Mitchell, Morgan W. and Kwiat, Paul G. and Bienfang, Joshua C. and Mirin, Richard P. and Knill, Emanuel and Nam, Sae Woo},
  journal = {Phys. Rev. Lett.},
  volume = {115},
  issue = {25},
  pages = {250402},
  numpages = {10},
  year = {2015},
  month = {Dec},
  publisher = {American Physical Society},
  doi = {10.1103/PhysRevLett.115.250402},
  url = {https://link.aps.org/doi/10.1103/PhysRevLett.115.250402}
}

@article{Rosenfeld_2017,
  title = {Event-Ready Bell Test Using Entangled Atoms Simultaneously Closing Detection and Locality Loopholes},
  author = {Rosenfeld, Wenjamin and Burchardt, Daniel and Garthoff, Robert and Redeker, Kai and Ortegel, Norbert and Rau, Markus and Weinfurter, Harald},
  journal = {Phys. Rev. Lett.},
  volume = {119},
  issue = {1},
  pages = {010402},
  numpages = {6},
  year = {2017},
  month = {Jul},
  publisher = {American Physical Society},
  doi = {10.1103/PhysRevLett.119.010402},
  url = {https://link.aps.org/doi/10.1103/PhysRevLett.119.010402}
}

@article{Li_2018,
  title = {Test of Local Realism into the Past without Detection and Locality Loopholes},
  author = {Li, Ming-Han and Wu, Cheng and Zhang, Yanbao and Liu, Wen-Zhao and Bai, Bing and Liu, Yang and Zhang, Weijun and Zhao, Qi and Li, Hao and Wang, Zhen and You, Lixing and Munro, W. J. and Yin, Juan and Zhang, Jun and Peng, Cheng-Zhi and Ma, Xiongfeng and Zhang, Qiang and Fan, Jingyun and Pan, Jian-Wei},
  journal = {Phys. Rev. Lett.},
  volume = {121},
  issue = {8},
  pages = {080404},
  numpages = {7},
  year = {2018},
  month = {Aug},
  publisher = {American Physical Society},
  doi = {10.1103/PhysRevLett.121.080404},
  url = {https://link.aps.org/doi/10.1103/PhysRevLett.121.080404}
}

@article{Vidick_2014,
  title = {Fully Device-Independent Quantum Key Distribution},
  author = {Vazirani, Umesh and Vidick, Thomas},
  journal = {Phys. Rev. Lett.},
  volume = {113},
  issue = {14},
  pages = {140501},
  numpages = {6},
  year = {2014},
  month = {Sep},
  publisher = {American Physical Society},
  doi = {10.1103/PhysRevLett.113.140501},
  url = {https://link.aps.org/doi/10.1103/PhysRevLett.113.140501}
}

@inbook{Bachor2019,
publisher = {John Wiley \& Sons, Ltd},
isbn = {9783527695805},
title = {Homodyne detection},
booktitle = {A Guide to Experiments in Quantum Optics},
chapter = {9},
pages = {303-376},
doi = {https://doi.org/10.1002/9783527695805.ch9},
url = {https://onlinelibrary.wiley.com/doi/abs/10.1002/9783527695805.ch9},
year = {2019},

}

@article{braunstein_quantum_2005,
	title = {Quantum information with continuous variables},
	volume = {77},
	url = {https://link.aps.org/doi/10.1103/RevModPhys.77.513},
	doi = {10.1103/RevModPhys.77.513},
	pages = {513--577},
	number = {2},
	journal = {Rev. Mod. Phys.},
	author = {Braunstein, Samuel L. and van Loock, Peter},
	urldate = {2025-05-08},
	date = {2005-06-29},
    year={2005},
	note = {Publisher: American Physical Society},
}

@article{Treps2020,
  title = {Modes and states in quantum optics},
  author = {Fabre, C. and Treps, N.},
  journal = {Rev. Mod. Phys.},
  volume = {92},
  issue = {3},
  pages = {035005},
  numpages = {40},
  year = {2020},
  month = {Sep},
  publisher = {American Physical Society},
  doi = {10.1103/RevModPhys.92.035005},
  url = {https://link.aps.org/doi/10.1103/RevModPhys.92.035005}
}

@article{Ra2020,
   author = {Ra, Young-Sik and Dufour, Adrien and Walschaers, Mattia and Jacquard, Clément and Michel, Thibault and Fabre, Claude and Treps, Nicolas},
   title = {Non-{G}aussian quantum states of a multimode light field},
   journal = {Nat. Phys.},
   volume = {16},
   number = {2},
   pages = {144-147},
   ISSN = {1745-2481},
   DOI = {10.1038/s41567-019-0726-y},
   url = {https://doi.org/10.1038/s41567-019-0726-y},
   year = {2020},
   type = {Journal Article}
}

@misc{andrei2025,
      title={Photon catalysis for general multimode multi-photon quantum state preparation}, 
      author={Andrei Aralov and Émilie Gillet and Viet Nguyen and Andrea Cosentino and Mattia Walschaers and Massimo Frigerio},
      year={2025},
      eprint={2507.19397},
      archivePrefix={arXiv},
      primaryClass={quant-ph},
      url={https://arxiv.org/abs/2507.19397}, 
}

@article{GKP_2001,
  title = {Encoding a qubit in an oscillator},
  author = {Gottesman, Daniel and Kitaev, Alexei and Preskill, John},
  journal = {Phys. Rev. A},
  volume = {64},
  issue = {1},
  pages = {012310},
  numpages = {21},
  year = {2001},
  month = {Jun},
  publisher = {American Physical Society},
  doi = {10.1103/PhysRevA.64.012310},
  url = {https://link.aps.org/doi/10.1103/PhysRevA.64.012310}
}

@article{Xanadu_2025,
	title = {Integrated photonic source of Gottesman–Kitaev–Preskill qubits},
	volume = {642},
	rights = {2025 The Author(s)},
	issn = {1476-4687},
	url = {https://www.nature.com/articles/s41586-025-09044-5},
	doi = {10.1038/s41586-025-09044-5},
	pages = {587--591},
	number = {8068},
	journal = {Nature},
	author = {Larsen, M. V. and Bourassa, J. E. and Kocsis, S. and Tasker, J. F. and Chadwick, R. S. and González-Arciniegas, C. and Hastrup, J. and Lopetegui-González, C. E. and Miatto, F. M. and Motamedi, A. and Noro, R. and Roeland, G. and Baby, R. and Chen, H. and Contu, P. and Di Luch, I. and Drago, C. and Giesbrecht, M. and Grainge, T. and Krasnokutska, I. and Menotti, M. and Morrison, B. and Puviraj, C. and Rezaei Shad, K. and Hussain, B. and {McMahon}, J. and Ortmann, J. E. and Collins, M. J. and Ma, C. and Phillips, D. S. and Seymour, M. and Tang, Q. Y. and Yang, B. and Vernon, Z. and Alexander, R. N. and Mahler, D. H.},
	urldate = {2025-11-12},
	year = {2025},
	langid = {english},
	note = {Publisher: Nature Publishing Group},
}

@article{Kudra2022,
  title = {Robust Preparation of Wigner-Negative States with Optimized SNAP-Displacement Sequences},
  author = {Kudra, Marina and Kervinen, Mikael and Strandberg, Ingrid and Ahmed, Shahnawaz and Scigliuzzo, Marco and Osman, Amr and Lozano, Daniel P\'erez and Thol\'en, Mats O. and Borgani, Riccardo and Haviland, David B. and Ferrini, Giulia and Bylander, Jonas and Kockum, Anton Frisk and Quijandr\'{\i}a, Fernando and Delsing, Per and Gasparinetti, Simone},
  journal = {PRX Quantum},
  volume = {3},
  issue = {3},
  pages = {030301},
  numpages = {12},
  year = {2022},
  month = {Jul},
  publisher = {American Physical Society},
  doi = {10.1103/PRXQuantum.3.030301},
  url = {https://link.aps.org/doi/10.1103/PRXQuantum.3.030301}
}

@article{Touzard2018,
  title = {Coherent Oscillations inside a Quantum Manifold Stabilized by Dissipation},
  author = {Touzard, S. and Grimm, A. and Leghtas, Z. and Mundhada, S. O. and Reinhold, P. and Axline, C. and Reagor, M. and Chou, K. and Blumoff, J. and Sliwa, K. M. and Shankar, S. and Frunzio, L. and Schoelkopf, R. J. and Mirrahimi, M. and Devoret, M. H.},
  journal = {Phys. Rev. X},
  volume = {8},
  issue = {2},
  pages = {021005},
  numpages = {7},
  year = {2018},
  month = {Apr},
  publisher = {American Physical Society},
  doi = {10.1103/PhysRevX.8.021005},
  url = {https://link.aps.org/doi/10.1103/PhysRevX.8.021005}
}

@article{fluhmann_encoding_2019,
	title = {Encoding a qubit in a trapped-ion mechanical oscillator},
	volume = {566},
	copyright = {2019 The Author(s), under exclusive licence to Springer Nature Limited},
	issn = {1476-4687},
	url = {https://www.nature.com/articles/s41586-019-0960-6},
	doi = {10.1038/s41586-019-0960-6},
	language = {en},
	number = {7745},
	urldate = {2025-01-30},
	journal = {Nature},
	author = {Flühmann, C. and Nguyen, T. L. and Marinelli, M. and Negnevitsky, V. and Mehta, K. and Home, J. P.},
	month = feb,
	year = {2019},
	note = {Publisher: Nature Publishing Group},
	pages = {513--517},
}

@article{
Yang_2024,
author = {Yu Yang  and Igor Kladarić  and Maxwell Drimmer  and Uwe von Lüpke  and Daan Lenterman  and Joost Bus  and Stefano Marti  and Matteo Fadel  and Yiwen Chu },
title = {A mechanical qubit},
journal = {Science},
volume = {386},
number = {6723},
pages = {783-788},
year = {2024},
doi = {10.1126/science.adr2464},
URL = {https://www.science.org/doi/abs/10.1126/science.adr2464}}

@article{Strandberg_2024,
  title = {Digital Homodyne and Heterodyne Detection for Stationary Bosonic Modes},
  author = {Strandberg, Ingrid and Eriksson, Axel M. and Royer, Baptiste and Kervinen, Mikael and Gasparinetti, Simone},
  journal = {Phys. Rev. Lett. },
  volume = {133},
  issue = {6},
  pages = {063601},
  numpages = {8},
  year = {2024},
  month = {Aug},
  publisher = {American Physical Society},
  doi = {10.1103/PhysRevLett.133.063601},
  url = {https://link.aps.org/doi/10.1103/PhysRevLett.133.063601}
}

@article{Gilchrist1998,
  title = {Contradiction of Quantum Mechanics with Local Hidden Variables for Quadrature Phase Amplitude Measurements},
  author = {Gilchrist, A. and Deuar, P. and Reid, M. D.},
  journal = {Phys. Rev. Lett.},
  volume = {80},
  issue = {15},
  pages = {3169--3172},
  numpages = {0},
  year = {1998},
  month = {Apr},
  publisher = {American Physical Society},
  doi = {10.1103/PhysRevLett.80.3169},
  url = {https://link.aps.org/doi/10.1103/PhysRevLett.80.3169}
}

@article{Munro1999,
  title = {Optimal states for Bell-inequality violations using quadrature-phase homodyne measurements},
  author = {Munro, W. J.},
  journal = {Phys. Rev. A},
  volume = {59},
  issue = {6},
  pages = {4197--4201},
  numpages = {0},
  year = {1999},
  month = {Jun},
  publisher = {American Physical Society},
  doi = {10.1103/PhysRevA.59.4197},
  url = {https://link.aps.org/doi/10.1103/PhysRevA.59.4197}
}

@article{Fred_2003,
  title = {Maximal violation of Bell inequalities using continuous-variable measurements},
  author = {Wenger, J\'er\^ome and Hafezi, Mohammad and Grosshans, Fr\'ed\'eric and Tualle-Brouri, Rosa and Grangier, Philippe},
  journal = {Phys. Rev. A},
  volume = {67},
  issue = {1},
  pages = {012105},
  numpages = {7},
  year = {2003},
  month = {Jan},
  publisher = {American Physical Society},
  doi = {10.1103/PhysRevA.67.012105},
  url = {https://link.aps.org/doi/10.1103/PhysRevA.67.012105}
}

@article{RGP1,
  title = {Proposal for a Loophole-Free Bell Test Using Homodyne Detection},
  author = {Garc\'{\i}a-Patr\'on, R. and Fiur\'a\ifmmode \check{s}\else \v{s}\fi{}ek, J. and Cerf, N. J. and Wenger, J. and Tualle-Brouri, R. and Grangier, Ph.},
  journal = {Phys. Rev. Lett.},
  volume = {93},
  issue = {13},
  pages = {130409},
  numpages = {4},
  year = {2004},
  month = {Sep},
  publisher = {American Physical Society},
  doi = {10.1103/PhysRevLett.93.130409},
  url = {https://link.aps.org/doi/10.1103/PhysRevLett.93.130409}
}

@article{RGP2,
  title = {Loophole-free test of quantum nonlocality using high-efficiency homodyne detectors},
  author = {Garc\'{\i}a-Patr\'on, Ra\'ul and Fiur\'a\ifmmode \check{s}\else \v{s}\fi{}ek, Jarom\'{\i}r and Cerf, Nicolas J.},
  journal = {Phys. Rev. A},
  volume = {71},
  issue = {2},
  pages = {022105},
  numpages = {14},
  year = {2005},
  month = {Feb},
  publisher = {American Physical Society},
  doi = {10.1103/PhysRevA.71.022105},
  url = {https://link.aps.org/doi/10.1103/PhysRevA.71.022105}
}

@article{Acin2009,
  title = {Tests of multimode quantum nonlocality with homodyne measurements},
  author = {Ac\'{\i}n, Antonio and Cerf, Nicolas J. and Ferraro, Alessandro and Niset, Julien},
  journal = {Phys. Rev. A},
  volume = {79},
  issue = {1},
  pages = {012112},
  numpages = {9},
  year = {2009},
  month = {Jan},
  publisher = {American Physical Society},
  doi = {10.1103/PhysRevA.79.012112},
  url = {https://link.aps.org/doi/10.1103/PhysRevA.79.012112}
}

@article{Etesse_2014,
doi = {10.1088/1367-2630/16/5/053001},
url = {https://doi.org/10.1088/1367-2630/16/5/053001},
year = {2014},
month = {may},
publisher = {IOP Publishing},
volume = {16},
number = {5},
pages = {053001},
author = {Etesse, Jean and Blandino, Rémi and Kanseri, Bhaskar and Tualle-Brouri, Rosa},
title = {Proposal for a loophole-free violation of Bell's inequalities with a set of single photons and homodyne measurements},
journal = {New Journal of Physics}
}

@misc{Enky_Gael_2024,
      title={Realistic {Bell} tests with homodyne measurements}, 
      author={Enky Oudot and Gaël Massé and Xavier Valcarce and Antonio Acín},
      year={2024},
      eprint={2402.01530},
      archivePrefix={arXiv},
      primaryClass={quant-ph},
      url={https://arxiv.org/abs/2402.01530}, 
}

@article{Cavalcanti_2007,
  title = {Bell Inequalities for Continuous-Variable Correlations},
  author = {Cavalcanti, E. G. and Foster, C. J. and Reid, M. D. and Drummond, P. D.},
  journal = {Phys. Rev. Lett.},
  volume = {99},
  issue = {21},
  pages = {210405},
  numpages = {4},
  year = {2007},
  month = {Nov},
  publisher = {American Physical Society},
  doi = {10.1103/PhysRevLett.99.210405},
  url = {https://link.aps.org/doi/10.1103/PhysRevLett.99.210405}
}

@article{Abramsky_2011,
doi = {10.1088/1367-2630/13/11/113036},
url = {https://doi.org/10.1088/1367-2630/13/11/113036},
year = {2011},
month = {nov},
publisher = {IOP Publishing},
volume = {13},
number = {11},
pages = {113036},
author = {Abramsky, Samson and Brandenburger, Adam},
title = {The sheaf-theoretic structure of non-locality and contextuality},
journal = {New Journal of Physics}
}

@article{Fine1982,
  title = {Hidden Variables, Joint Probability, and the Bell Inequalities},
  author = {Fine, Arthur},
  journal = {Phys. Rev. Lett.},
  volume = {48},
  issue = {5},
  pages = {291--295},
  numpages = {0},
  year = {1982},
  month = {Feb},
  publisher = {American Physical Society},
  doi = {10.1103/PhysRevLett.48.291},
  url = {https://link.aps.org/doi/10.1103/PhysRevLett.48.291}
}

@article{Abramsky2017,
  title = {Contextual Fraction as a Measure of Contextuality},
  author = {Abramsky, Samson and Barbosa, Rui Soares and Mansfield, Shane},
  journal = {Phys. Rev. Lett.},
  volume = {119},
  issue = {5},
  pages = {050504},
  numpages = {6},
  year = {2017},
  month = {Aug},
  publisher = {American Physical Society},
  doi = {10.1103/PhysRevLett.119.050504},
  url = {https://link.aps.org/doi/10.1103/PhysRevLett.119.050504}
}

@article{Moradi_2024,
doi = {10.1088/1367-2630/ad2d40},
url = {https://doi.org/10.1088/1367-2630/ad2d40},
year = {2024},
month = {mar},
publisher = {IOP Publishing},
volume = {26},
number = {3},
pages = {033019},
author = {Moradi, Morteza and Camilo López Carreño, Juan and Buraczewski, Adam and McDermott, Thomas and Elisabeth Asenbeck, Beate and Laurat, Julien and Stobińska, Magdalena},
title = {CHSH Bell tests for optical hybrid entanglement},
journal = {New Journal of Physics }
}

@article{Cavailles2018,
  title = {Demonstration of Einstein-Podolsky-Rosen Steering Using Hybrid Continuous- and Discrete-Variable Entanglement of Light},
  author = {Cavaill\`es, A. and Le Jeannic, H. and Raskop, J. and Guccione, G. and Markham, D. and Diamanti, E. and Shaw, M. D. and Verma, V. B. and Nam, S. W. and Laurat, J.},
  journal = {Phys. Rev. Lett.},
  volume = {121},
  issue = {17},
  pages = {170403},
  numpages = {6},
  year = {2018},
  month = {Oct},
  publisher = {American Physical Society},
  doi = {10.1103/PhysRevLett.121.170403},
  url = {https://link.aps.org/doi/10.1103/PhysRevLett.121.170403}
}

@article{
Laurat2020,
author = {Giovanni Guccione  and Tom Darras  and Hanna {Le Jeannic}  and Varun B. Verma  and Sae Woo Nam  and Adrien Cavaillès  and Julien Laurat },
title = {Connecting heterogeneous quantum networks by hybrid entanglement swapping},
journal = {Science Advances},
volume = {6},
number = {22},
pages = {eaba4508},
year = {2020},
doi = {10.1126/sciadv.aba4508},
URL = {https://www.science.org/doi/abs/10.1126/sciadv.aba4508}}

@article{deGliniasty2024spinopticalquantum,
  doi = {10.22331/q-2024-07-24-1423},
  url = {https://doi.org/10.22331/q-2024-07-24-1423},
  title = {A {S}pin-{O}ptical {Q}uantum {C}omputing {A}rchitecture},
  author = {de Gliniasty, Gr{\'{e}}goire and Hilaire, Paul and Emeriau, Pierre-Emmanuel and Wein, Stephen C. and Salavrakos, Alexia and Mansfield, Shane},
  journal = {{Quantum}},
  issn = {2521-327X},
  publisher = {{Verein zur F{\"{o}}rderung des Open Access Publizierens in den Quantenwissenschaften}},
  volume = {8},
  pages = {1423},
  month = jul,
  year = {2024}
}

@article{Hacker2019,
  author    = {Hacker, Bastian and Welte, Stephan and Daiss, Severin and Shaukat, Armin and Ritter, Stephan and Li, Lin and Rempe, Gerhard},
  title     = {Deterministic creation of entangled atom--light Schr\"odinger-cat states},
  journal   = {Nature Photonics},
  volume    = {13},
  number    = {2},
  pages     = {110--115},
  year      = {2019},
  month     = {February},
  doi       = {10.1038/s41566-018-0339-5},
  publisher = {Springer Nature},
}

@software{MrMustard,
  author       = {},
  title        = {{MrMustard}: A differentiable library for continuous-variable quantum optics},
  year         = {2025},
  url          = {https://github.com/XanaduAI/MrMustard},
  note         = {Version 0.7.3}
}

@article{Elitzur_1992,
title = {Quantum nonlocality for each pair in an ensemble},
journal = {Physics Letters A},
volume = {162},
number = {1},
pages = {25-28},
year = {1992},
issn = {0375-9601},
doi = {https://doi.org/10.1016/0375-9601(92)90952-I},
url = {https://www.sciencedirect.com/science/article/pii/037596019290952I},
author = {Avshalom C. Elitzur and Sandu Popescu and Daniel Rohrlich}
}

@article{Cabello_2002,
  title = {Bell's theorem with and without inequalities for the three-qubit Greenberger-Horne-Zeilinger and W states},
  author = {Cabello, Ad\'an},
  journal = {Phys. Rev. A},
  volume = {65},
  issue = {3},
  pages = {032108},
  numpages = {4},
  year = {2002},
  month = {Feb},
  publisher = {American Physical Society},
  doi = {10.1103/PhysRevA.65.032108},
  url = {https://link.aps.org/doi/10.1103/PhysRevA.65.032108}
}

@article{Barrett2006,
  title = {Maximally Nonlocal and Monogamous Quantum Correlations},
  author = {Barrett, Jonathan and Kent, Adrian and Pironio, Stefano},
  journal = {Phys. Rev. Lett.},
  volume = {97},
  issue = {17},
  pages = {170409},
  numpages = {4},
  year = {2006},
  month = {Oct},
  publisher = {American Physical Society},
  doi = {10.1103/PhysRevLett.97.170409},
  url = {https://link.aps.org/doi/10.1103/PhysRevLett.97.170409}
}

@article{Aolita_2012,
  title = {Fully nonlocal quantum correlations},
  author = {Aolita, Leandro and Gallego, Rodrigo and Ac\'{\i}n, Antonio and Chiuri, Andrea and Vallone, Giuseppe and Mataloni, Paolo and Cabello, Ad\'an},
  journal = {Phys. Rev. A},
  volume = {85},
  issue = {3},
  pages = {032107},
  numpages = {8},
  year = {2012},
  month = {Mar},
  publisher = {American Physical Society},
  doi = {10.1103/PhysRevA.85.032107},
  url = {https://link.aps.org/doi/10.1103/PhysRevA.85.032107}
}

@book{Boyd_Vandenberghe_2004, place={Cambridge}, title={Convex Optimization}, publisher={Cambridge University Press}, author={Boyd, Stephen and Vandenberghe, Lieven}, year={2004}}

@misc{Braccini2025,
      title={Superpositions of Quantum {G}aussian Processes}, 
      author={Lorenzo Braccini and Sougato Bose and Alessio Serafini},
      year={2025},
      eprint={2510.01156},
      archivePrefix={arXiv},
      primaryClass={quant-ph},
      url={https://arxiv.org/abs/2510.01156}, 
}

@misc{Fede2026,
      title={Gaussian superpositions for bosonic encodings}, 
      author={Federico Centrone and Juan Pablo Paz and Augusto Roncaglia},
      year={2026},
      eprint={2603.15258},
      archivePrefix={arXiv},
      primaryClass={quant-ph},
      url={https://arxiv.org/abs/2603.15258}, 
}

@article{Wang2016,
author = {Chen Wang  and Yvonne Y. Gao  and Philip Reinhold  and R. W. Heeres  and Nissim Ofek  and Kevin Chou  and Christopher Axline  and Matthew Reagor  and Jacob Blumoff  and K. M. Sliwa  and L. Frunzio  and S. M. Girvin  and Liang Jiang  and M. Mirrahimi  and M. H. Devoret  and R. J. Schoelkopf },
title = {A Schrödinger cat living in two boxes},
journal = {Science},
volume = {352},
number = {6289},
pages = {1087-1091},
year = {2016},
doi = {10.1126/science.aaf2941},
URL = {https://www.science.org/doi/abs/10.1126/science.aaf2941},
eprint = {https://www.science.org/doi/pdf/10.1126/science.aaf2941}}

@misc{Fontbote2026,
      title={Error Correction of Beamsplitter-Generated Entangled {GKP} States}, 
      author={Moritz Fontboté-Schmidt and Jeremy Metzner and Florence Berterottière and Ivan Rojkov and Alexander Ferk and Martin Stadler and Bahadir Dönmez and Ralf Berner and Stephan Welte and Daniel Kienzler and Jonathan P. Home},
      year={2026},
      eprint={2605.08009},
      archivePrefix={arXiv},
      primaryClass={quant-ph},
      url={https://arxiv.org/abs/2605.08009}, 
}

@misc{Loke2026,
      title={Demonstration of tripartite cat states in two distinct classes of entanglement}, 
      author={May Chee Loke and Jonathan Schwinger and Kehui Yu and Yingshan Zhang and Amon M. Kasper and Tanjung Krisnanda and Yvonne Y. Gao},
      year={2026},
      eprint={2607.12317},
      archivePrefix={arXiv},
      primaryClass={quant-ph},
      url={https://arxiv.org/abs/2607.12317}, 
}

@article{Andersen2015,
  author    = {Andersen, Ulrik L. and Neergaard-Nielsen, Jonas S. and van Loock, Peter and Furusawa, Akira},
  title     = {Hybrid discrete- and continuous-variable quantum information},
  journal   = {Nature Physics},
  volume    = {11},
  number    = {9},
  pages     = {713--719},
  year      = {2015},
  month     = {September},
  doi       = {10.1038/nphys3410},
  publisher = {Nature Publishing Group},
}

@article{Hybrid_tutorial,
  title = {Hybrid Oscillator-Qubit Quantum Processors: Instruction Set Architectures, Abstract Machine Models, and Applications},
  author = {Liu, Yuan and Singh, Shraddha and Smith, Kevin C. and Crane, Eleanor and Martyn, John M. and Eickbusch, Alec and Schuckert, Alexander and Li, Richard D. and Sinanan-Singh, Jasmine and Soley, Micheline B. and Tsunoda, Takahiro and Chuang, Isaac L. and Wiebe, Nathan and Girvin, Steven M.},
  journal = {PRX Quantum},
  volume = {7},
  issue = {1},
  pages = {010201},
  numpages = {166},
  year = {2026},
  month = {Jan},
  publisher = {American Physical Society},
  doi = {10.1103/4rf7-9tfx},
  url = {https://link.aps.org/doi/10.1103/4rf7-9tfx}
}

@incollection{Albert2025,
  author    = {Albert, Victor V.},
  title     = {Bosonic coding: introduction and use cases},
  booktitle = {Quantum Fluids of Light and Matter -- {QFLM} 2022},
  series    = {Proceedings of the International School of Physics ``{Enrico Fermi}''},
  volume    = {209},
  editor    = {Bramati, Alberto and Carusotto, Iacopo and Ciuti, Cristiano},
  publisher = {IOS Press},
  year      = {2025},
  doi       = {10.3254/ENFI250007},
  isbn      = {978-1-64368-574-8},
  note      = {School held in Varenna, Italy, 2--7 July 2022},
}

\newpage

\appendix

\section{Introduction to continuous-variable quantum optics}
\label{app:CV}

In this section we introduce the framework of CV quantum information processing, used to describe multimode bosonic systems. The state of such systems exists in an infinite-dimensional Hilbert space $\mathcal{H}$. These states can be described in two complementary pictures. On the one hand, there is the Fock space description of the state, with the number of particles as the main observable. On the other hand, there is a continuous description, for which the phase-space quadratures are the main observables. These refer for example to the position and momentum of a harmonic oscillator, or to the amplitude and phase quadratures of an electromagnetic field. \par 
Any bosonic pure state can be written as a superposition of states with a well-defined particle number, called Fock states $\{|\mathbf n\rangle\,|\, \mathbf n\in \mathbb N^M\}$, where $M$ is the number of modes of the system. In the context of quantum optics, different modes are the different solutions to Maxwell's equations. For mechanical systems, they refer to the different modes of oscillation. In the common setting of spatially separated systems, each subsystem can be seen as an independent mode. \par 
All Fock states of a multimode system can be obtained from the vacuum as
\begin{equation}
    |n_1\dots n_M\rangle=\frac{(\hat a_1^\dagger)^{n_1}\otimes \dots \otimes (\hat a_M^\dagger)^{n_M}}{\sqrt{n_1!\dots n_M!}}|\bm 0\rangle,
\end{equation}
where $\hat a_i^\dagger$ is the creation operator associated to mode $i$. Its Hermitian adjoint $\hat a_i$ is the annihilation operator of mode $i$, whose effect is given by 
\begin{equation}
    \hat a_i |n_i\rangle=\sqrt{n_i}|n_i-1\rangle.
\end{equation}
Together, they can be used to define the number operator $\hat n_i=\hat a_i^\dagger \hat a_i$, such that $\hat n_i|n_i\rangle=n_i|n_i\rangle$. Any pure quantum state can then be written as 
\begin{equation}
    |\psi\rangle=\sum_{\bm n:|\bm n|\geq 0}c_{\bm n}|\bm n\rangle.
\end{equation}
Notice that this involves, in principle, an infinite number of terms. Mixed states, on the other hand, must be described using the density operator 
\begin{equation}
    \hat \rho=\sum_{\bm n,\bm m}\rho_{\bm n\bm m}|\bm n\rangle\langle\bm m|.
\end{equation}
The density matrix $\hat\rho$ is infinite-dimensional, which implies that any practical use of it requires the choice of a finite cutoff on the number of photons per mode. This is well motivated by the assumption that every physical experiment has a finite amount of energy.  \par 

The CV description on the other hand is based on the position and momentum observables, or amplitude and phase quadratures, which can be defined in terms of the creation and annihilation operators of each mode as 
\begin{equation}
    \begin{split}
        & \hat x_j=\hat a_j+\hat a^\dagger_j \\
        & \hat p_j=\frac{\hat a_j-\hat a^\dagger_j}{i},
    \end{split}
\end{equation}
which satisfy the canonical commutation relation $\left[\hat x_k,\hat p_l\right]=2i \delta_{k,l}$. The outcomes of these observables define the different points in a phase space with coordinates $\vec \xi=(x_1,x_2...,x_M,p_1,...,p_M)$, with an associated symplectic structure 
\begin{equation*}
    \Omega=\left(\begin{matrix}
        \mathbb 0 & -\Id \\
        \Id & \mathbb 0 
    \end{matrix}\right),
\end{equation*}
where each block is of dimension $M\times M$. We have $\left[\hat \xi_k,\hat \xi_l\right]=2 i \Omega_{k,l}$. Rotations in the phase space are equivalent to the action of phase rotations (which occur from the free evolution of the harmonic oscillator), or, when mixing different modes, from the action of optical beam splitters. At the single-mode level we can always define rotated quadratures 
\begin{equation}
    \hat q_i^{\varphi}=\hat x_i \cos \varphi -\hat p_i \sin\phi.   
\end{equation}
Homodyne detection, the main type of measurements that we are concerned with in this work, corresponds to measurements of these rotated quadratures, and can be seen as projective measurements into their eigenstates, so that 
\begin{equation}\label{eq:homodyne_statistics}
p(q_1^{\phi_1},...,q_M^{\phi_M})=\Tr\left(\hat \rho |q_1^{\phi_1},...,q_M^{\phi_M}\rangle\langle q_1^{\phi_1},...,q_M^{\phi_M} |\right).
\end{equation}
Different possibilities are available to efficiently describe these systems, depending on the specific states at hand. There is, naturally, the wave function description in terms of quadrature eigenstates. Yet, a generally convenient description is provided by quasiprobability distributions in phase space, and most notably the Wigner function, which is defined as the Fourier transform of the expectation values of displacement operators 
\begin{equation}
    W(\vec q=\{x_1,...,x_M,p_1,...p_M\})=\frac{1}{(2\pi)^M}\int_{\mathbb R^{2M}}d\vec \xi\,  \chi(\vec \xi) e^{-i \vec \xi^T \vec q },
\end{equation}
where 
\begin{equation*}
    \chi(\vec \xi)=\Tr(\hat \rho e^{i \vec \xi \Omega \vec{\hat q}}),
\end{equation*}
is the expectation value of a displacement operator, also known as characteristic function of $\hat \rho$. The Wigner function is not a well-defined probability distribution, since it can reach negative values in phase space; nevertheless, it is normalized, and all its marginals that involve only commuting quadratures provide the probability distributions that correspond to the measurement statistics of homodyne detection along the selected quadratures. This implies that we can compute the probability distributions in \eqref{eq:homodyne_statistics}, which are the central objects in this work, as 
\begin{equation}
    p(q_1^{\phi_1},...,q_M^{\phi_M})=\int \prod_{i=1}^M dq_i^{\phi_i+\pi/2} W(O_{\vec \phi}^T \vec q),
\end{equation}
where $O_{\vec \phi}$ corresponds to a rotation of the quadratures in phase space. Notice that the integration takes place along the set of quadratures that are complementary to each of the quadratures of interest.\par 
Moreover, the description in phase space allows for a classification of quantum states based on the properties of their Wigner function. Notably, there are the Gaussian states, whose phase space description, and in particular the Wigner function, is given by a Gaussian function. On the other hand there are states that are described through non-Gaussian Wigner functions. The Wigner function of these states can potentially reach negative values, which is a necessary requirement for the violation of some Bell inequality using homodyne data, given that otherwise the Wigner function could be used to construct a local hidden variable model. This dichotomy also allows one to separate quantum operations between Gaussian and non-Gaussian operations. Gaussian operations act as Bogoliubov transformations on the phase space coordinates, thus never introducing any non-Gaussian features. These include the squeezing operations, whose action on the quadratures is given by  
\begin{equation}
    \begin{split}
    &\hat S(r) \hat x \hat S(r)^\dagger=e^{-r} \hat x\\
    & \hat S(r) \hat p \hat S(r)^\dagger=e^{r} \hat p.
    \end{split}
\end{equation}
The displacement operators act as follws
\begin{equation}
    \begin{split}
    &\hat D(\alpha)\hat x\hat D^\dagger(\alpha)=\hat x-2 \text{Re}(\alpha)\\
    & \hat D(\alpha)\hat p\hat D^\dagger(\alpha)=\hat p-2 \text{Im}(\alpha). 
    \end{split}
\end{equation}
Single-mode rotations $\hat R(\phi)$ are responsible for the transformation from the canonical $\hat x,\hat p$ to the rotated quadratures $\hat q_\phi$. The other set of Gaussian operations that we considered, and that suffice, together with the ones considered so far, to generate any Gaussian operation, is the two-mode beam splitter $\hat U_{ij}(\theta)$, whose action consists in mixing equivalent quadratures of modes $i,j$ as
\begin{equation}
    \begin{split}
        &\hat U_{ij}(\theta) \hat q_i^{\phi}\hat U_{ij}^\dagger(\theta)=\hat q_i^{\phi} \cos \theta -\hat q_j^{\phi} \sin \theta\\
        &\hat U_{ij}(\theta) \hat q_j^{\phi}\hat U_{ij}^\dagger(\theta)=\hat q_j^{\phi} \cos \theta +\hat q_i^{\phi} \sin \theta.
    \end{split}
\end{equation}
Finally, non-Gaussian operations form a much wilder set, which we do not aim to fully characterize here. The non-Gaussian states considered in the main text of the manuscript are constructed in three main ways. There are the ones provided by finite superposition of Fock states, which are non-Gaussian by definition, but are not simply related to specific non-Gaussian operations. We also consider the case of photon-subtracted states, which are obtained by de-Gaussifying Gaussian states through a coherent subtraction of a single photon, which mathematically is equivalent to the application of the annihilation operator on an arbitrary mode, that can always be written as a linear combination of annihilation operators of the different modes of the system. Its effect on phase space could be thought of as taking a derivative of the phase space distribution with respect to some axis. Finally, the other main set of states that we considered is obtained by coherent superpositions of displacement operations. The paradigmatic example is the cat state 
\begin{equation}
    |\text{cat}(\alpha)\rangle=\frac{\hat D(\alpha/2)|0\rangle +\hat D(-\alpha/2)|0\rangle}{\mathcal N(\alpha)},
\end{equation}
whose Wigner function displays negative values, that arise from the interference between the two coherent states. 
\section{Moment-based SDP}\label{app:SDP_definition}
The primal program at level $k$ in the hierarchy is given by 
\begin{equation}\label{eq:primal_program_order_k}
   (SP_k) \begin{cases}
        \begin{aligned}
            & \sup_{\bm{y}\in \mathrm R^{s(k)}} y_0\\
            \text{s.t.} & \forall C\in \mathcal M, M_{k}(y^{e,C}-y|_C)\succeq 0\\
            & M_k(y)\succeq 0\\
            & \forall j \in \{1,...,m\}, M_{k-r_j}(P_{j} y)\succeq 0,
        \end{aligned}
    \end{cases}
\end{equation}
where $P_j$ is the set of algebraic conditions defining the compact set in which the problem is evaluated. In the case in which we solve the problem in $\mathrm R^{|X|}$ this set of constraints is not present. The dual program can be written, after some manipulations, as 
\begin{equation}
    (SD_k) \begin{cases}
        \begin{aligned}
            \text{Find} \quad & (f_{c})\subset \Sigma^{2}\mathrm R\left[x\right]_k, (\sigma_{j})\subset \Sigma^{2}\mathrm R\left[x\right]_{k-r_j} \\
             \inf  \quad & \sum_{C}\int_{O_C}f_C de_C\\
            \text{s.t.} \quad & \sum_{C}f_C-\bm{1}=\sigma_0+\sum_{j=1}^{m}\sigma_j P_j.
        \end{aligned}
    \end{cases}
\end{equation}
Here $\sigma_j$ can be dropped if we consider optimization on $\mathrm R^{|X|}$. $\Sigma^{2}\mathrm R[x]_{k}$ refers to the set of SOS polynomials of order $2k$ \cite{PE_2022}. From here it is possible to recover again a Bell inequality. If the empirical model is non-contextual the constraint implies that 
\begin{equation}
\begin{aligned}
    B(e_C)& =1-\sum_{C}\int_{O_C}f_C de_C\\
    & = 1-\int_{O_C}\left(1+\sigma_0+\sum_{j=1}^{m}\sigma_j P_j\right) de_C\\
    & = -\int_{O_C}\left(\sigma_0+\sum_{j=1}^{m}\sigma_j P_j\right) de_C \leq 0.
\end{aligned}
\end{equation}
More generally the Bell inequality is just given by the first line above, which can be extracted from the program. 

\subsection{Results}
Our exploration of this tool on CV systems started from using the moment-based SDP \eqref{eq:primal_program_order_k}. Nevertheless, so far we have not managed to observe any non-zero contextual fraction for any quantum behavior. On the other hand, it is interesting to notice that contextual behaviors beyond the set of quantum correlations can be detected using this program. 
\par
Popescu-Rohrlich boxes \cite{popescu_quantum_1994} provide a maximally contextual empirical model in discrete variables. CV generalizations of those probabilistic behaviors can be constructed quite naturally. Below we can see the table \\
\begin{figure}[h!]

\begin{minipage}{0.5\textwidth}
    \centering
     \begin{tabular}{|c|c|c|c|c|}
    \hline
         & 00 & 01 & 10 & 11  \\
         \hline
       ($a_1,b_1$)  & 1/2 & 0 & 0 & 1/2 \\
       \hline
       ($a_1,b_2$)  & 1/2 & 0 & 0 & 1/2 \\
       \hline
       ($a_2,b_1$)  & 1/2 & 0 & 0 & 1/2 \\
       \hline
       ($a_2,b_2$)  & 0 & 1/2 & 1/2 & 0 \\
       \hline
    \end{tabular}
    \includegraphics[width=0.7\linewidth]{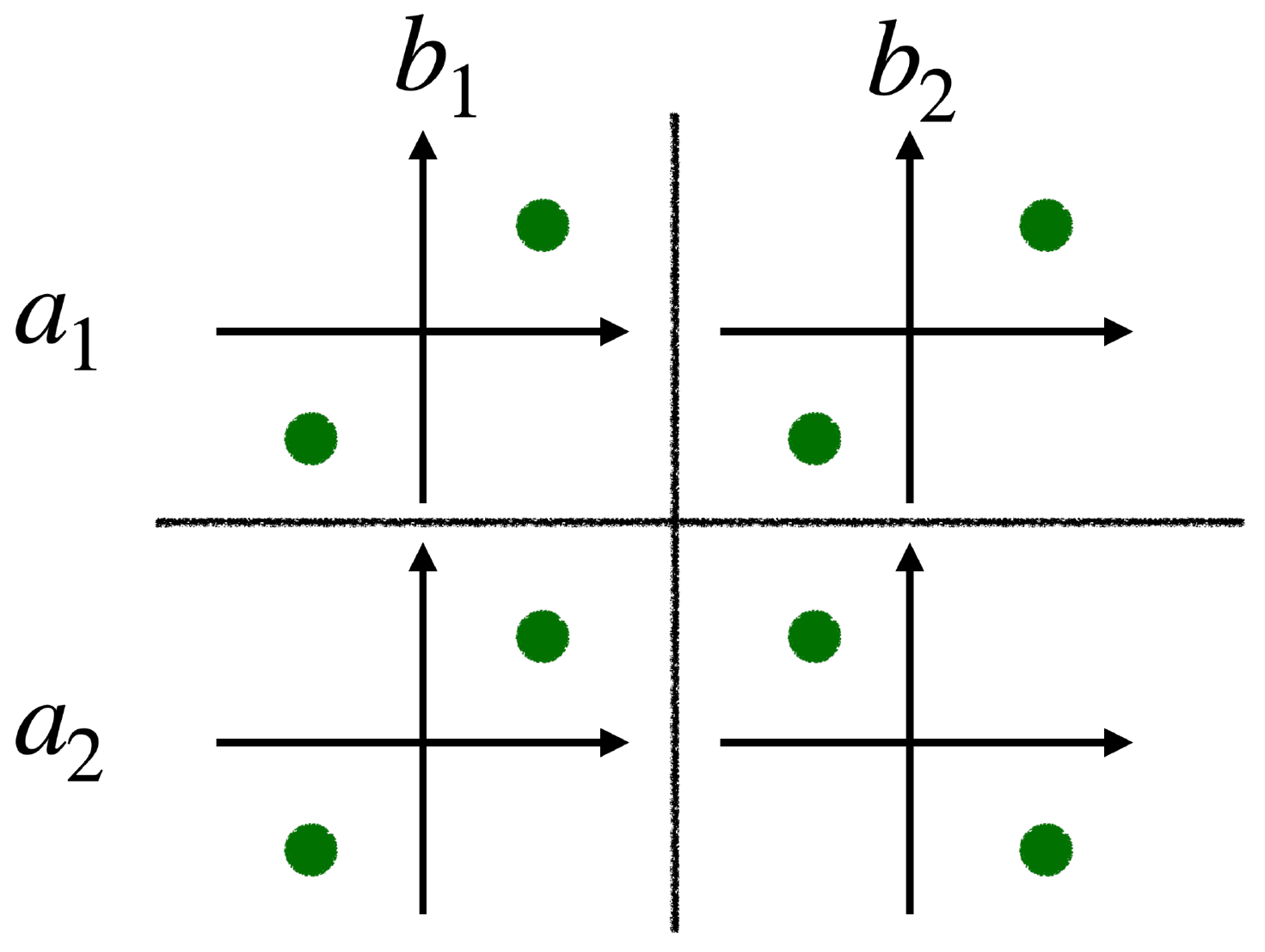}
\end{minipage}
\caption{PR behavior in its discrete and continuous generalization. }
\label{fig:PR_box_def}
\end{figure}
\par
\begin{figure}[htbp]
\centering
\includegraphics[width =0.9\linewidth]{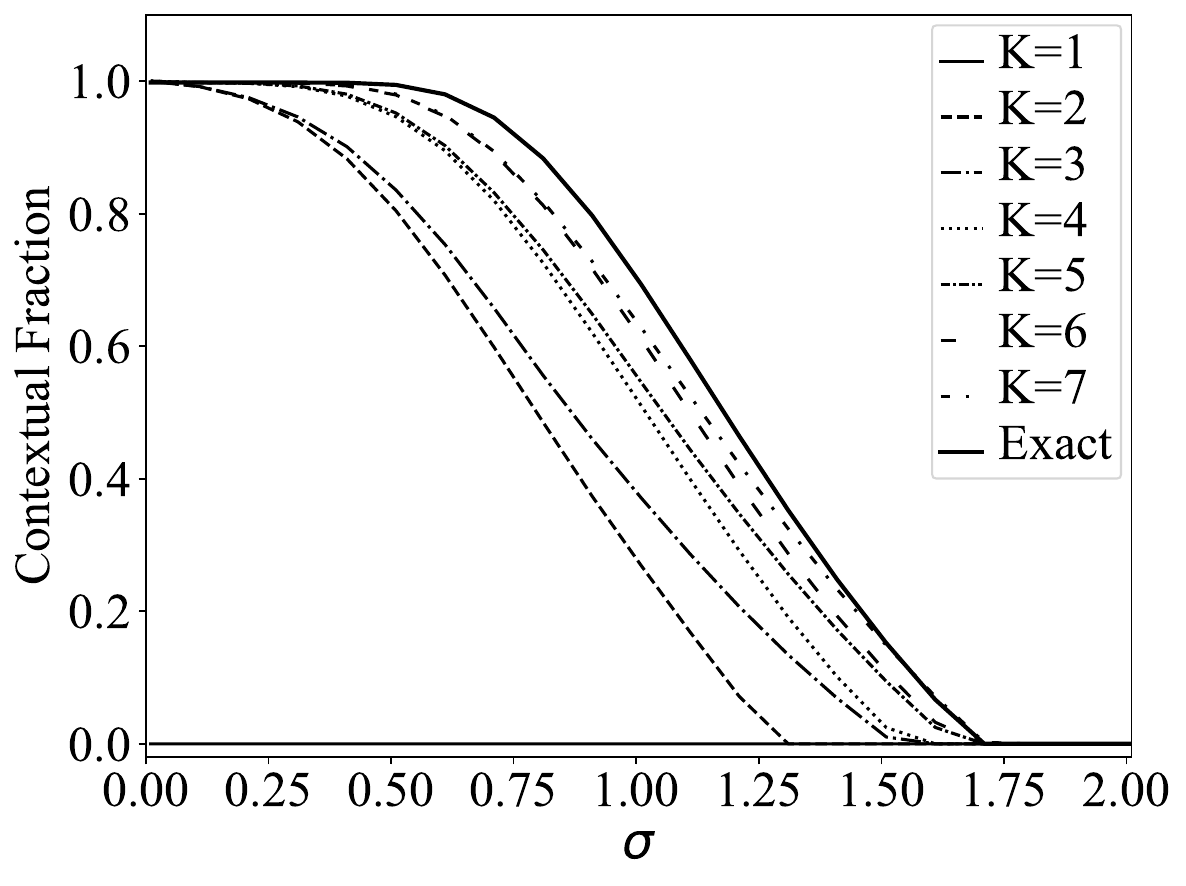}
\caption{Each line in the plot represents the value of the contextual fraction of the empirical behavior associated to a PR box, evaluated at different levels $K$ of the hierarchy of SDPs. The plots are evaluated against the variance $\sigma$ of the peaks of the distribution. All peaks are located at $( \pm 2,\pm 2)$. Even at the first non-trivial order in the hierarchy ($K=2$), the SDP converges to the exact value, for small $\sigma$. For large values of $\sigma$, on the other hand, there is a finite gap between the exact value, computed using the $LP$ implementation, and the values obtained using $K\leq 6$.}
\label{fig:CF_PR_vs_sigma}
\end{figure}
\begin{figure}
    \centering
    \includegraphics[width=0.8\linewidth]{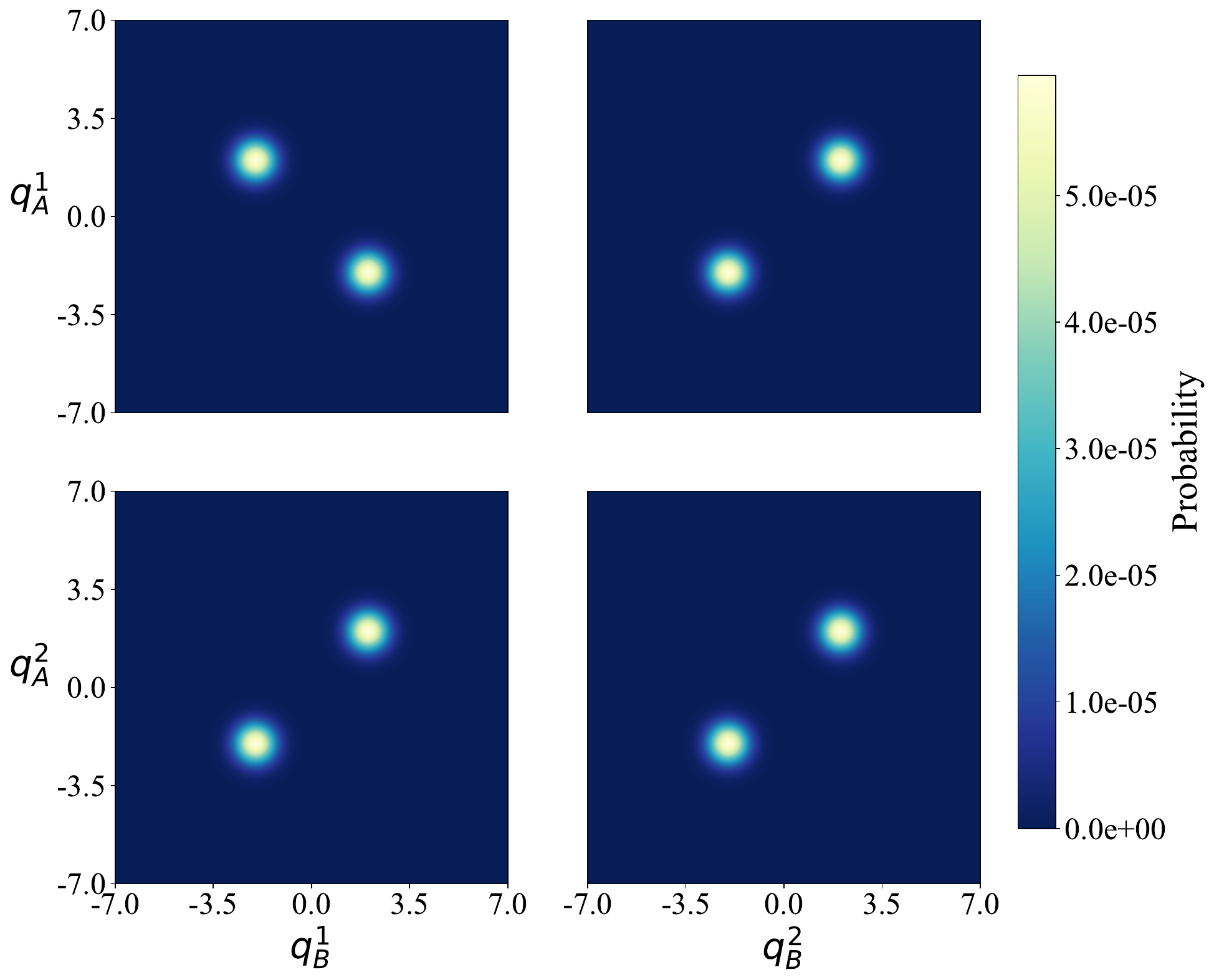}
    \includegraphics[width=0.8\linewidth]{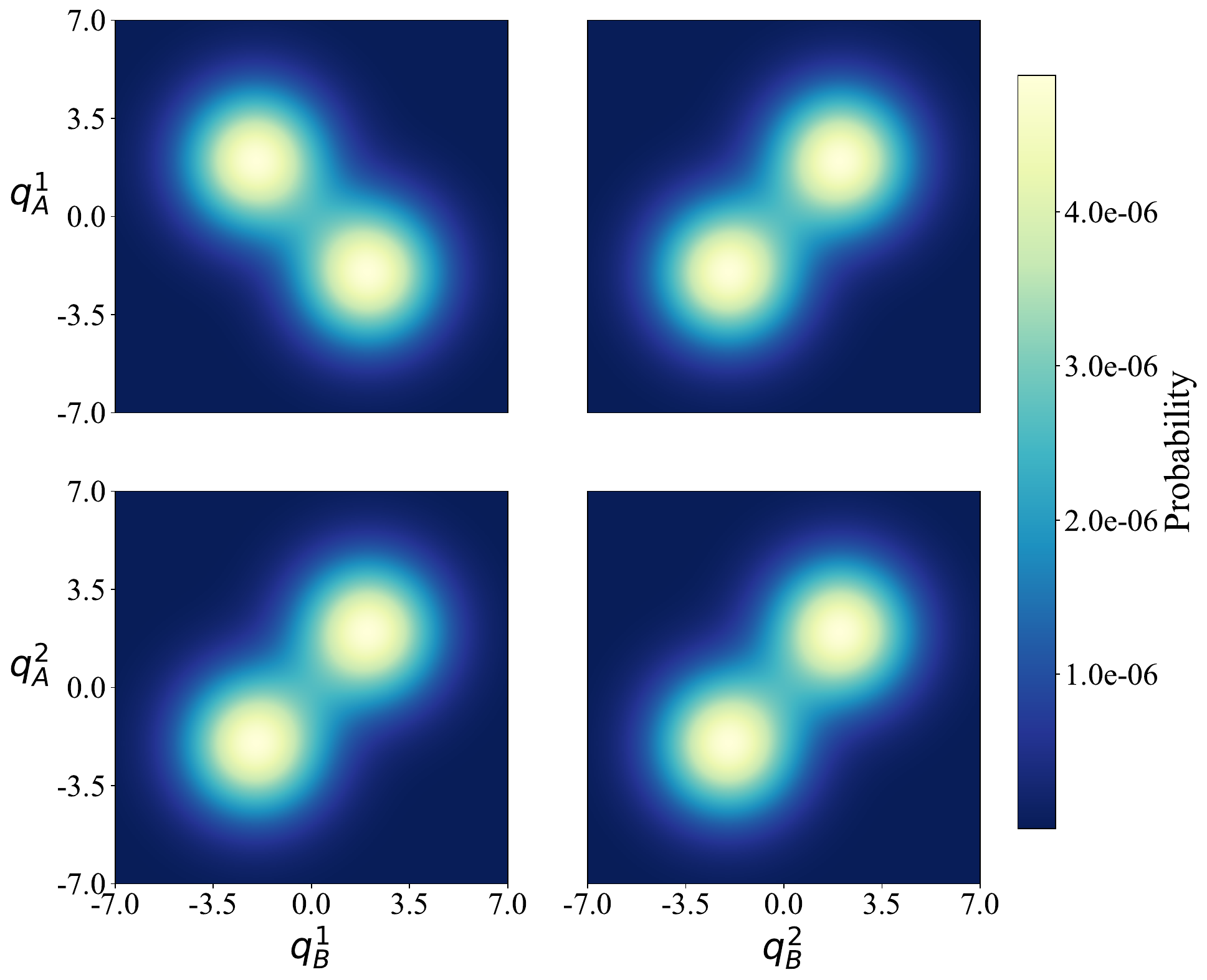}
    \caption{Continuous extensions of the PR empirical models, for two different peak widths, \ie different values of $\sigma$ in equation \eqref{eq:continuous_PR}. The values correspond to the probability for each bin, provided the accuracy with which the plot is constructed ($\delta q=1.95 \times10^{-2}$). The left plots correspond to $\sigma=0.5$, for which maximum contextuality is obtained. The right plots correspond to $\sigma=1.75$, slightly above the limit beyond which no contextuality is observed. Notice that this starts happening when the overlap between the peaks is different from zero. Nevertheless, the exact value of the contextual fraction is a high one. }
\label{fig:distributions_PR_different_sigmas}
\end{figure}

The continuous generalization of the PR box, which can be observed in Fig.~\ref{fig:PR_box_def}, can be parametrized as 
\begin{equation}\label{eq:continuous_PR}
    p_{pr}^{C}(\bm x_C)=\frac{1}{4 \pi \sigma^2}\left(e^{-\frac{|(\bm x_C+\left(a,(-1^{ij})a\right))|^2}{2 \sigma^2}}+e^{-\frac{|(\bm x_C-\left(a,(-1^{ij})a\right))|^2}{2 \sigma^2}}\right),
\end{equation}
where $a$ represents the center of the peaks of the distributions, and $\sigma$ their variance, which we consider homogeneous along both directions. When evaluating the contextual fraction of such states, for small $\sigma$ we obtain $CF=1$, and such a behavior persists up to large variances, as can be observed in Fig.~\ref{fig:CF_PR_vs_sigma}. The plot corresponds to the case in which we pick $a=2$. Naturally, as $\sigma$ increases, the value of the contextual fraction decreases. This happens beyond some point, at which the peaks of the distributions start to overlap. The threshold beyond which contextuality is no longer observed increases with $K$ for small values of $K$. In Fig.~\ref{fig:distributions_PR_different_sigmas}, we can observe the histograms for values of $\sigma$ below and above the threshold. Moreover, we can observe that the threshold decreases with the spacing between the peaks of the distribution, for similar reason.

\par 

For each run of the program that returns a non-zero contextual fraction, we obtain a polynomial Bell inequality, following the recipe described in Section \ref{app:SDP_definition}. These moment-based Bell inequalities can be optimized within the set of quantum states with a finite cutoff on Fock space. Nevertheless, all the Bell inequalities obtained are trivial for the set of quantum states up to the cutoff considered $N_{max}=50$. Another degree of freedom to consider is the homodyne settings with respect to which the inequalities are optimized. Nevertheless, an exhaustive exploration over this degree of freedom provided no positive results. \par 
Other finite behaviors, previously reported to display contextuality as discrete probability measures, were also extended to the continuous case, following a similar procedure as for the PR boxes. None of the Bell inequalities obtained provided any non-zero contextual fraction for a quantum state. Other extensions were considered by mixing the continuous versions of discrete non-local behaviors with CV behaviors expected to be non-local. In this way, one would expect to obtain non-trivial Bell inequalities, yet the results proved negative as well. \par
Finally, we would like to stress that this hierarchy of SDP's is well suited for the study of Bell experiments on DV systems, which can be treated exactly, as can be seen in the following section.   
\section{All previously identified binary strategies are
essentially optimal}\label{app:previous_examples}
In what follows, we briefly discuss several examples previously considered in the literature for violating Bell inequalities with homodyne detection, for which our hierarchy produces the optimal Bell inequality. Despite searching for optimal Bell inequalities for all  cases in an automatic way, we found no improvement over the previous results, after normalizing the Bell inequality so that its maximum value is $1$.

\subsubsection{Photon-subtracted states}

\begin{figure}
\centering
\includegraphics[width =0.75\linewidth]{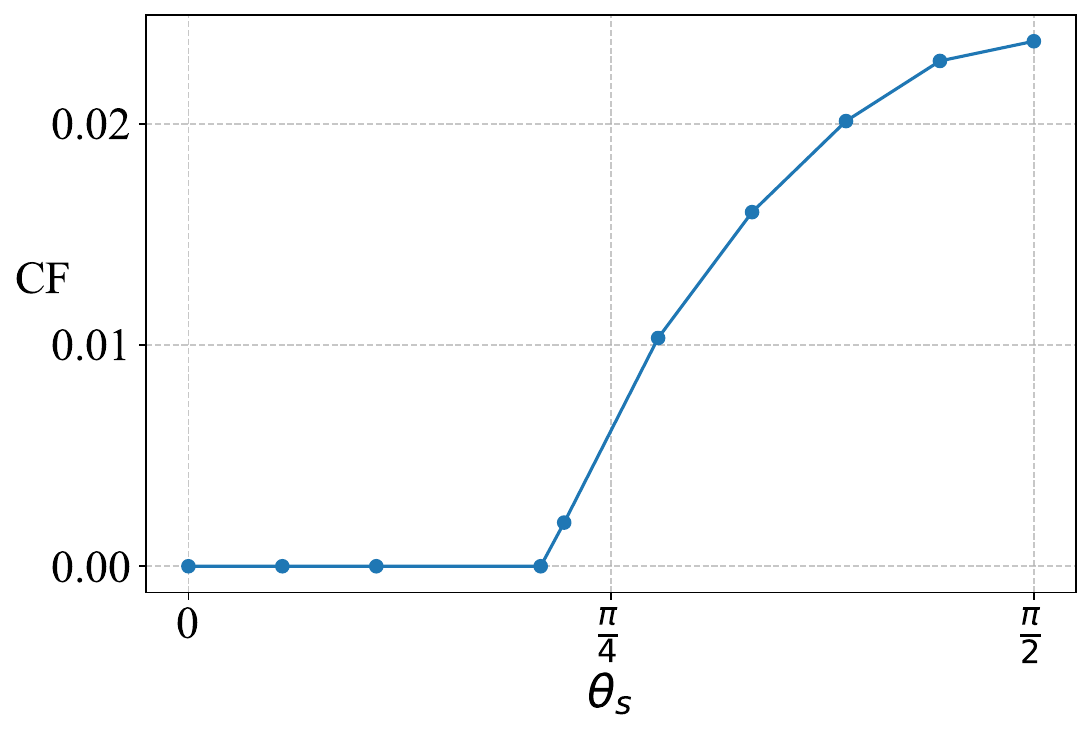}
\caption{Contextual fraction associated to the measurement statistics of two photon-subtracted states from the family described in equation \eqref{eq:two_photon_subtracted_family1}. In all cases, a search was performed over different configurations of the homodyne detector settings to obtain the maximum contextual fraction possible. All non-zero values correspond to measurements with $\phi_1\in \{0,\pi/2\}$ and $\phi_2\in\{-\pi/4,\pi/4\}$. }
\label{fig:CF_photon_subtr_vs_theta}
\end{figure}

In \cite{RGP1,RGP2}, Garc\'ia-Patr\'on et al.\ proposed a scheme for violating Bell inequalities with photon-subtracted states \cite{Ra2020}, using homodyne detection statistics. They employ the CHSH inequality, after sign binning the outcomes of the homodyne detection. That is, if Alice measures a quadrature value $q_A^{\phi}$, then they will record the value $a_{\phi}=\text{sign}(q_A^{\phi})$. \par
We probed a similar setup with the aim of finding the optimal Bell inequalities for this setting. It turns out that, as we show below, the strategy employed in \cite{RGP1,RGP2} is the optimal one. This claim was conjectured in the paper based on numerical explorations and analysis of the features observed on the probability distributions.

\begin{figure}[htbp]
\centering
\includegraphics[width =0.47\linewidth]{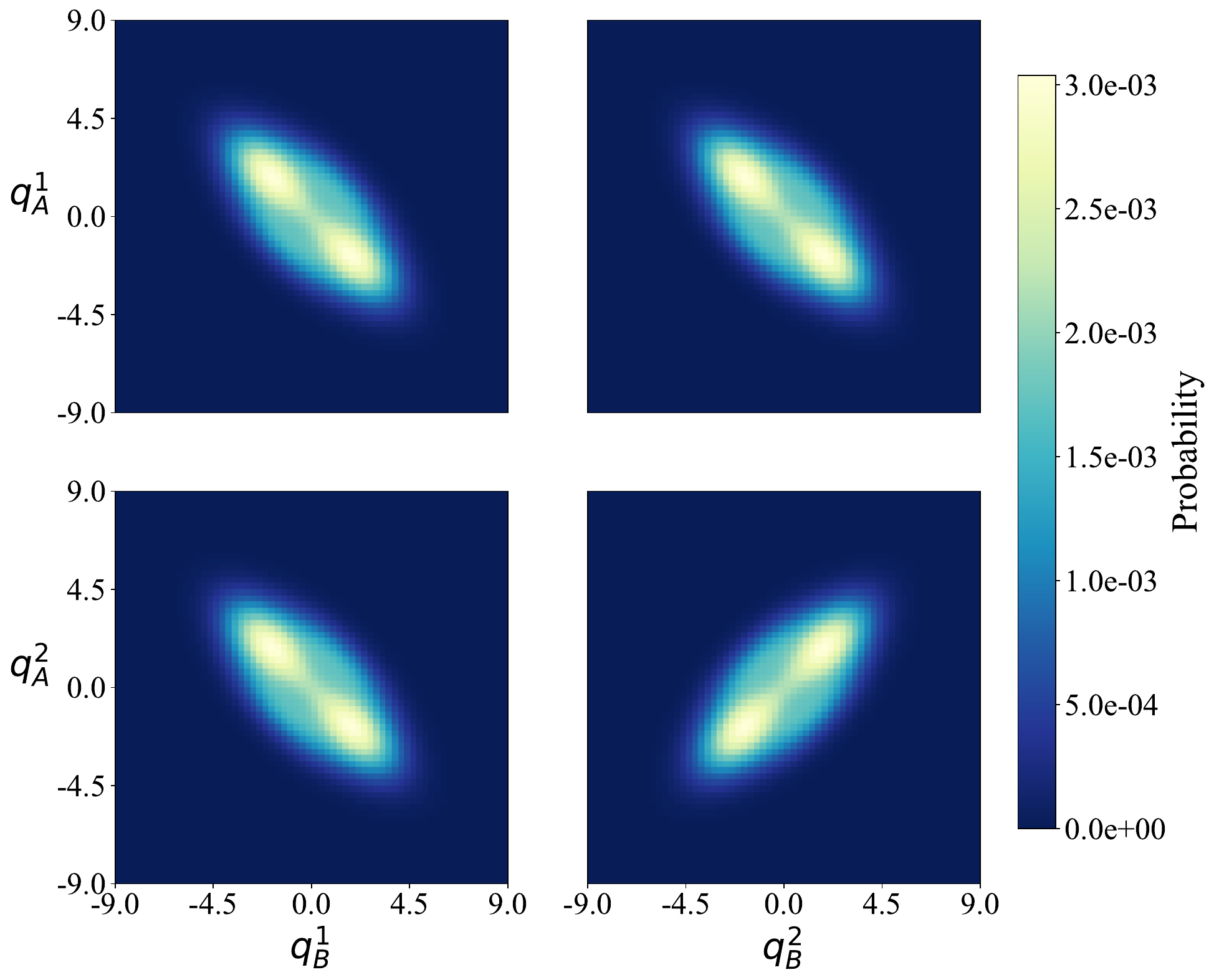}\includegraphics[width =0.45\linewidth]{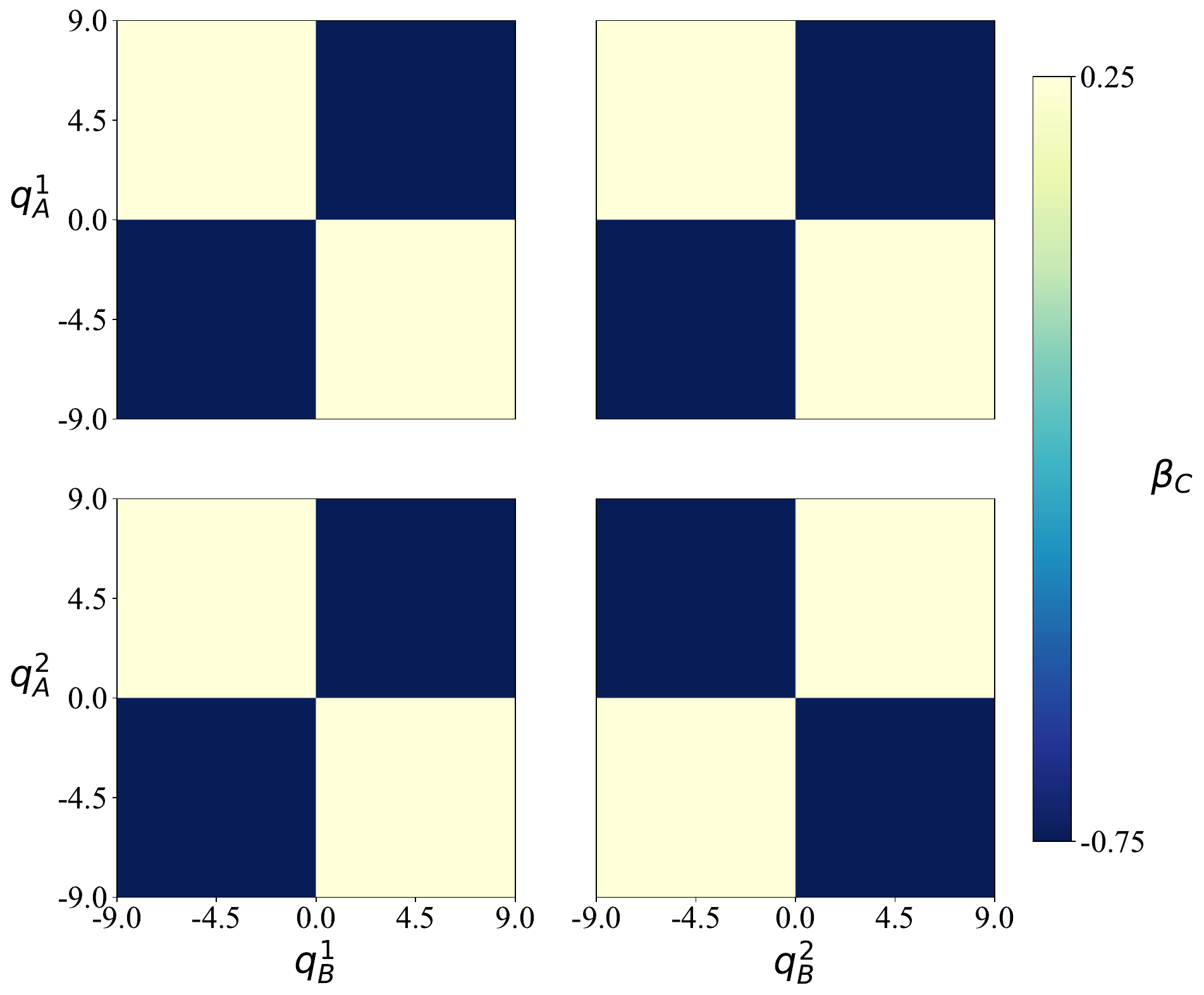}
\caption{On the left the empirical distribution corresponding to the homodyne statistics obtained by measuring settings $\phi_A\in \{0,\pi/2\}$ and $\phi_B\in\{-\pi/4,\pi/4\}$, on a two photon-subtracted state, from the family described in equation \eqref{eq:two_photon_subtracted_family1}. This corresponds to the case $r_1=-r_2=0.68$, $\theta=\pi/2$. On the right: the \textit{filter} functions, $\beta_C$, that define the optimal Bell inequality for the empirical model considered. }
\label{fig:empirical_model_and_bell_filters_phton_subtr}
\end{figure}

More precisely, we consider the following family of states: 
\begin{equation}\label{eq:two_photon_subtracted_family1}
   |\psi\rangle=\left(\hat a_1 \cos \theta+ \hat a_2\sin \theta\right)\hat a_1\hat O(\pi/4) \hat S_1(r_1)\hat S_2(r_2)|0\rangle,
\end{equation}
where $\hat O(\pi/4)$ represents a balanced beam splitter operation, $\hat S_i(r_i)$ represents single-mode squeezers \cite{Bachor2019}, and $\hat a_i$ stands for the annihilation operator corresponding to mode $i$ \cite{Treps2020}. In Fig.~\ref{fig:CF_photon_subtr_vs_theta} we can observe the behavior of the contextual fraction for these states, as the mode on which the second photon is subtracted is changed, according to the parameter $\theta$. This allows us to interpolate between the case in which both photons are subtracted on the same mode $\theta=0$, and the one in which they are subtracted in complementary modes $\theta=\pi/2$. The results obtained for those extreme cases coincide with those reported in \cite{RGP1,RGP2}. There, the Bell value for the CHSH for the case of $\theta=\pi/2$ is $B=2.046$, which corresponds to a fractional violation $\mu=\frac{B-2}{B_M-2}=0.023$, similar to the value we obtain. The plot is constructed using $r_1=-r_2=0.68$, which provides the maximum Bell violation in \cite{RGP1,RGP2}.

Further, in Fig.~\ref{fig:empirical_model_and_bell_filters_phton_subtr}
we can observe the empirical distributions for which the maximum contextual fraction was obtained. Alongside we show the \textit{filter} functions $\beta_C$, that parametrize the corresponding Bell inequality. We can see that the distributions bear a strong resemblance to PR distributions, and thus the corresponding Bell inequality is the same. Such filters precisely correspond, after some renormalization steps, to the sign binning procedure employed in \cite{RGP1,RGP2}, which thus appears to be optimal, as they had conjectured.

Several other examples of two-photon-subtracted states were considered, without major improvement over the results described in \cite{RGP1,RGP2}. Moreover, we explored the case of three-photon subtraction, for which no Bell inequality violation was reported in that paper, and we were not able to find any numerically significant contextual fraction. The fact that specifically three photon subtractions not only do not increase the contextual fraction for homodyne measurements, but actually reduce it to zero, is a rather counterintuitive result, for which we do not have a convincing explanation and which may deserve further study.

\subsubsection{Multi-peak states}
In what follows we consider multi-peak states, described in \cite{Fred_2003}. Several states were considered in that paper, of the form:
\begin{equation}\label{eq:Fred_state_general}
|\psi\rangle=\frac{|ff\rangle+e^{\frac{i\pi}{4}}|gg\rangle}{\sqrt{2}},
\end{equation}
where $|f\rangle, |g\rangle$ are two orthogonal (or approximately orthogonal) state vectors. In particular, finite photon number superpositions were considered, taking:
\begin{equation}\label{eq:Fred_state_basis}
    \begin{split}
& |f\rangle= \sqrt{0.459} |0\rangle -\sqrt{0.491}|4\rangle-\sqrt{0.008}
|8\rangle-\sqrt{0.042}|12\rangle \\
& |g\rangle= \sqrt{0.729} |1\rangle +\sqrt{0.155}|5\rangle-\sqrt{0.107}
|9\rangle-\sqrt{0.009}|13\rangle.
\end{split}
\end{equation}

\begin{figure}[htbp]
\centering
\includegraphics[width =0.45\linewidth]{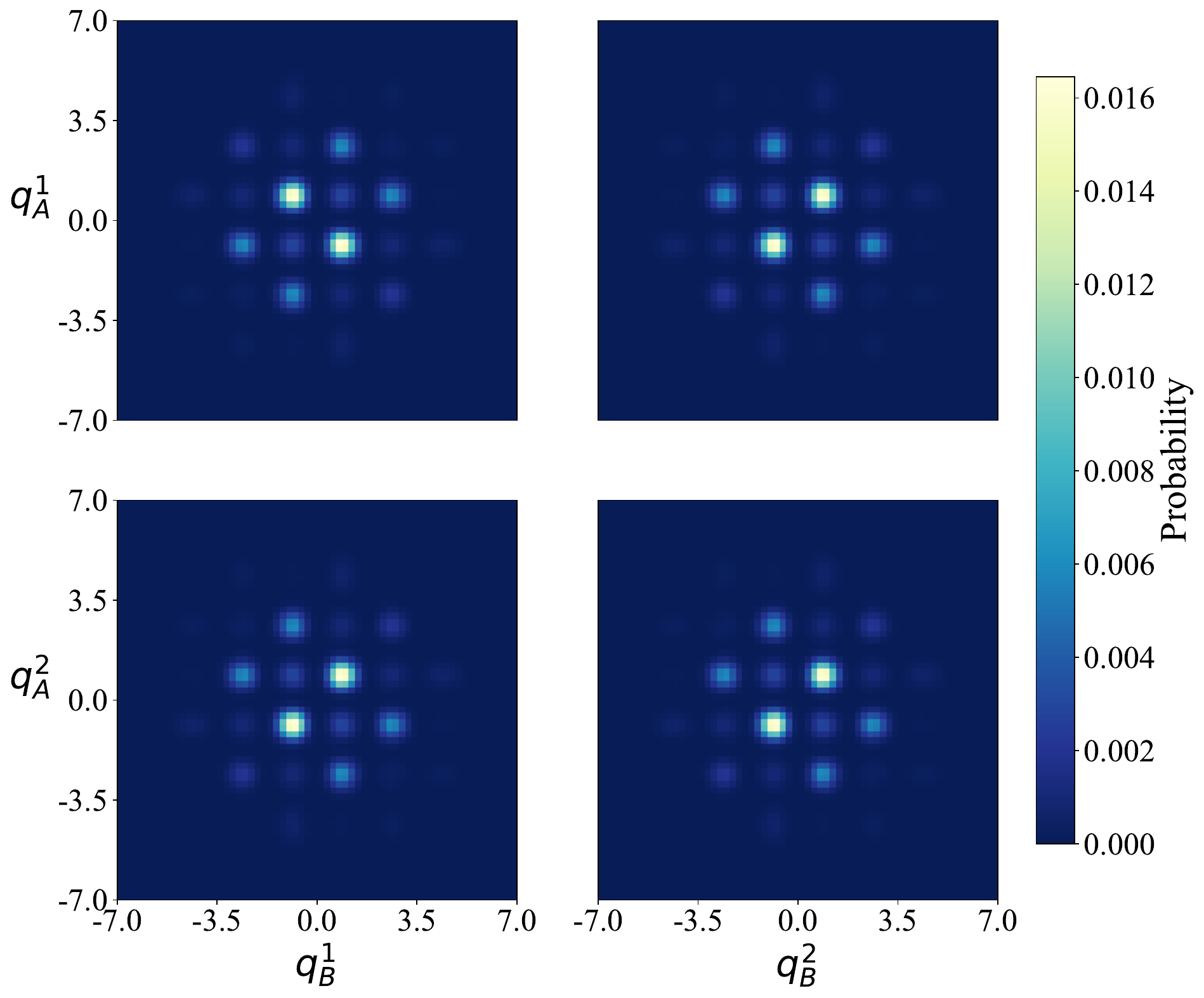}\includegraphics[width =0.45\linewidth]{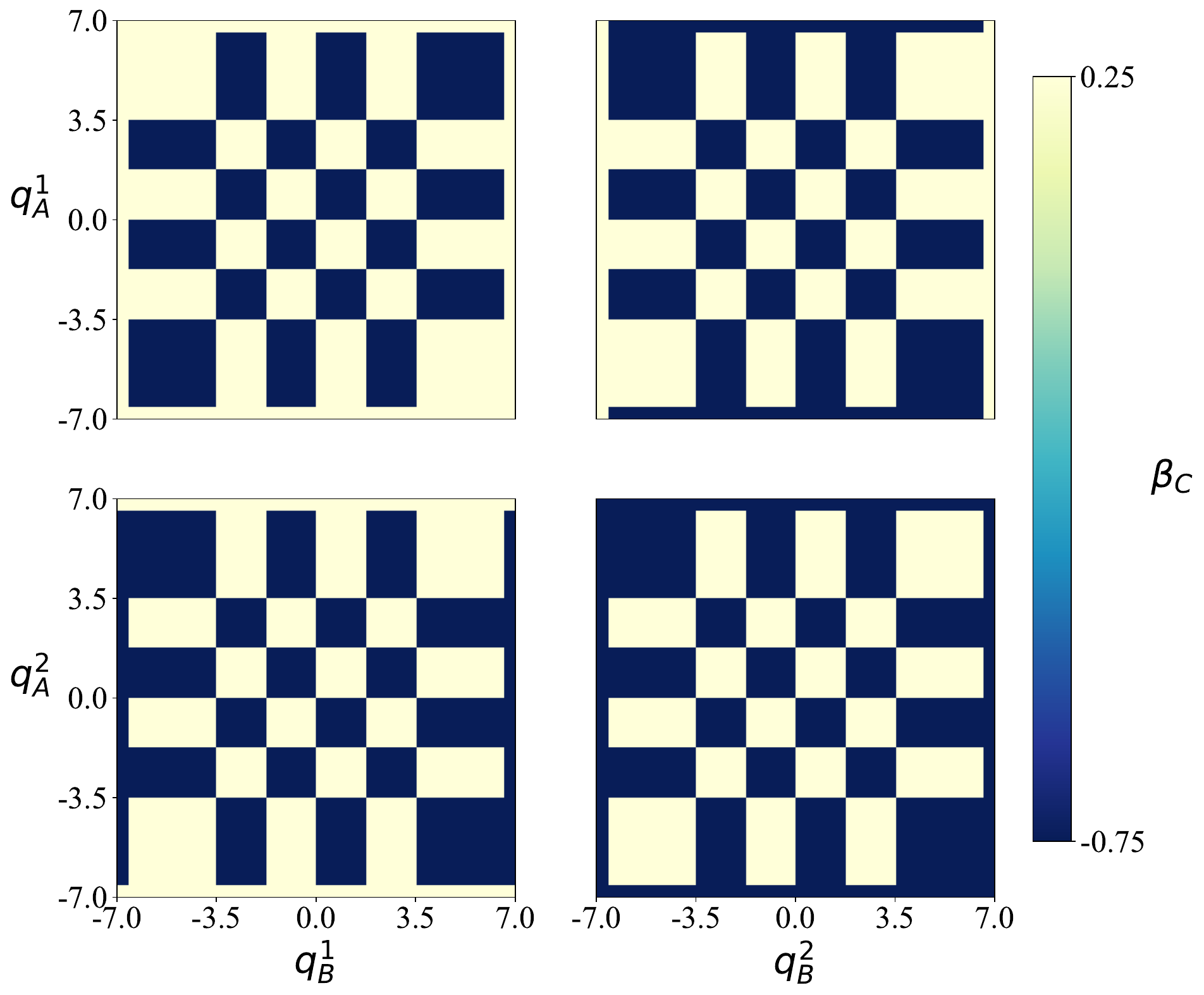}
\caption{On the left the empirical distribution corresponding to the homodyne statistics obtained by measuring settings $\phi_A\in \{0,\pi/2\}$ and $\phi_B\in\{0,\pi/2\}$, on the state described jointly by equations \eqref{eq:Fred_state_general} and \eqref{eq:Fred_state_basis}. On the right: the \textit{filter} functions, $\beta_C$, that define the optimal Bell inequality for the empirical model considered. }
\label{fig:empirical_model_and_Bell_filter_Fred_state}
\end{figure}
In Fig.~\ref{fig:empirical_model_and_Bell_filter_Fred_state}, we show the empirical behavior and the Bell filters obtained through the implementation of the linear program. The contextual fraction obtained is $\approx0.404$, remarkably close to the Tsirelson bound \cite{tsirelson_quantum_1987}. Moreover this value is similar to the value obtained in \cite{Fred_2003}, implying that the Bell inequality used was optimal, as it coincides with the Bell inequality parametrized by the filter functions $\beta_C$ in Fig.~\ref{fig:empirical_model_and_Bell_filter_Fred_state}. Further simplifications of the states described above, with a smaller number of photons, which were already discussed in \cite{Fred_2003}, still lead to a significant contextual fraction. Such states could be produced in the lab, using Boson Sampling-like strategies, following the results in \cite{andrei2025}. 
\par 
Other states with a finite number of photons, such as those considered in \cite{Enky_Gael_2024} have been explored, yielding results similar to those reported by relying on CHSH-based strategies. We also analyzed the three-mode states considered in \cite{Acin2009}, for which a binary strategy also turns out to be optimal. 
\section{A basic introduction to GKP states}\label{app:GKP_states}
Only a few notions about the encoding of qubits into states of an oscillator using GKP states \cite{GKP_2001} are necessary for our discussion. The basic idea behind this kind of encoding is to exploit the infinite-dimensional Hilbert space of the bosonic system to redundantly encode the information of a finite-dimensional Hilbert space. This allows us to encode qubit states  as well as qudit states. Other bosonic codes can be considered \cite{Albert2025}, but we focus on GKP states, for which, as we will see, quadrature measurements correspond to measuring Pauli observables. \par 

The proposal of Gottesman, Kitaev and Preskill \cite{GKP_2001} consisted in encoding the logical codewords $|0\rangle$ and $|1\rangle$ into infinite superpositions of position eigenstates, as 
\begin{equation}\label{eq:GKP_encoding}
    \begin{split}
        & |0\rangle_{L}\propto \sum_{k\in \mathbb Z} |k \alpha\rangle_q\\ 
        & |1\rangle_L\propto \sum_{k\in \mathbb Z} \left|(2k+1)\frac{\alpha}{2}\right>_q,
    \end{split}
\end{equation}
where $|\beta\rangle_q$ are position eigenstates, \ie $\hat q |\beta\rangle_q=\beta |\beta\rangle_q$. These states are not normalizable and are thus not physically realizable. However, they can be regularized to finite-energy versions \cite{GKP_2001}. Both for $|0\rangle_L$ and $|1\rangle_L$ the spacing between different peaks is given by the parameter $\alpha$, with the $|1\rangle_L$ state offset by $\alpha/2$ with respect to the zero logical state (see Fig.~\ref{fig:CF_vs_Delta}D for example). The $q$ quadrature wave-function is given by 
\begin{equation}\label{eq:GKP_q_wave_functions}
\begin{split}
    &\psi_0(q)\propto \sum_k \delta(q-k\alpha) \\
    & \psi_1(q)\propto \sum_k \delta\left(q-(2k+1)\alpha/2\right),
\end{split}
\end{equation}
where $\delta(q-\beta)$ is the Dirac delta distribution centered at the value $\beta$. The corresponding $p$-quadrature wave functions, given by their Fourier transforms, are
\begin{equation}\label{eq:p_wave_functions}
    \begin{split}
    &\tilde\psi_0(p)\propto \sum_k \delta(p-k\frac{2\pi}{\alpha}) \\
    & \tilde\psi_1(p)\propto \sum_k (-1)^k \delta\left(p-k \frac{2 \pi}{\alpha}\right),
\end{split}
\end{equation}
where, for convenience, we have assumed $\left[\hat q, \hat p \right]=i$.
\par
Logical codewords are eigenstates of the Pauli operator $Z$, such that $Z|n\rangle=(-1)^{n}|n\rangle$. The bosonic equivalent of the Pauli $Z$ operator is given by 
\begin{equation}\label{eq:Pauli_Z}
    \hat Z_L=e^{2 \pi i \frac{\hat q}{\alpha} },
\end{equation}
such that 
\begin{equation*}
\begin{split}
    \hat Z_L |j\rangle_L&=\sum_k e^{\pi i (2k+j)\alpha} |(2k+j)\alpha/2\rangle\\
    &=(-1)^j|j\rangle_L.
\end{split}
\end{equation*}
 Notice that the logical operator $\hat Z_L$ corresponds to a displacement along the $p$-quadrature by $\frac{2\pi}{\alpha}$. This directly hints at the fact that the Pauli $X$ eigenstates, $(|0\rangle \pm |1\rangle)/\sqrt{2}$, are given by superpositions of $p$-eigenstates spaced by $\frac{4 \pi}{\alpha}$. Indeed, from Eq.\eqref{eq:p_wave_functions} we have 
\begin{equation}\label{eq:Pauli_X_eigenstates}
    \begin{split}
        &\tilde \psi_0(p)+\tilde \psi_1(p)=\sum_k \delta(p-2 k \frac{2\pi}{\alpha})\\
        & \tilde \psi_0(p)-\tilde \psi_1(p)=\sum_k \delta(p-(2 k+1) \frac{2\pi}{\alpha}),
    \end{split}
\end{equation}
and thus the logical Pauli $X$ eigenstates are given by 
\begin{equation}
    |X=j\rangle\propto\sum_k \left|(2k+j)\frac{2\pi}{\alpha}\right>_p,
\end{equation}
where $|\beta\rangle_p$ are the eigenstates of the $p$ quadrature operator, \ie $\hat p |\beta\rangle_p=\beta |\beta\rangle_p$. The spacing between the different peaks is given by $\frac{4\pi}{\alpha}$, and the two logical states are offset by a $\frac{2 \pi}{\alpha}$ displacement. The Pauli $X_L$ operator is, similarly to $Z_L$, given by 
\begin{equation}\label{eq:Pauli_X_GKP}
    X_L=e^{i \frac{\alpha\hat p}{2}},
\end{equation}
which corresponds to a displacement along the $\hat q$ quadrature by $\alpha/2$. The definitions of the Pauli logical operators on the GKP encoding through displacements in phase space, make it possible to define the logical Pauli $Y$ operator, which is given by
\begin{equation}\label{eq:Pauli_Y_GKP}
    Y_L=i X_L Z_L=ie^{i \frac{\alpha \hat p }{2}}e^{2 \pi i \frac{ \hat q }{\alpha}}=e^{i \left(\frac{\alpha \hat p}{2}+2 \pi \frac{\hat q}{\alpha}\right)},
\end{equation}
which corresponds to a displacement by $\delta=\sqrt{\frac{4 \pi^2}{\alpha^2}+\frac{\alpha^2}{4}}$, along a rotated quadrature $\hat q_\phi$, with $\phi=\arctan{\frac{\alpha^2}{4 \pi^2}}$. Like the Pauli $X$ and $Z$ operators, the Pauli $Y$ operator can be measured through homodyne detection of the quadrature $\hat q_{\phi^\prime}$, with $\phi^\prime=\arctan{\frac{\alpha^2}{4 \pi^2}}$. For the canonical choice of GKP qubit encoding, given by $\alpha=2\sqrt{\pi}$, that treats the $q$ and $p$ quadratures symmetrically, the corresponding quadrature is $\hat q_{\pi/4}=(\hat q+\hat p)/\sqrt{2}$. Notice that in this case the inter-peak spacing in both quadratures $\hat q$ and $\hat p$ is given by $2\sqrt{\pi}$, and the spacing along the rotated quadrature is given by $\delta=2\sqrt{2\pi}$. \par 
The ideal states we have described in this section are not normalized, and therefore are unphysical. Finite-energy approximations of these states can be constructed in different ways, all of which basically substitute quadrature eigenstates by finitely squeezed states, and enforce some damping on the coefficients of each peak as their amplitude increases. In Eq.\eqref{eq:finite_energy_GKP} we present one choice of description of finite-energy GKP states, which was considered in the original GKP paper \cite{GKP_2001}. 
\section{A family of two-mode entangled GKP states for homodyne-based Bell inequality violation}\label{app:entangled_GKP}
In the main text we considered a specific example (given in Eq.~\eqref{eq:entangled_GKP}) of a two-mode entangled GKP state for probing non-locality with homodyne detection. Nevertheless, as we mentioned there, it is only one of many different possible choices. A neat characterization of this family of states is provided in this section. \par 
The construction we describe here is based on three main observations. 
\begin{itemize}
    \item All maximally entangled two-qubit states can be obtained by local rotations from Bell states $|\Phi_+\rangle$.
    \item Bell states saturate Tsirelson's bound for measurements that lie along orthogonal axes of the Bloch spehere, and can be jointly rotated to a pair of Pauli measurements. 
    \item The equivalent of Pauli measurements $\{Z,X,Y\}$ in GKP encoded states are quadrature measurements $\{\hat q, \hat p, \hat q_{\pi/4}=\frac{\hat q+\hat p}{\sqrt{2}}\}$ respectively. 
 \end{itemize}
 If we consider the Bell state 
 \begin{equation}
     |\Phi_+\rangle=\frac{|00\rangle+|11\rangle}{\sqrt{2}},
 \end{equation}
 it saturates the Tsirelson bound for a measurement scenario where both parties measure 
 \begin{equation}\label{eq:Bell_optimal_meas}
     X\cos \phi_i +Y\sin \phi_i,
 \end{equation}
 choosing the rotation angle $\phi_i\in \{\pi/8,5 \pi/8\}$. These can be unitarily related to any pair of main axes of the Bloch sphere $(X,Y), (X,Z), (Z,Y)$. Applying the corresponding unitary also to the Bell state $|\Phi_+\rangle$ allows us to find a state that saturates the Tsirelson bound for the corresponding choice of Pauli measurements. We consider two choices for illustration 
 \begin{equation}
 \begin{split}
     &M_1=\{X,Z\}\\
     & M_2=\{X,Y\}.
\end{split}
 \end{equation}
The transformations that map the choices of measurements in Eq.~\eqref{eq:Bell_optimal_meas} to each of these choices are denoted $U_{M_1}$ and $U_{M_2}$, and are given by 
\begin{equation}
    \begin{split}
        &U_{M_1}=e^{i\frac{\pi}{4}X}e^{-i \frac{\pi}{16} Z} \\
        & U_{M_2}=e^{-i \frac{\pi}{16} Z}.
    \end{split}
\end{equation}
The optimal states corresponding to each side measuring the same settings, either from $M_1$ or from $M_2$, are given by 
\begin{equation}
    |\psi_i\rangle=U_{M_i}|\Phi_+\rangle,
\end{equation}
which results in 
\begin{equation}\label{eq:entangled_GKP_app}
\begin{split}
    &|\psi_1\rangle = \frac{1}{\sqrt{2}}\left(\sin(\frac{\pi}{8}) (|00\rangle-|11\rangle)+\cos(\frac{\pi}{8})(|01\rangle+|10\rangle)\right) \\
    &|\psi_2\rangle=\frac{|00\rangle +e^{-i\pi/4}|11\rangle}{\sqrt{2}}.
\end{split}
\end{equation}

\begin{figure}[htbp]
\centering
\includegraphics[width =0.85\linewidth]{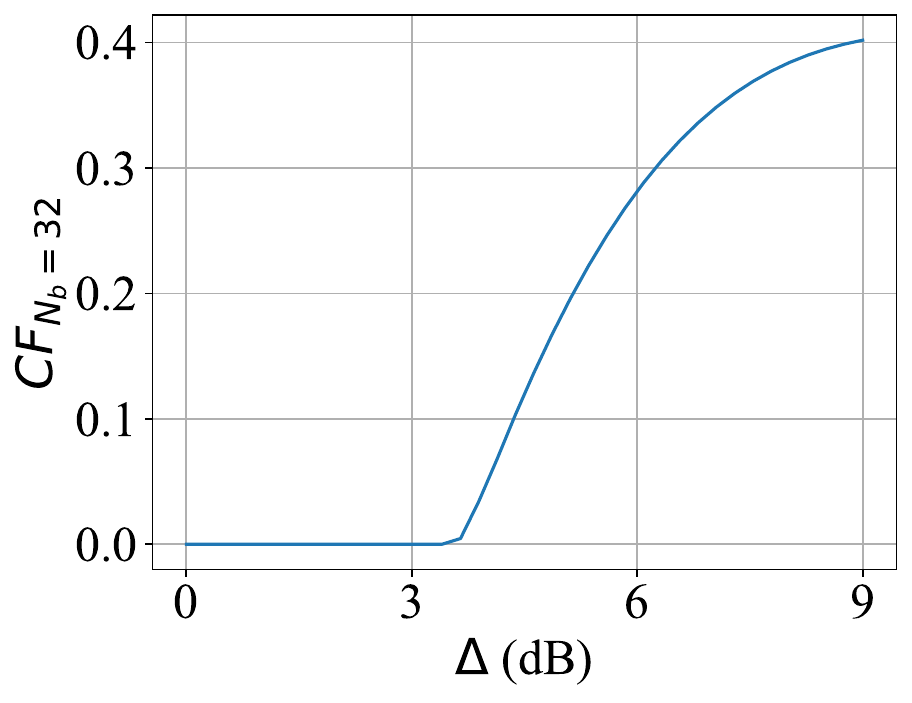}
\caption{Contextual fraction obtained as a function of the parameter $\Delta$ that governs the quality of the GKP encoding, for the entangled GKP state $|\psi_2\rangle$ in  Eq.\eqref{eq:entangled_GKP_app}. We have considered in this plot the hierarchy at the level $k=5$, \ie considering $N_b = 32$ bins per quadrature. }
\label{fig:CF_vs_Delta_psi_2}
\end{figure}

It is the former of these states that we consider in the main text. Nevertheless, it is the latter that has the simplest description. As we can observe by comparing Fig.~\ref{fig:CF_vs_Delta_psi_2} to Fig.~\ref{fig:CF_vs_Delta}, this comes at the cost of requiring encoded states with  a better quality. Indeed, for a fixed spacing of the GKP encoding, here $\alpha=2\sqrt{\pi}$, the onset of non-locality for $|\psi_2\rangle$ happens for states with a larger squeezing $\Delta_{dB}$, compared to $|\psi_1\rangle$. Nevertheless, the simpler structure of the state makes it relevant for experiments where already very high quality GKP states can be prepared like \cite{Fontbote2026}. 

\section{Optimal qutrit state for CGLMP Bell inequality using Pauli measurements}\label{app:qutrit}
In \cite{CGLMP2002}, the authors propose explicit expressions for states and qudit measurement results. They show that these states violate some Bell inequalities that they propose. The measurements that they propose to perform are not Pauli measurements, which would be desired when trying to find a mapping to a CV encoding. Also, the Bell inequality is not in operator form, but rather constructed in a way similar to the one we use in this paper, by making the filters that parametrize the Bell inequality explicit for each outcome. We divide the procedure to find an optimal CV realization into the following steps: 
\begin{enumerate}
    \item Find an explicit operator expression for the CGLMP Bell inequality for the optimal states and measurements. 
    \item Use the same expression for constructing the Bell operator but using generalized Pauli operators $X,Z$, instead of the optimal measurements, and construct the ``Pauli" Bell operator. For these operators, find the maximum eigenvalue and its corresponding eigenvector, which parametrizes the optimal state for this Bell inequality. 
    \item Construct a CV-encoded version of this state using a qutrit GKP state. 
\end{enumerate}
The procedure described above is implemented in one of the examples in the repository \cite{repo}. For the first step, we used the moment-based SDP on the optimal distributions considered in \cite{CGLMP2002}, which becomes exact once moments up to order 6 are considered, in the cases of qutrit systems and two measurements per party. The explicit expression for the qudit state obtained is 
\[
\begin{aligned}
|\psi\rangle
= {} &
(-0.22081098 + 0.382455814\,i)\,|0,0\rangle \\
&+ (0.01050768 - 0.164565746\,i)\,|0,1\rangle \\
&+ (0.31141613 + 0.118874310\,i)\,|0,2\rangle \\
&+ (0.31141611 + 0.118874303\,i)\,|1,0\rangle \\
&+ 0.44162192\,|1,1\rangle \\
&+ (0.13726426 + 0.0913827838\,i)\,|1,2\rangle \\
&+ (0.01050767 - 0.164565737\,i)\,|2,0\rangle \\
&+ (-0.25865624 + 0.210257131\,i)\,|2,1\rangle \\
&+ (0.44162195)\,|2,2\rangle.
\end{aligned}
\]
\end{document}